\documentclass[12pt]{article}
\usepackage{epsfig}
\usepackage{epstopdf}
\usepackage{graphicx}
\usepackage{upgreek}
\usepackage{amssymb}
\usepackage{amsmath,amsfonts,amsthm}
\usepackage[margin=1in]{geometry}
\usepackage{array}
\usepackage{caption}
\usepackage{subfig}
\usepackage{cite} 

\newcommand{\q}{\mathfrak{q}}

\newtheorem{Thm}{Theorem}[section]
\newtheorem{proposition}{Proposition}[section]

\newtheorem{Lem}{Lemma}[section]
\newtheorem{Def}{Definition}[section]

\usepackage[english]{babel}
\usepackage[font=small,format=plain,labelfont=bf]{caption}

\usepackage{hyperref}       
\hypersetup{       
   pdftex,                  
   colorlinks=true,        
   pdfstartview=FitH,      
   allcolors=[rgb]{0,0.3,0.6},        
   bookmarks=true,          
	 bookmarksdepth=section, 
   bookmarksopen=false,     
   bookmarksnumbered=true, 
	 backref, pagebackref   
}

\def\begeq{\begin{equation}}
\def\endeq{\end{equation}}

\def\begeqar{\begin{eqnarray}}
\def\endeqar{\end{eqnarray}}

\def\a{\alpha}
\def\g{\gamma}
\def\<{\langle}
\def\>{\rangle}

\def\Vir{{Vir}\otimes \overline{Vir}}
\newcommand{\be}{\begin{equation}}
\newcommand{\ba}{\begin{eqnarray}}
\newcommand{\ea}{\end{eqnarray}}
\newcommand{\ee}{\end{equation}}

\newcommand{\half}{\frac{1}{2}}

\newcommand{\Z}{\mathsf{Z}(\mathsf{aTL})}

\newcommand{\XXX}[1]{}

\newcommand{\atl}[1]{\mathsf{aTL}_{#1}}
\newcommand{\tl}[1]{\mathsf{TL}_{#1}}

\newcommand{\aTL}{\mathsf{aTL}}

\def\begeq{\begin{equation}}
\def\endeq{\end{equation}}

\def\begeqar{\begin{eqnarray}}
\def\endeqar{\end{eqnarray}}

\def\a{\alpha}

\def\r{\vec{r}}

\usepackage{tikz}
\usetikzlibrary{decorations.pathreplacing}
\usetikzlibrary{decorations.markings}

\newcounter{braid}
\newcounter{strands}
\pgfkeyssetvalue{/tikz/braid height}{1cm}
\pgfkeyssetvalue{/tikz/braid width}{1cm}
\pgfkeyssetvalue{/tikz/braid start}{(0,0)}
\pgfkeyssetvalue{/tikz/braid colour}{black}
\pgfkeys{/tikz/strands/.code={\setcounter{strands}{#1}}}

\makeatletter
\def\cross{%
  \@ifnextchar^{\message{Got sup}\cross@sup}{\cross@sub}}

\def\cross@sup^#1_#2{\render@cross{#2}{#1}}

\def\cross@sub_#1{\@ifnextchar^{\cross@@sub{#1}}{\render@cross{#1}{1}}}

\def\cross@@sub#1^#2{\render@cross{#1}{#2}}

\def\render@cross#1#2{
  \def\strand{#1}
  \def\crossing{#2}
  \pgfmathsetmacro{\cross@y}{-\value{braid}*\braid@h}
  \pgfmathtruncatemacro{\nextstrand}{#1+1}
  \foreach \thread in {1,...,\value{strands}}
  {
    \pgfmathsetmacro{\strand@x}{\thread * \braid@w}
    \ifnum\thread=\strand
    \pgfmathsetmacro{\over@x}{\strand * \braid@w + .5*(1 - \crossing) * \braid@w}
    \pgfmathsetmacro{\under@x}{\strand * \braid@w + .5*(1 + \crossing) * \braid@w}
    \draw[braid] \pgfkeysvalueof{/tikz/braid start} +(\under@x pt,\cross@y pt) to[out=-90,in=90] +(\over@x pt,\cross@y pt -\braid@h);
    \draw[braid] \pgfkeysvalueof{/tikz/braid start} +(\over@x pt,\cross@y pt) to[out=-90,in=90] +(\under@x pt,\cross@y pt -\braid@h);
    \else
    \ifnum\thread=\nextstrand
    \else
     \draw[braid] \pgfkeysvalueof{/tikz/braid start} ++(\strand@x pt,\cross@y pt) -- ++(0,-\braid@h);
    \fi
   \fi
  }
  \stepcounter{braid}
}

\tikzset{braid/.style={double=\pgfkeysvalueof{/tikz/braid colour},double distance=1pt,line width=2pt,white}}

\newcommand{\ffrac}[2]{\mbox{\footnotesize$\displaystyle\frac{#1}{#2}$}}

\newcommand{\braid}[2][]{%
  \begingroup
  \pgfkeys{/tikz/strands=2}
  \tikzset{#1}
  \pgfkeysgetvalue{/tikz/braid width}{\braid@w}
  \pgfkeysgetvalue{/tikz/braid height}{\braid@h}
  \setcounter{braid}{0}
  \let\g=\cross
  #2
  \endgroup
}
\makeatother

\newcommand{\utiles}[1]{
\begin{tikzpicture}[scale=1/2]
\filldraw[white] (0,0) -- (1,1) -- (2 ,0) -- (1 , -1 ) -- (0, 0);
\draw[line width = 2] (0,0) -- (1,1) -- (2 ,0) -- (1 , -1 ) -- (0, 0);
\draw (0.2357,  - 0.2357) arc (-45:45:1/3);
\node (center) at (1 , 0) {$\scriptstyle{#1}$};
\end{tikzpicture}
}
\newcommand{\etiles}{
\begin{tikzpicture}[scale=1/2]
\filldraw[white] (0,0) -- (1,1) -- (2 ,0) -- (1 , -1 ) -- (0, 0);
\draw[line width = 2] (0,0) -- (1,1) -- (2 ,0) -- (1 , -1 ) -- (0, 0);
\draw[line width = 1] (.5,-.5) arc (-45:45:.707);
\draw[line width = 1] (1.5,-.5) arc (225:135:.707);
\end{tikzpicture}
}
\newcommand{\idtiles}{
\begin{tikzpicture}[scale=1/2,rotate=90]
\filldraw[white] (0,0) -- (1,1) -- (2 ,0) -- (1 , -1 ) -- (0, 0);
\draw[line width = 2] (0,0) -- (1,1) -- (2 ,0) -- (1 , -1 ) -- (0, 0);
\draw[line width = 1] (.5,-.5) arc (-45:45:.707);
\draw[line width = 1] (1.5,-.5) arc (225:135:.707);
\end{tikzpicture}
}

\newcommand{\rutiles}[2]{
\begin{tikzpicture}[scale=1/2,rotate = #2]
\filldraw[white] (0,0) -- (1,1) -- (2 ,0) -- (1 , -1 ) -- (0, 0);
\draw[line width = 2] (0,0) -- (1,1) -- (2 ,0) -- (1 , -1 ) -- (0, 0);
\draw (0.2357,  - 0.2357) arc (-45:45:1/3);
\node (center) at (1 , 0) {$\scriptstyle{#1}$};
\end{tikzpicture}
}
\newcommand{\correcttiles}[3]{
\begin{tikzpicture}[scale=#3,rotate = #2]
\filldraw[white] (0,0) -- (1,1) -- (2 ,0) -- (1 , -1 ) -- (0, 0);
\draw[line width = 2] (0,0) -- (1,1) -- (2 ,0) -- (1 , -1 ) -- (0, 0);
\draw (0.2357,  - 0.2357) arc (-45:45:1/3);
\node (center) at (1 , 0) {$\scriptstyle{#1}$};
\end{tikzpicture}
}

\usepackage{color}

\def\({\left(}		\def\){\right)}
\def\[{\left[}          \def\]{\right]}

\begin{document}
\thispagestyle{empty}

\begin{center}

\Large{Topological defects in lattice models\\ and affine Temperley--Lieb algebra}

\vskip 1cm

\normalsize{J. Bellet\^ete\,$^{1,7}$, A.M. Gainutdinov\,$^{2,6}$, J.L. Jacobsen\,$^{1,3,4}$, H. Saleur\,$^{1,5}$, T.S. Tavares\,$^1$}

\vspace{1.0cm}

{\sl\small $^1$  Institut de Physique Th\'eorique, Universit\'e Paris Saclay, CEA, CNRS, 91191 Gif-sur-Yvette, France\\}

{\sl\small $^2$
Institut Denis Poisson, CNRS, Universit\'e de Tours, Universit\'e d'Orl\'eans, \\Parc de Grandmont, 37200 Tours, France\\}

{\sl\small $^3$
Laboratoire de Physique Th\'eorique, D\'epartement de Physique de l'ENS,
  \'Ecole Normale Sup\'erieure, \\ Sorbonne Universit\'e, CNRS,
 PSL Research University, 75005 Paris, France\\}
 
{\sl\small $^4$
Sorbonne Universit\'e, \'Ecole Normale Sup\'erieure, CNRS, \\
 Laboratoire de Physique Th\'eorique (LPT ENS), 75005 Paris, France \\}

{\sl\small $^5$
Department of Physics,
  University of Southern California, Los Angeles, CA 90089-0484,
   USA \\}
 
 {\sl\small $^6$  
National Research University Higher School of Economics, Myasnitskaya Ulitsa, 20, Moscow, Russia \\}
 
 {\sl\small $^7$
Laboratoire de Physique Th\'eorique et Mod\'elisation, CY Cergy Paris Universit\'e, CNRS, F-95302 Cergy-Pontoise, France \\}
\end{center}


\begin{abstract}
This paper is the first in a series where we attempt to define defects in critical  lattice models  that give rise to conformal field theory topological defects in the continuum limit. We focus mostly on models based on the Temperley-Lieb algebra, with future applications to restricted solid-on-solid (also called anyonic chains) models, as well as non-unitary models like percolation or self-avoiding walks. Our approach is essentially algebraic and focusses on the defects from two points of view: the ``crossed channel" where the defect is seen as an operator acting on the Hilbert space of the models, and the ``direct channel" where it corresponds to a modification of the basic Hamiltonian with some sort of impurity. Algebraic characterizations and constructions are proposed in both points of view. In the crossed channel,
 this leads us to new results about the center of the affine Temperley-Lieb algebra; in particular we find there a special basis  with non-negative integer structure constants that are interpreted as fusion rules of defects. 
In the direct channel, meanwhile, this leads to the introduction of fusion products and fusion quotients, with interesting algebraic properties that allow to describe representations content of the lattice model with a defect, and  to describe its spectrum.
\end{abstract}


\newpage 

\tableofcontents

\newpage
\section{Introduction}

A defect---or interface---in conformal field theory is  generally defined as a non-contractible line separating two a priori different conformal field theories (CFTs), with matching conditions between the two sides of the line. Various situations can be encountered in this general context. We will restrict here to the case of so-called topological defects, where the two CFTs are identical, and the stress-energy tensor is continuous across the defect line. In this case, correlation functions for fields inserted away from the defect line are unchanged when the line is continuously deformed, as long as the line is not taken across the field insertions: hence the name ``topological''.

Defects  in CFT appear in a variety of physical problems, both in two-dimensional statistical mechanics, e.g.\ in the context of Kramers--Wannier duality~\cite{FFRS,FFRS1}, and in imaginary-time one-dimensional quantum mechanics, e.g.\ in the context of quantum impurity problems such as the Kondo problem~\cite{BG,OA}. The problem of classifying topological defects   has received considerable attention, in particular in the case of rational CFTs~\cite{PZ,Petkova}. For such theories with diagonal modular invariants, for instance, it has been shown that the set of defects is in bijection with the set of representations of the chiral algebra. Many results for non-diagonal invariants, or for non-rational unitary theories such as Liouville \cite{Sark} are also known. 

\smallskip
Meanwhile, the general question of relating structures within the CFTs with properties of underlying lattice models has also attracted much attention. Work in this direction has included attempts to define lattice versions of the Virasoro algebra \cite{KooSaleur,Vidal,ZW}, to define fusion of primary fields in terms or representation theory of lattice algebras \cite{ReadSaleur,GV,GS,GJS,BSA, BelleteteFusion}, to calculate modular transformations  from   lattice partition functions \cite{PS}, and to build topological defect lines directly on the lattice \cite{PearceMerkat, Fendley,tensor-net}. Many of these attempts drew from the pioneering works \cite{KadanoffCeva,AGR}.

\smallskip

The present work is motivated by our interest in non-unitary (in particular, logarithmic) conformal field theory (LCFT)~\cite{SpecialIssue}. Decisive progress has been realized in this difficult subject by turning to lattice models---in particular, to understand better the indecomposable properties of the Virasoro-algebra representations involved. In view of the close relationship between defects, primary fields and fusion in the unitary case, it is natural to continue the program set out in  \cite{Ourreview} by trying to define topological defects on the lattice using an algebraic approach. While such endeavor has been partially completed in the case of restricted solid-on-solid models---whose associated CFTs are rational, and which are closely related to ``anyonic'' spin chains \cite{Anyons, Anyons2,Fendley}---we will be interested here in the profoundly different case of  loop models, which provide regularizations of the simplest known LCFTs. This paper will discuss the first part of our study, where we will focus on the definition and mathematical properties of a certain kind of lattice topological defects.

\smallskip

\begin{figure}
\begin{center}
	$\hbox{crossed channel}  \qquad \qquad$
	\begin{tikzpicture}[scale = 1/3, baseline = {(current bounding box.center)}]
		\draw[black, line width = 1pt ] (-2,-2) -- (-2,2);
		\draw[black, line width = 1pt ] (2,-2) -- (2,2);
		\draw[black, line width = 1pt ] (-2,-2) .. controls (-2,-3) and (2,-3) .. (2,-2);
		\draw[black, line width = 1pt ] (-2,2) .. controls (-2,1) and (2,1) .. (2,2);
		\draw[black, line width = 1pt ] (-2,2) .. controls (-2,3) and (2,3) .. (2,2);
		\draw[black, line width = 1pt, ->] (0,-2.5) -- (0,-1);
		\draw[blue, line width = 1pt, dotted] (-2,0.25) .. controls (-2,-.75) and (2,-.75) .. (2,.25);
	\end{tikzpicture} $ \quad \equiv \quad $
	\begin{tikzpicture}[scale = 4/9, baseline = {(current bounding box.center)}]
		\def\a{3.2};
		\def\b{1.5};
		\def\Pi{3.14159265359};
		\draw[line width = 1pt, domain = 0:2*\Pi] plot ({\a*cos(\x r)}, {\b*sin(\x r)});
		\draw[line width = 1pt, domain = \Pi/4:3*\Pi/4] plot ({.5*\a*cos(\x r)}, {.5*\b*sin(\x r)-.5});
		\draw[line width = 1pt, domain =  -.5+5*\Pi/4: .5 + 7*\Pi/4] plot ({.5*\a*cos(\x r)}, {.5*\b*sin(\x r)+.5});
		\draw[dotted, blue, line width = 1pt] (0,-.25) .. controls (.5,-.5) and (-.5, -1) .. (0,-1.5);
		\draw[black, line width = 1pt, ->] (-2, 0) .. controls (-2,-.25) and (-1,-.75) .. (-.5,-.75);  
	\end{tikzpicture} \\
	$\hbox{direct channel}  \qquad \qquad$ \begin{tikzpicture}[scale = 1/3, baseline = {(current bounding box.center)}]
		\draw[black, line width = 1pt ] (-2,-2) -- (-2,2);
		\draw[black, line width = 1pt ] (2,-2) -- (2,2);
		\draw[blue, line width = 1pt, dotted] (0,-3) .. controls (-1,-1) and (1,1) .. (0,3);
		\filldraw[white] (-2,-2) .. controls (-2,-3) and (2,-3) .. (2,-2) -- (2,-3.5) -- (-2,-3.5) -- (-2,2);
		\filldraw[white] (-2,2) .. controls (-2, 1) and (2,1) .. (2,2) -- (2,3.5) -- (-2,3.5) -- (-2,2);
		\draw[black, line width = 1pt ] (-2,-2) .. controls (-2,-3) and (2,-3) .. (2,-2);
		\draw[black, line width = 1pt ] (-2,2) .. controls (-2,1) and (2,1) .. (2,2);
		\draw[black, line width = 1pt ] (-2,2) .. controls (-2,3) and (2,3) .. (2,2);
		\draw[black, line width = 1pt, ->] (1,-1.5) -- (1,.5);
	\end{tikzpicture} 
	$ \quad \equiv \quad$
	\begin{tikzpicture}[scale = 4/9, baseline = {(current bounding box.center)}]
		\def\a{3.2};
		\def\b{1.5};
		\def\Pi{3.14159265359};
		\draw[line width = 1pt, domain = 0:2*\Pi] plot ({\a*cos(\x r)}, {\b*sin(\x r)});
		\draw[line width = 1pt, domain = \Pi/4:3*\Pi/4] plot ({.5*\a*cos(\x r)}, {.5*\b*sin(\x r)-.5});
		\draw[line width = 1pt, domain =  -.5+5*\Pi/4: .5 + 7*\Pi/4] plot ({.5*\a*cos(\x r)}, {.5*\b*sin(\x r)+.5});
		\draw[dotted, blue, line width = 1pt] (0,-.25) .. controls (.5,-.5) and (-.5, -1) .. (0,-1.5);
		\draw[black, line width = 1pt, ->] (.5, -1.25) --  (.5,-.5);  
	\end{tikzpicture}
     \caption{The two possible geometries for a defect line after mapping the plane to the cylinder.  }\label{TopoZ}
\end{center}
\end{figure}
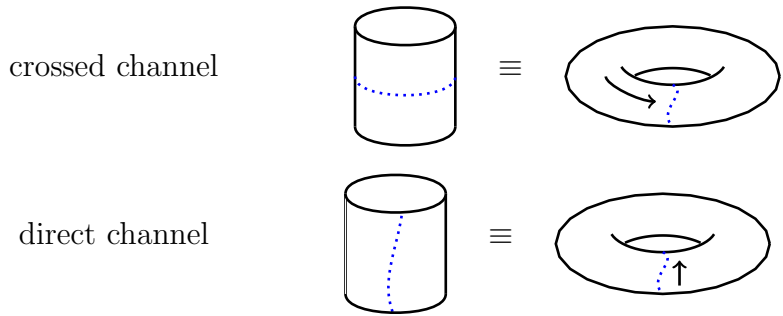

The correspondence between CFTs and lattice models is often best handled by thinking of the CFT in radial quantization, where, after  the usual logarithmic mapping, (imaginary) time propagation occurs along the axis of a cylinder, and space is periodic. In this point of view, the  non-contractible line for  the topological defect  can either run along the infinite cylinder, or be a non-contractible loop winding around it. We will refer to these two situations as a defect in the ``direct''  or in the ``crossed'' channel, respectively, see Fig.~\ref{TopoZ}.

In the crossed channel, the defect can be associated with an operator $X$ acting on the Hilbert space of the bulk CFT. 
The defect is topological if  $X$ commutes with the chiral $Vir$ and the anti-chiral $\overline{Vir}$ Virasoro generators~\cite{PZ,Petkova}:
\begin{equation}
[L_n,X]=0=[\bar{L}_n,X]\label{centVir}\ .
\end{equation}
%
Our strategy to identify the possible choices of operators $X$ is based on the identification of (representations of) the Virasoro algebra via the continuum limit of the Temperley-Lieb (TL) algebra---an idea that has been used in several works on related topics \cite{KooSaleur, GRS3,GRSV1}. This ``identification'' must be qualified. First, since we are dealing with bulk CFTs, we must think of the product of the chiral and anti-chiral Virasoro algebras, $\Vir$. Similarly, since the lattice models are defined on a cylinder, the proper lattice algebra is a ``periodicized'' version---the affine Temperley-Lieb algebra~$\aTL$: strictly speaking, the continuum limit of this algebra is known to be larger than $\Vir$, and has been identified as the ``interchiral algebra'' in~\cite{GRS3}. 

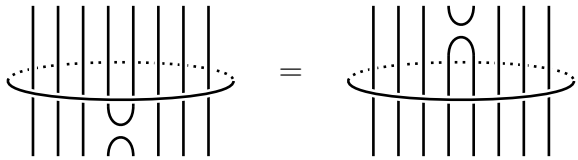
\begin{figure}
\begin{center}
	\begin{tikzpicture}[scale = 1/3, baseline = {(current bounding box.center)}]
	\foreach \i in {1,2,3,6,7,8}{
	\draw[line width = 1pt, black] (\i,-3) -- (\i,0);
	};
	\draw[line width = 1pt, black] (4,-1) -- (4,0);
	\draw[line width = 1pt, black] (5,-1) -- (5,0);
	\draw[line width = 1pt, black] (4,-3) .. controls (4,-2) and (5,-2) .. (5,-3);
	\draw[line width = 1pt, black] (4,-1) .. controls (4,-2) and (5,-2) .. (5,-1);
	\draw[line width = 3pt, white] (9,0) .. controls (9,-1) and (0,-1) .. (0,0);
	\draw[line width = 1pt, black] (9,0) .. controls (9,-1) and (0,-1) .. (0,0);
	\draw[line width = 1pt, black , dotted] (0,0) .. controls (0,1) and (9,1) .. (9,0);
	\foreach \i in {1,2,3,4,5,6,7,8}{
	\draw[line width = 3pt, white] (\i,3) -- (\i,0);	
	\draw[line width = 1pt, black] (\i,3) -- (\i,0);
	};
	\end{tikzpicture}  $\quad  = \quad $
	 \begin{tikzpicture}[scale = 1/3, baseline = {(current bounding box.center)}]
	\foreach \i in {1,2,3,4,5,6,7,8}{
	\draw[line width = 1pt, black] (\i,-3) -- (\i,0);
	};
	\draw[line width = 1pt, black , dotted] (0,0) .. controls (0,1) and (9,1) .. (9,0);
	\foreach \i in {1,2,3,6,7,8}{
	\draw[line width = 3pt, white] (\i,3) -- (\i,0);	
	\draw[line width = 1pt, black] (\i,3) -- (\i,0);
	};
	\draw[line width = 3pt, white] (4,0) -- (4,1);	
	\draw[line width = 1pt, black] (4,0) -- (4,1);
	\draw[line width = 3pt, white] (5,0) -- (5,1);	
	\draw[line width = 1pt, black] (5,0) -- (5,1);
	\draw[line width = 3pt, white] (4,1) .. controls (4,2) and (5,2) .. (5,2);
	\draw[line width = 1pt, black] (4,3) .. controls (4,2) and (5,2) .. (5,3);
	\draw[line width = 1pt, black] (4,1) .. controls (4,2) and (5,2) .. (5,1);
	\draw[line width = 3pt, white] (9,0) .. controls (9,-1) and (0,-1) .. (0,0);
	\draw[line width = 1pt, black] (9,0) .. controls (9,-1) and (0,-1) .. (0,0);
	\end{tikzpicture} 
     \caption{Commutativity of defect operator $Y$ with $e_j$ generators of $\aTL$.}\label{fig:Y-commTL}
\end{center}
\end{figure}

In the typical physical interpretation of the (affine) Temperley-Lieb algebras on $n$ sites, the nodes on the top and bottom of the TL diagrams should be interpreted as a chain of $n$ subsystems whose interactions are determined by the TL generators, but whose internal sub-structure is not---it is determined by the specific model chosen, which also fixes the $\atl{}$ representation corresponding to the chain.
 The simplest examples of these are the various kinds of spin-chains, like the twisted XXZ model. 
We will therefore start our search for lattice analogues of topological defects by demanding the closest lattice equivalent of~\eqref{centVir}, that is by looking for operators $X$  on the lattice that commute with the interactions in the chain, or, in a model-independent setting, that are central in $\aTL$. 
We will follow this  model-independent point of view on lattice defects as central elements satisfying certain nice properties, e.g.\ having a well-defined fusion. 
This is discussed in Section~\ref{sec:3} after the algebraic preliminaries of Section~\ref{sec:2} where we recall the usual definition of $\aTL$ together with a less standard formulation using a blobbed set of generators. In this last formulation, the lattice meaning of~$X$ turns out to be very simple: 
it just consists in passing a  line ``above'' or ``below'' the non-contractible loops by using solutions of the spectral-parameter independent Yang-Baxter equation exchanging spin-$1/2$ (the value relevant for bulk loops) and spin-$j$ (the value relevant for the defect lines) representations. The topological nature of this defect is obvious, as the Yang-Baxter equation allows one to move and deform  the defect line at will without changing neither the partition function, nor the correlation functions if operators are inserted.
The simplest example of such a defect operator $X$ is given by a diagram corresponding to a single non-contractible loop passing \textit{over} the bulk, see Fig.~\ref{fig:Y-commTL} where we denote this operator by~$Y$. This operator and its powers are manifestly in the center of the affine TL algebra. We define similarly operators~$\bar{Y}$ where the non-contractible loop  is passing  \textit{under} the bulk. The two operators generate an interesting algebra of defects.

Let us describe this type of defect operators in more precise mathematical terms. First of all, the $\aTL$ algebras depend on $n$ (the number of sites) and a loop parameter $\q+\q^{-1}$. One of our main mathematical results is Theorem~\ref{thm:main} stating  that  for any non-zero value of $\q$, that is not a root of unity, the two central elements $Y$ and $\bar{Y}$ generate the center $\Z$ of the affine TL algebra $\aTL$. As a by-product we obtain an expression of symmetric Laurent polynomials in the Jucys-Murphy elements in terms of $Y$ and $\bar{Y}$: the power-sum polynomials have a simple relation to the Chebychev polynomials in $Y$ and $\bar{Y}$ of 1st kind,  this is discussed in Section~\ref{sec:JM-elements}. Moreover, we show that for $\q$ a generic complex number\footnote{i.e.\ for $\q$ not a root of unity, we leave the root of unity case discussion for a forthcoming paper.} products of Chebyshev polynomials in $Y$ and $\bar{Y}$ provide a ``canonical" basis in $\Z$ with \textsl{non-negative integer} structure constants. The spin-$j$ defect operators mentioned above are defined as the $2j$-th Chebyshev polynomials and they thus form a special basis in the center, i.e.\ a product of two defect operators is decomposed onto defect operators again, and with non-negative multiplicities. These multiplicities are interpreted as fusion rules of the defects.

\smallskip

Of course, the line passing above or below the loops can as well be taken to run along the axis of the cylinder, i.e.\ along the time direction. This corresponds to having  the defect in the direct channel. In this setting, the presence of the defect line leads to a modified Hilbert space where an extra representation of spin $j$ is introduced, together with a Hamiltonian suitably modified by corresponding ``defect" terms. This is discussed in Section~\ref{sec:4} with the main result in Theorem~\ref{eq:thm-H}, where we relate spectral properties of such a modified Hamiltonian (which is hard to study directly) to a clear and precise algebraic construction within the representation theory of $\aTL$ algebras---namely the fusion product and fusion quotient. The first is based on a certain induction, while the second is dual to it and practically very convenient for actual calculations. In simple terms, the spectrum of the spin-$j$ defect Hamiltonian is given by the spectrum of the standard affine TL Hamiltonian with no defects however acting on 
the fusion quotient of an $\aTL$ representation by 
the spin-$j$ standard TL representation. The advantage of this construction is that it allows us to perform precise calculations, as we demonstrate in several examples, including the case of the twisted XXZ model. 

\smallskip

 In the last section~\ref{sec:5}, we provide conclusions and discuss a CFT interpretation together with further steps that will be discussed in the next papers, like the analysis of modular $S$-transformation in infinite lattices and the continuum limit from a more physical point of view. In Section~\ref{sec:5}, we also make an attempt to give a precise mathematical definition of lattice defects studied in this work.
Finally, several appendices contain proofs of our mathematical results and auxiliary calculations, such as examples of fusion products and fusion quotients.

\newcommand{\one}{\textbf{1}}

\section{The affine TL algebra and its center}\label{sec:2}
In this section, we fix our notations and conventions. We first give a definition of the affine Temperley-Lieb algebra $\aTL$ in terms of generators, and in terms of diagrams. We give the definition both in terms of the translation generator, which is very standard, and a new one in terms of the so-called blob and hoop generators; the blob formulation is significantly more convenient when discussing topological defects. We then discuss the standard modules and give the eigenvalues of the central elements on them.	The key result of this section is Theorem~\ref{thm:main} where we describe the center of $\aTL$.

\subsection{Two definitions}
The affine Temperley-Lieb algebras $\lbrace \atl{n}(\q) \rbrace$ form a family of infinite dimensional associative $\mathbb{C}$-algebras, indexed by a positive integer $n$ -- number of sites -- and a non-zero complex number $\q$. They can be defined in many ways but we chose three particular presentations for their relevance in physics. Each of these are described in terms of generators with relations and were chosen because they lighten the notation in particular sub-sections of this work. 

The first set of generators, which we shall refer to as the \emph{periodic} set of generators, is the one appearing in the original literature on these algebras: two \emph{shift} generators $u, u^{-1}$, and $n$ \emph{arc} generators $e_{1}, \hdots, e_{n}$, with the defining relations ($n > 2$) \cite{MartinSaleur2}
\begin{align}
	e_{i}e_{i} & = (\q + \q^{-1})e_{i},\notag\\
	 e_{i}e_{i \pm 1}e_{i} & = e_{i} , \notag\\
	  e_{i}e_{j} & = e_{j}e_{i} \; \text{ if } |i-j| \geq 2, \label{eq:period-gen-1}\\
	 u e_{i} &= e_{i+1} u,\notag \\
	 u^{2}e_{n-1} & = e_{1} \hdots e_{n-1}, \notag
\end{align}
which stands for all $i$, and we defined $e_{0} \equiv e_{n}$, $e_{n+1} \equiv e_{1}$. If $n = 2$, one must remove the relations $e_{i}e_{i\pm 1} e_{i} = e_{i}$, but the other relations are unchanged. If $n=1$, one must remove all the arc generators, keeping only the shift generators with the defining relations $u u^{-1} = u^{-1}u = 1$. One notices immediately that this set of generators is not minimal, since for instance $e_{i} = u^{i-1} e_{1} u^{1-i}$ for all $i\geq 1$. Furthermore, the elements $u^{\pm n}$ are both obviously central. The sub-algebra generated by $\lbrace e_{1}, \hdots, e_{n} \rbrace$ is often called the \emph{periodic} Temperley-Lieb algebra, while the one generated by $\lbrace e_{1}, \hdots, e_{n-1} \rbrace$ is called the \emph{regular} Temperley-Lieb algebra.

The second set of generators, which we shall refer to as the \emph{blobbed} set of generators, is significantly less known: there are two \emph{blob} generators $b,b^{-1}$, and $n-1$ \emph{arc} generators $e_i$, $1\leq i \leq n-1$ (so if $n= 1$ there are no arc generators), with defining relations
\begin{align}
	e_{i}e_{i} & = (\q + \q^{-1})e_{i}, \notag\\
	e_{i}e_{i \pm 1}e_{i} &= e_{i}, \notag\\
    e_{i}e_{j}&= e_{j}e_{i} \qquad \text{ if } |i-j| \geq 2,\label{eq:blob-gen-1}\\
    	e_{i} b &= b e_{i} \qquad\;\; \text{ if } i \geq 2, \notag\\
	e_{1}be_{1} &= (\underbrace{\q b + \q^{-1}b^{-1}}_{\equiv -Y})e_{1} = e_{1}(\q b + \q^{-1}b^{-1}), \notag
\end{align}
which stands for all $i$ such that these expressions make sense; note that in this case we have no generator~$e_n$.  We also note that the element $Y \equiv -\q b - \q^{-1}b^{-1} $  introduced in the above relations is  central, it will be called the \emph{hoop} operator\footnote{The name will be justified via its diagrammatic presentation that we discuss below.}. The epithet \emph{blob} for the generators $b^{\pm1}$ is due to  the relation with the so-called \emph{blob algebra}\cite{MartinSaleur}, which is a finite-dimensional algebra.
 This latter algebra is  obtained by taking the quotient of $\atl{n}(\q)$ by the two-sided ideal 
 $\atl{n}(\q)\cdot (Y - y\mathbb{I})$
for some $y \in \mathbb{C}$ such that $y^{2} \neq 4$, or in simple words the blob algebra is obtained via fixing the eigenvalue of $Y$.  We must however point out that our invertible blob operator is \emph{not} the pre-image of the blob operator appearing in the blob algebra, 
 which is an idempotent, see also a similar discussion of relations between the affine Hecke and type $B$ Hecke algebras in~\cite{Halverson}.
 In more details, for the quotient with 
$Y = (\delta + \delta^{-1}) \mathbb{I}$
for $\delta \neq \pm 1$ the idempotent blob is $ \tilde{b} = \frac{ \q b + \delta }{\delta - \delta^{-1}} $, and the blob weight is 
 \begin{equation}\label{eq:blobweightfromdelta}
	 \tilde{y}= \frac{\delta \q^{-1} - \q \delta^{-1}}{\delta - \delta^{-1}},
 \end{equation} 
 i.e.\ after the quotient we have $\tilde{b}^2 = \tilde{b}$ and $e_1 \tilde{b}  e_1 = \tilde{y} e_1$.
 The special case with $\delta=\pm1$ (or $y = \pm 2$) corresponds to the ``generalized" blob algebra~\cite{GJSV} with defining relations ${(b')}^2 = 0$ and $e_1 {b'}  e_1 =\mp(\q - \q^{-1}) e_1$ where we set $b' =  \q b \pm 1$.
  It is also  related to the boundary seam algebra~\cite{MDRasRidout}.

We note that the connection of the blob type generators with the first description, i.e.\ in terms of ``periodic" type generators is (here, we place periodic type generators in RHS)
	\begin{align}
	e_i &= e_i , \qquad 1\leq i\leq n-1 ,\notag\\
		 b  &= (-\q)^{-3/2} g_{1}^{-1}\hdots g_{n-1}^{-1}u^{-1}, \label{eq:blob-ug-1}\\
		b^{-1} &= (-\q)^{3/2} u\, g_{n-1}\hdots g_{1},\label{eq:blob-ug-2}
	\end{align}
where we introduced the \emph{braid} generators
\be\label{eq:braids-g}
g^{\pm1}_{i} = (-\q)^{\pm 1/2}1 + (-\q)^{\mp 1/2} e_{i}.
\ee 
It is straightforward to check the braid relations
\be
g_i g_{i\pm1}g_i = g_{i\pm1}g_i  g_{i\pm1}.
\ee
The normalization\footnote{The normalization used here  for $g_i$ will also become useful when doing graphical calculations.} in~\eqref{eq:braids-g} was chosen such that  
\begin{eqnarray}
g_i^{\pm 1} g_{i+1}^{\pm 1} e_i &=& e_{i+1} g_i^{\pm 1} g_{i+1}^{\pm 1} = e_{i+1} e_i \,, \nonumber \\
g_{i+1}^{\pm 1} g_i^{\pm 1} e_{i+1} &=& e_i g_{i+1}^{\pm 1} g_i^{\pm 1} = e_i e_{i+1} \,. \label{gge_rels}
	\end{eqnarray}
These relations are used to prove the equivalence of the relations in equations~\eqref{eq:blob-gen-1} and~\eqref{eq:period-gen-1}, specifically when verifying those involving $b$, $u$, or $e_{n}$.

We note that an expression of periodic generators in terms of the blobbed  ones is obtained as follows: the shift generators $u^{\pm1}$ are obtained multiplying both sides of~\eqref{eq:blob-ug-1}-\eqref{eq:blob-ug-2} with appropriate $g_i^{\pm1}$'s, then the generator $e_n$ is formally defined as $u^{-1}e_1 u$. It is then rather straightforward, however tedious, to show that the defining relations~\eqref{eq:period-gen-1} are equivalent to those in~\eqref{eq:blob-gen-1}. We give one example of such computations as the others are all quite similar; recall the proposed form for $b$ in~\eqref{eq:blob-ug-1}, we then verify that 
\begin{align*}
	(\q b)^{2} e_{1} 	& = (-\q)^{-1} u g^{-1}_{n-1} \hdots g_{1}^{-1} u g_{n-1}^{-1}\hdots \underbrace{g_{1}^{-1}e_{1}}_{ = -(-\q)^{3/2} e_{1}},\\
						& =  - u g^{-1}_{n-1} \hdots g_{2}^{-1} \left( 1 - \q e_{1}\right)u g_{n-1}^{-1}\hdots g_{2}^{-1}e_{1} ,\\
						& = - u^{2}\underbrace{g_{n-2}^{-1} \hdots g_{1}^{-1}g_{n-2}^{-1} \hdots g_{2}^{-1} e_{1}}_{= e_{n-1}e_{n-2}\hdots e_{1}}  + \q b e_{1}b e_{1} ,\\
						& = - u^{2}e_{n-1} e_{n-2}\hdots e_{1} + \q b e_{1} b e_{1},\\
						& = - e_{1} + \q b e_{1}b e_{1},
\end{align*}
and then multiplying both sides by $b^{-1}$ from the left
 yields the identity $e_{1}b e_{1} = (\q b + \q^{-1}b^{-1})e_{1}$ from the list in~\eqref{eq:blob-gen-1}. 
We also note that in the context of blob algebras (recalled above as the quotients), the relation between periodic and blobbed generators reflects what was called ``braid translation" in~\cite{MartinSaleur,BGJSR}.

\medskip
We will also use another relation between the periodic and blobbed set of generators: Because the algebra is invariant under the substitution $b \to b^{-1}$, $\q \to \q^{-1} $, i.e. it provides an algebra automorphism, there is a second way to write the blob generators in terms of the generators of ``periodic" type:
	\be\label{eq:b-bar-def}
	\begin{split}
		\bar{b} & = (-\q)^{-3/2} u\, g_{n-1}^{-1}\hdots g_{1}^{-1}, \\
		\bar{b}^{-1}  &= (-\q)^{3/2} g_{1}\hdots g_{n-1}u^{-1} .
		\end{split}
	\ee

We turn now to introduction of diagrammatical presentations for both types of generators, and it is much easier to check such an equivalence (or isomorphism of the two algebras) by doing standard diagram calculations.

\subsection{Diagrammatic presentation}\label{sec:diag-present}
We now introduce the graphical presentation of the algebra \cite{FanGreen}, which can be used to write words in the algebra in a very compact and intuitive form. Each of the classical generators gets associated to a diagram with $2n$ nodes connected by $n$ \emph{strands}, or \emph{lines}:
\begin{align}
	e_{i} = \;
	\begin{tikzpicture}[scale = 1/3, baseline ={(current bounding box.center)}]
	\draw[line width = 1pt, black] (0.5,1) -- (8.5,1);
	\draw[line width = 1pt, black] (0.5,3) -- (8.5,3);
	\draw[line width = 1pt, black] (1,1) -- (1,3);
	\draw[line width = 1pt, black] (3,1) -- (3,3);
	\draw[line width = 1pt, black] (4,1) .. controls (4,2) and (5,2) .. (5,1);
	\draw[line width = 1pt, black] (4,3) .. controls (4,2) and (5,2) .. (5,3);
	\draw[line width = 1pt, black] (6,1) -- (6,3);
	\draw[line width = 1pt, black] (8,1) -- (8,3);
	\node[anchor = north] at (2,2.5) {$\hdots$};
	\draw[decorate, decoration = {brace, mirror, amplitude = 2 pt}, yshift = -3pt] (1,0.5) -- (3,0.5) node [midway,yshift = -7pt] {\footnotesize{i-1}};
	\node[anchor = north] at (7,2.5) {$\hdots$};
	\draw[decorate, decoration = {brace, mirror, amplitude = 2 pt}, yshift = -3pt] (6,0.5) -- (8,0.5) node [midway,yshift = -7pt] {\footnotesize{n-i-1}};
	\end{tikzpicture}\; ,& \qquad
	e_{n} = \;
	\begin{tikzpicture}[scale = 1/3, baseline ={(current bounding box.center)}]
	\draw[line width = 1pt, black] (0.5,1) -- (8.5,1);
	\draw[line width = 1pt, black] (0.5,3) -- (8.5,3);
	\foreach \s in {2,3,6,7}{
		\draw[line width = 1pt, black ] (\s,1) -- (\s, 3);
	}
	\draw[line width = 1pt, black] (1,1) .. controls (1,2) and (0,2) .. (0,1);
	\draw[line width = 1pt, black] (1,3) .. controls (1,2) and (0,2) .. (0,3);
	\draw[line width = 1pt, black] (8,1) .. controls (8,2) and (9,2) .. (9,1);
	\draw[line width = 1pt, black] (8,3) .. controls (8,2) and (9,2) .. (9,3);
	\filldraw[line width = 2pt, white] (0,1) -- (0.5,1) -- (0.5,3) -- (0,3) -- (0,1);
	\filldraw[line width = 2pt, white] (9,1) -- (8.5,1) -- (8.5,3) -- (9,3) -- (9,1);
	\node[anchor = north] at (4.5,2.5) {$\hdots$};
	\draw[decorate, decoration = {brace, mirror, amplitude = 2 pt}, yshift = -3pt] (2,0.5) -- (7,0.5) node [midway,yshift = -7pt] {\footnotesize{n-2}};
	\end{tikzpicture}\;, & i = 1, \hdots, n-1,\\
	u = \;
	\begin{tikzpicture}[scale = 1/3, baseline ={(current bounding box.center)}, yscale= -1]
	\draw[line width = 1pt, black] (0.5,1) -- (8.5,1);
	\draw[line width = 1pt, black] (0.5,3) -- (8.5,3);
	\foreach \s in {1,2,3,6,7,8,9}{
		\draw[line width = 1pt, black ] (\s,1) .. controls (\s, 2) and (\s -1, 2) .. (\s-1, 3);
	};
	\filldraw[line width = 2pt, white] (0,1) -- (0.5,1) -- (0.5,3) -- (0,3) -- (0,1);
	\filldraw[line width = 2pt, white] (9,1) -- (8.5,1) -- (8.5,3) -- (9,3) -- (9,1);
	\node[anchor = south] at (4,2.5) {$\hdots$};
	\draw[decorate, decoration = {brace, mirror, amplitude = 2 pt}, yshift = 3pt] (1,3.5) -- (8,3.5) node [midway,yshift = -7pt] {\footnotesize{n}};
	\end{tikzpicture}\;, & \qquad
	u^{-1} = \;
	\begin{tikzpicture}[scale = 1/3, baseline ={(current bounding box.center)}, yscale= -1]
	\draw[line width = 1pt, black] (0.5,1) -- (8.5,1);
	\draw[line width = 1pt, black] (0.5,3) -- (8.5,3);
	\foreach \s in {1,2,3,6,7,8,9}{
		\draw[line width = 1pt, black ] (\s -1,1) .. controls (\s -1, 2) and (\s, 2) .. (\s, 3);
	};
	\filldraw[line width = 2pt, white] (0,1) -- (0.5,1) -- (0.5,3) -- (0,3) -- (0,1);
	\filldraw[line width = 2pt, white] (9,1) -- (8.5,1) -- (8.5,3) -- (9,3) -- (9,1);
	\node[anchor = south] at (4,2.5) {$\hdots$};
	\draw[decorate, decoration = {brace, mirror, amplitude = 2 pt}, yshift = 3pt] (1,3.5) -- (8,3.5) node [midway,yshift = -7pt] {\footnotesize{n}};
	\end{tikzpicture}, & 
	1_{\atl{n}} = 
	\begin{tikzpicture}[scale = 1/3, baseline ={(current bounding box.center)}]
	\draw[line width = 1pt, black] (0.5,1) -- (8.5,1);
	\draw[line width = 1pt, black] (0.5,3) -- (8.5,3);
	\draw[line width = 1pt, black] (1,1) -- (1,3);
	\draw[line width = 1pt, black] (3,1) -- (3,3);
	\draw[line width = 1pt, black] (4,1) --  (4,3);
	\draw[line width = 1pt, black] (5,1) -- (5,3);
	\draw[line width = 1pt, black] (6,1) -- (6,3);
	\draw[line width = 1pt, black] (8,1) -- (8,3);
	\node[anchor = north] at (2,2.5) {$\hdots$};
	\draw[decorate, decoration = {brace, mirror, amplitude = 2 pt}, yshift = -3pt] (1,0.5) -- (3,0.5) node [midway,yshift = -7pt] {\footnotesize{i-1}};
	\node[anchor = north] at (7,2.5) {$\hdots$};
	\draw[decorate, decoration = {brace, mirror, amplitude = 2 pt}, yshift = -3pt] (6,0.5) -- (8,0.5) node [midway,yshift = -7pt] {\footnotesize{n-i-1}};
	\end{tikzpicture},\label{eq:u-inv-diag}
\end{align}
	where the opposing vertical sides are identified, so these drawings should be imagined as being drawn on a cylinder, with the top and bottom black lines resting on its top and bottom edge, respectively. Strands that connects both edges of the cylinder are called \emph{through lines}. One can show that every diagram which can be drawn on this cylinder with $n$ non-intersecting strings represents a non-zero element of the algebra, and every such element is represented by a unique diagram, up to isotopy of the strands which is ambient on the boundary. Sums of elements of the algebra can be understood as formal sums of diagrams, and products in the algebra are computed using \emph{diagram composition}\footnote{In this work, product of operators are read from right to left, and diagrams are read from top to bottom. In some of the authors previous work, for instance in \cite{BSA}, the opposite convention is used so operators were multiplied left to right and diagrams read from bottom to top.}: the diagram $a b$ is defined by putting the diagram for $a$ on top of the diagram for $b$ and joining the strands that meet. A closed arc that is homotopic to a point is simply removed and replaced by a factor $\q + \q^{-1}$. For instance, here are some of the defining relations of the algebra in the diagrammatic presentation (for $n = 3$):
	\begin{equation}
		e_{1}e_{1} = \;
	\begin{tikzpicture}[scale = 1/3, baseline ={(current bounding box.center)}]
	\draw[line width = 1pt, black] (0.5,1) -- (3.5,1);
	\draw[line width = 1pt, black] (1,1) .. controls (1,2) and (2,2) .. (2,1);
	\draw[line width = 1pt, black] (1,3) .. controls (1,2) and (2,2) .. (2,3);
	\draw[line width = 1pt, black] (0.5,5) -- (3.5,5);
	\draw[line width = 1pt, black] (1,3) .. controls (1,4) and (2,4) .. (2,3);
	\draw[line width = 1pt, black] (1,5) .. controls (1,4) and (2,4) .. (2,5);
	\draw[line width = 1pt, black] (3,1) -- (3,5);
	\end{tikzpicture}
		\; = (\q + \q^{-1})e_{1}, \qquad
		e_{1}e_{2}e_{1} = \;
	\begin{tikzpicture}[scale = 1/3, baseline ={(current bounding box.center)}]
	\draw[line width = 1pt, black] (0.5,1) -- (3.5,1);
	\draw[line width = 1pt, black] (1,1) .. controls (1,2) and (2,2) .. (2,1);
	\draw[line width = 1pt, black] (1,3) .. controls (1,2) and (2,2) .. (2,3);
	\draw[line width = 1pt, black] (3,1) -- (3,3);
	\draw[line width = 1pt, black] (0.5,7) -- (3.5,7);
	\draw[line width = 1pt, black] (3,5) -- (3,7);
	\draw[line width = 1pt, black] (1,7) .. controls (1,6) and (2,6) .. (2,7);
	\draw[line width = 1pt, black] (1,5) .. controls (1,6) and (2,6) .. (2,5);
	\draw[line width = 1pt, black] (3,3) .. controls (3,4) and (2,4) .. (2,3);
	\draw[line width = 1pt, black] (3,5) .. controls (3,4) and (2,4) .. (2,5);
	\draw[line width = 1pt, black] (1,3) -- (1,5);
	\end{tikzpicture}\; = \;
	\begin{tikzpicture}[scale = 1/3, baseline ={(current bounding box.center)}]
	\draw[line width = 1pt, black] (0.5,1) -- (3.5,1);
	\draw[line width = 1pt, black] (1,1) .. controls (1,2) and (2,2) .. (2,1);
	\draw[line width = 1pt, black] (1,3) .. controls (1,2) and (2,2) .. (2,3);
	\draw[line width = 1pt, black] (0.5,3) -- (3.5,3);
	\draw[line width = 1pt, black] (3,1) -- (3,3);
	\end{tikzpicture}\; = e_{1}.
	\end{equation}

\begin{figure}
\begin{align*}
	&\begin{tikzpicture}[scale = 1/3, baseline ={(current bounding box.center)}, yscale= -1]
	\draw[line width = 1pt, black] (1,1) .. controls (1,2) and (3,2) .. (3,3);
	\draw[line width = 3pt, white] (3,1) .. controls (3,2) and (1,2) .. (1,3);
	\draw[line width = 1pt, black] (3,1) .. controls (3,2) and (1,2) .. (1,3);
	\end{tikzpicture} \; \equiv (-\q)^{\half}\;
	\begin{tikzpicture}[scale = 1/3, baseline ={(current bounding box.center)}, rotate = 90]
	\draw[line width = 1pt, black] (1,1) .. controls (1,2) and (3,2).. (3,1);
	\draw[line width = 1pt, black] (1,3) .. controls (1,2) and (3,2) .. (3,3);
	\end{tikzpicture} \; + (-\q)^{-\half} \;
	\begin{tikzpicture}[scale = 1/3, baseline ={(current bounding box.center)}]
	\draw[line width = 1pt, black] (1,1) .. controls (1,2) and (3,2).. (3,1);
	\draw[line width = 1pt, black] (1,3) .. controls (1,2) and (3,2) .. (3,3);
	\end{tikzpicture}\\
	g_{i} & = (-\q)^{\half}\one + (-\q)^{-\half} e_{i} = \; 
	\begin{tikzpicture}[scale = 1/3,baseline={(current bounding box.center)}, yscale= -1] 
	\foreach \r in {1,3,6,8}{
		\draw[black, line width = 1pt] (\r, 1) -- (\r,3);
	};
	\draw[black, line width = 1pt] (4,1) .. controls (4,2) and (5,2) .. (5,3);
	\draw[white, line width = 3pt] (5,1) .. controls (5,2) and (4,2) .. (4,3);
	\draw[black, line width = 1pt] (5,1) .. controls (5,2) and (4,2) .. (4,3);
	\node[anchor = south] at (2,1.5) {$\hdots $};
	\node[anchor = south] at (7,1.5) {$\hdots $};
	\draw[black, line width = 2pt] (.5,1) -- (8.5,1);
	\draw[black, line width = 2pt] (.5,3) -- (8.5,3);
	\draw[decorate, decoration = {brace, mirror, amplitude = 2 pt}, yshift = 3pt] (.5,3.5) -- (3.5,3.5) node [midway,yshift = -7pt] {\footnotesize{i-1}};
	\draw[decorate, decoration = {brace, mirror, amplitude = 2 pt}, yshift = 3pt] (5.5,3.5) -- (8.5,3.5) node [midway,yshift = -7pt] {\footnotesize{n-i-1}};
	\end{tikzpicture}\\
	g^{-1}_{i} & = (-\q)^{-\half}\one + (-\q)^{\half} e_{i} = \; 
	\begin{tikzpicture}[scale = 1/3,baseline={(current bounding box.center)}, yscale= -1] 
	\foreach \r in {1,3,6,8}{
		\draw[black, line width = 1pt] (\r, 1) -- (\r,3);
	};
	\draw[black, line width = 1pt] (5,1) .. controls (5,2) and (4,2) .. (4,3);
	\draw[white, line width = 3pt] (4,1) .. controls (4,2) and (5,2) .. (5,3);
	\draw[black, line width = 1pt] (4,1) .. controls (4,2) and (5,2) .. (5,3);
	\node[anchor = south] at (2,1.5) {$\hdots $};
	\node[anchor = south] at (7,1.5) {$\hdots $};
	\draw[black, line width = 2pt] (.5,1) -- (8.5,1);
	\draw[black, line width = 2pt] (.5,3) -- (8.5,3);
	\draw[decorate, decoration = {brace, mirror, amplitude = 2 pt}, yshift = 3pt] (.5,3.5) -- (3.5,3.5) node [midway,yshift = -7pt] {\footnotesize{i-1}};
	\draw[decorate, decoration = {brace, mirror, amplitude = 2 pt}, yshift = 3pt] (5.5,3.5) -- (8.5,3.5) node [midway,yshift = -7pt] {\footnotesize{n-i-1}};
	\end{tikzpicture} 
\end{align*}
     \caption{Braid notations}\label{fig:br-diag}
\end{figure}

For  graphical presentation of the blobbed generators, we introduce first the \emph{braid notation} for the overlapping strands in Fig.~\ref{fig:br-diag}, as well as
 the diagram presentation of  $g^{\pm1}_{i}$ introduced in~\eqref{eq:braids-g}.
Then using~\eqref{eq:u-inv-diag} we get by stacking the diagrams:
\begin{equation}
	 g_{1}^{-1}\hdots g_{n-1}^{-1} u^{-1} \quad = 
	\begin{tikzpicture}[scale = 1/3, baseline = {(current bounding box.center)}, yscale= -1]
		\foreach \i in {1,2,3,5,6,7,8}{
			\draw[black, line width = 1pt] (\i,3) .. controls (\i, 4) and (\i+1,4) .. (\i+1, 5);
		};
		\filldraw[white] (1.5,2) -- (-.5,2) -- (-.5,5) -- (1.5,5) -- (1.5,2); 
		\filldraw[white] (8.5,2) -- (9.5,2) -- (9.5,5) -- (8.5,5) -- (8.5,2); 
		\foreach \i in {2,3,5,6,7}{
			\draw[ black, line width = 1pt] ( \i+1,1) .. controls (\i +1, 2) and  (\i, 2) .. (\i, 3);
		};
		\draw[white, line width = 3pt] (2,1) .. controls (2,2) and (8,2) .. (8,3);
		\draw[black, line width = 1pt] (2,1) .. controls (2,2) and (8,2) .. (8,3);
		\node[anchor = south] at (5,2) {$\hdots$};
		\node[anchor = north] at (4,2.75) {$\hdots$};
		\draw[decorate, decoration = {brace, mirror, amplitude = 2 pt}, yshift = 3pt] (3,5.5) -- (8,5.5) node [midway,yshift = -7pt] {\footnotesize{n-1}};
		\draw[black, line width = 2pt] (1.5,1) -- (8.5,1);
		\draw[black, line width = 2pt] (1.5,5) -- (8.5,5);
	\end{tikzpicture} \; = 
	\begin{tikzpicture}[scale = 1/3, baseline = {(current bounding box.center)}, yscale= -1]
		\foreach \i in {2,3,5,6,7}{
			\draw[ black, line width = 1pt] ( \i,1) -- (\i, 3);
		};
		\draw[line width = 3pt, white] (0,2) -- (8,2);
		\draw[line width = 1pt, black] (2,2) -- (8,2);
		\draw[line width = 1pt, black] (1,1) .. controls (1,1.5) and (1.5,2)  .. (2,2);
		\draw[line width = 1pt, black] (1,3) .. controls (1,2.5) and (0.5,2)  .. (0,2);
		\node[anchor = south] at (4,1.75) {$\hdots$};
		\node[anchor = north] at (4,2.25) {$\hdots$};
		\draw[decorate, decoration = {brace, mirror, amplitude = 2 pt}, yshift = 3pt] (2,3.5) -- (7,3.5) node [midway,yshift = -7pt] {\footnotesize{n-1}};
		\filldraw[white] (-.5,1) -- (.5,1) -- (.5,3) -- (-.5,3) -- (-.5,1); 
		\draw[black, line width = 2pt] (.5,1) -- (8,1);
		\draw[black, line width = 2pt] (.5,3) -- (8,3);
	\end{tikzpicture}
\end{equation}
and a similar calculation for $u  g_{n-1}\hdots g_{1}$.
Therefore, the blob generators $b$ and $b^{-1}$ from~\eqref{eq:blob-ug-1}-\eqref{eq:blob-ug-2} can be represented as
	\begin{equation}\label{eq:b-diag}
		b = (-\q)^{-3/2}\;
	\begin{tikzpicture}[scale = 1/3, baseline ={(current bounding box.center)}, yscale= -1]
	\foreach \s in {2,4,5}{
		\draw[line width = 1pt, black ] (\s, 1) -- (\s, 3);
	};
	\draw[line width = 1pt, black] (1,3) .. controls (1,2) and (0,2) .. (0,1);
	\draw[line width = 3pt, white] (6,2) .. controls (1,2) .. (1,1);
	\draw[line width = 1pt, black] (6,2) .. controls (1,2) .. (1,1);
	\filldraw[line width = 1pt, white] (0,1) -- (.5,1) -- (.5,3) -- (0,3) -- (0,1); 
	\filldraw[line width = 1pt, white] (6,1) -- (5.5,1) -- (5.5,3) -- (6,3) -- (6,1);
	\node[anchor = south] at (3,2.75) {$\hdots$};
	\draw[decorate, decoration = {brace, mirror, amplitude = 2 pt}, yshift = 3pt] (2,3.5) -- (5,3.5) node [midway,yshift = -7pt] {\footnotesize{n-1}};
	\draw[line width = 2pt, black] (0.5,1) -- (5.5,1);
	\draw[line width = 2pt, black] (0.5,3) -- (5.5,3);
	\end{tikzpicture}\;  , \qquad 
		b^{-1} = (-\q)^{3/2}\;
	\begin{tikzpicture}[scale = 1/3, baseline ={(current bounding box.center)}, yscale= -1]
	\foreach \s in {2,4,5}{
		\draw[line width = 1pt, black ] (\s, 1) -- (\s, 3);
	};
	\draw[line width = 1pt, black] (1,1) .. controls (1,2) and (0,2) .. (0,3);
	\draw[line width = 3pt, white] (6,2) .. controls (1,2) .. (1,3);
	\draw[line width = 1pt, black] (6,2) .. controls (1,2) .. (1,3);
	\filldraw[line width = 1pt, white] (0,1) -- (.5,1) -- (.5,3) -- (0,3) -- (0,1); 
	\filldraw[line width = 1pt, white] (6,1) -- (5.5,1) -- (5.5,3) -- (6,3) -- (6,1);
	\node[anchor = south] at (3,2.75) {$\hdots$};
	\draw[decorate, decoration = {brace, mirror, amplitude = 2 pt}, yshift = 3pt] (2,3.5) -- (5,3.5) node [midway,yshift = -7pt] {\footnotesize{n-1}};
	\draw[line width = 2pt, black] (0.5,1) -- (5.5,1);
	\draw[line width = 2pt, black] (0.5,3) -- (5.5,3);
	\end{tikzpicture}\;.
	\end{equation}
It is then straightforward to check the relations~\eqref{eq:blob-gen-1} using the standard graphical manipulations together with the relations~\eqref{gge_rels}.

\medskip

We  recall the central element $Y= -(\q b + \q^{-1}b^{-1})$. In the diagram basis, it can be written as
\begin{center}
$Y = $ 	  $ (-\q)^{-\half} \;$
	\begin{tikzpicture}[scale = 1/3, baseline = {(current bounding box.center)}, yscale= -1]
		\foreach \i in {2,3,5,6,7}{
			\draw[ black, line width = 1pt] ( \i,1) -- (\i, 3);
		};
		\draw[line width = 3pt, white] (0,2) -- (8,2);
		\draw[line width = 1pt, black] (2,2) -- (8,2);
		\draw[line width = 1pt, black] (1,1) .. controls (1,1.5) and (1.5,2)  .. (2,2);
		\draw[line width = 1pt, black] (1,3) .. controls (1,2.5) and (0.5,2)  .. (0,2);
		\node[anchor = south] at (4,2) {$\hdots$};
		\node[anchor = north] at (4,2) {$\hdots$};
		\draw[decorate, decoration = {brace, mirror, amplitude = 2 pt}, yshift = 3pt] (2,3.5) -- (7,3.5) node [midway,yshift = -7pt] {\footnotesize{n-1}};
		\filldraw[line width = 1pt, white] (0,1) -- (.5,1) -- (.5,3) -- (0,3) -- (0,1); 
		\filldraw[line width = 1pt, white] (8,1) -- (7.5,1) -- (7.5,3) -- (8,3) -- (8,1);
		\draw[black, line width = 2pt] (.5,1) -- (7.5,1);
		\draw[black, line width = 2pt] (.5,3) -- (7.5,3);
	\end{tikzpicture} $ \;+ \;(-\q)^{\half} \;$
	\begin{tikzpicture}[scale = 1/3, baseline = {(current bounding box.center)}, yscale= -1]
		\foreach \i in {2,3,5,6,7}{
			\draw[ black, line width = 1pt] ( \i,1) -- (\i, 3);
		};
		\draw[line width = 3pt, white] (0,2) -- (8,2);
		\draw[line width = 1pt, black] (2,2) -- (8,2);
		\draw[line width = 1pt, black] (1,1) .. controls (1,1.5) and (0.5,2)  .. (0,2);
		\draw[line width = 1pt, black] (1,3) .. controls (1,2.5) and (1.5,2)  .. (2,2);
		\node[anchor = south] at (4,2) {$\hdots$};
		\node[anchor = north] at (4,2) {$\hdots$};
		\draw[decorate, decoration = {brace, mirror, amplitude = 2 pt}, yshift = 3pt] (2,3.5) -- (7,3.5) node [midway,yshift = -7pt] {\footnotesize{n-1}};
		\filldraw[line width = 1pt, white] (0,1) -- (.5,1) -- (.5,3) -- (0,3) -- (0,1); 
		\filldraw[line width = 1pt, white] (8,1) -- (7.5,1) -- (7.5,3) -- (8,3) -- (8,1);
		\draw[black, line width = 2pt] (.5,1) -- (7.5,1);
		\draw[black, line width = 2pt] (.5,3) -- (7.5,3);
	\end{tikzpicture}
$\; = $
\begin{tikzpicture}[scale = 1/3, baseline = {(current bounding box.center)}]
		\foreach \i in {1,2,3,5,6,7}{
			\draw[ black, line width = 1pt] ( \i,1) -- (\i, 3);
		};
		\draw[line width = 3pt, white] (0,2) -- (8,2);
		\draw[line width = 1pt, black] (0,2) -- (8,2);
		\node[anchor = north] at (4,2) {$\hdots$};
		\node[anchor = south] at (4,2) {$\hdots$};
		\draw[decorate, decoration = {brace, mirror, amplitude = 2 pt}, yshift = -3pt] (1,0.5) -- (7,0.5) node [midway,yshift = -7pt] {\footnotesize{n}};
		\filldraw[line width = 1pt, white] (0,1) -- (.5,1) -- (.5,3) -- (0,3) -- (0,1); 
		\filldraw[line width = 1pt, white] (8,1) -- (7.5,1) -- (7.5,3) -- (8,3) -- (8,1);
		\draw[black, line width = 2pt] (.5,1) -- (7.5,1);
		\draw[black, line width = 2pt] (.5,3) -- (7.5,3);
	\end{tikzpicture} \ ,
	\end{center}
where for the last equality we also used the braid conventions in Fig.~\ref{fig:br-diag}.
That $Y$ is central is easy to check using the diagrammatic calculation as in Fig.~\ref{fig:Y-commTL}: generators $e_j$ obviously commute with the insertion of a line going ``above'' or ``under'' the system, the same applies for the commutativity with the shift operators where one just uses the braid relations.

	Recall  now the algebra automorphism $b \to b^{-1}$, $\q \to \q^{-1} $ discussed above~\eqref{eq:b-bar-def}. The diagram presentation for the second set of blobbed generators is
	\begin{equation}
		\bar{b} = (-\q)^{-3/2}\;
	\begin{tikzpicture}[scale = 1/3, baseline ={(current bounding box.center)}, yscale= -1]
	\draw[line width = 1pt, black] (1,1) .. controls (1,2) and (0,2) .. (0,3);
	\draw[line width = 1pt, black] (6,2) .. controls (1,2) .. (1,3);
		\foreach \s in {2,4,5}{
		\draw[line width = 3pt, white ] (\s, 1) -- (\s, 3);
		\draw[line width = 1pt, black ] (\s, 1) -- (\s, 3);
	};
	\draw[line width = 2pt, black] (0.5,1) -- (5.5,1);
	\draw[line width = 2pt, black] (0.5,3) -- (5.5,3);
	\filldraw[line width = 1pt, white] (0,1) -- (.5,1) -- (.5,3) -- (0,3) -- (0,1); 
	\filldraw[line width = 1pt, white] (6,1) -- (5.5,1) -- (5.5,3) -- (6,3) -- (6,1);
	\node[anchor = south] at (3,2) {$\hdots$};
	\draw[decorate, decoration = {brace, mirror, amplitude = 2 pt}, yshift = 3pt] (2,3.5) -- (5,3.5) node [midway,yshift = -7pt] {\footnotesize{n-1}};
	\end{tikzpicture}, \qquad 
		\bar{b}^{-1} = (-\q)^{3/2}\;
	\begin{tikzpicture}[scale = 1/3, baseline ={(current bounding box.center)}, yscale= -1]
	\draw[line width = 1pt, black] (1,3) .. controls (1,2) and (0,2) .. (0,1);
	\draw[line width = 1pt, black] (6,2) .. controls (1,2) .. (1,1);
		\foreach \s in {2,4,5}{
		\draw[line width = 3pt, white ] (\s, 1) -- (\s, 3);
		\draw[line width = 1pt, black ] (\s, 1) -- (\s, 3);
	};
	\draw[line width = 2pt, black] (0.5,1) -- (5.5,1);
	\draw[line width = 2pt, black] (0.5,3) -- (5.5,3);
	\filldraw[line width = 1pt, white] (0,1) -- (.5,1) -- (.5,3) -- (0,3) -- (0,1); 
	\filldraw[line width = 1pt, white] (6,1) -- (5.5,1) -- (5.5,3) -- (6,3) -- (6,1);
	\node[anchor = south] at (3,2) {$\hdots$};
	\draw[decorate, decoration = {brace, mirror, amplitude = 2 pt}, yshift = 3pt] (2,3.5) -- (5,3.5) node [midway,yshift = -7pt] {\footnotesize{n-1}};
	\end{tikzpicture}\;.
	\end{equation}
The second representative of the blob generators $\bar{b}$ and $\bar{b}^{-1}$ allows us to identify the second distinct central element $\bar{Y}$:
\begin{equation}
	\bar{Y} \equiv -(\q \bar{b} + \q^{-1}\bar{b}^{-1}) =  \;
	\begin{tikzpicture}[scale = 1/3, baseline ={(current bounding box.center)}]
	\draw[line width = 1pt, black] (0.5,2) -- (5.5,2);
	\foreach \s in {1,2,4,5}{
		\draw[line width = 3pt, white ] (\s, 1) -- (\s, 3);
		\draw[line width = 1pt, black ] (\s, 1) -- (\s, 3);
	};
	\draw[line width = 2pt, black] (0.5,1) -- (5.5,1);
	\draw[line width = 2pt, black] (0.5,3) -- (5.5,3);
	\filldraw[line width = 1pt, white] (0,1) -- (.5,1) -- (.5,3) -- (0,3) -- (0,1); 
	\filldraw[line width = 1pt, white] (6,1) -- (5.5,1) -- (5.5,3) -- (6,3) -- (6,1);
	\node[anchor = north] at (3,2) {$\hdots$};
	\node[anchor = south] at (3,2) {$\hdots$};
	\draw[decorate, decoration = {brace, mirror, amplitude = 2 pt}, yshift = -3pt] (1,0.5) -- (5,0.5) node [midway,yshift = -7pt] {\footnotesize{n}};
	\end{tikzpicture}.
	\end{equation}
We  will show in section \ref{sec:proof.center} that for any generic value of~$\q$ the two central elements $Y$ and  $\bar{Y}$ generate the center of  $\atl{n}(\q)$. 

Finally, note that the definition of the hoop operator is strikingly similar to that of a central element in the regular Temperley-Lieb algebra \cite{RidoutSA}, $F_n$, represented by the diagram:
\begin{equation}
	F_n = \;
	\begin{tikzpicture}[scale = 1/3, baseline ={(current bounding box.center)}]
	\draw[line width = 1pt, black] (0.5,2) -- (5.5,2);
	\foreach \s in {1,2,4,5}{
		\draw[line width = 3pt, white ] (\s, 1) -- (\s, 4);
		\draw[line width = 1pt, black ] (\s, 1) -- (\s, 4);
	};
	\draw[line width = 3pt, white] (0.5,3) -- (5.5,3);
	\draw[line width = 1pt, black] (0.5,3) -- (5.5,3);
	\draw[line width = 2pt, black] (0.5,1) -- (5.5,1);
	\draw[line width = 2pt, black] (0.5,4) -- (5.5,4);
	\draw[line width = 1pt, black] (.5,2) .. controls (0,2) and (0,3) .. (.5,3);
	\draw[line width = 1pt, black] (5.5,2) .. controls (6,2) and (6,3) .. (5.5,3);
	\node[anchor = north] at (3,2) {$\hdots$};
	\node[anchor = south] at (3,2) {$\hdots$};
	\node[anchor = south] at (3,3) {$\hdots$};
	\draw[decorate, decoration = {brace, mirror, amplitude = 2 pt}, yshift = -3pt] (1,0.5) -- (5,0.5) node [midway,yshift = -7pt] {\footnotesize{n}};
	\end{tikzpicture}.
\end{equation}
In terms of generators, it can also be written $F_n = - \q \tilde{b} - \q^{-1} \tilde{b}^{-1}$, with 
\begin{equation}
	\tilde{b}^{-1} = \q^{3} g_1g_2\hdots g_{n-1}g_{n-1}g_{n-2} \hdots g_1,
\end{equation}
and it satisfies the relations
\begin{equation}
	e_1 \tilde{b} e_1 = F_n e_1,
\end{equation}
\begin{equation}
	e_i \tilde{b} = \tilde{b} e_i \; \; i = 2, \hdots n-1.
\end{equation}
\subsection{Standard modules}

We present a brief overview of the most common class of $\atl{n} \equiv \atl{n}(\q)$ modules (see \cite{GL} for details): the \textit{standard} modules $\mathsf{W}_{k,z}(n)$; these are indexed by a non-negative integer $2k \leq n $ (so $k$ is a half-integer), of the same parity as $n$, and a non-zero complex number $z$. The simplest way of describing their basis is in terms of diagrams having $n$ ($2k$) nodes on their top (bottom) side, and having exactly $2k$ through lines. One simply takes the formal sums of every such diagrams, and use diagram composition to describe the action of the algebra (by stacking an algebra diagram on the top), understanding that if the composition produces a diagram with less than $2k$ through lines, it is identified with the zero element. For instance, in the case of $k=1$ we have 
\begin{equation}
	\begin{tikzpicture}[scale = 1/3, baseline ={(current bounding box.center)},yscale=-1]
	\draw[line width = 1pt, black] (0.5,1) -- (4.5,1);
	\draw[line width = 1pt, black] (0.5,3) -- (4.5,3);
	\draw[line width = 1pt, black] (1,1) .. controls (1,2) and (2,2) .. (2,1);
	\draw[line width = 1pt, black] (1,3) .. controls (1,2) and (2,2) .. (2,3);
	\draw[line width = 1pt, black] (3,1) -- (3,3);
	\draw[line width = 1pt, black] (4,1) -- (4,3);
	\draw[line width = 1pt, black] (0.5,5) -- (2.5,5);
	\draw[line width = 1pt, black] (1,3) -- (1,5);
	\draw[line width = 1pt, black] (4,3) ..controls (4,4) and (2,4) .. (2,5);
	\draw[line width = 1pt, black] (2,3) .. controls (2,4) and (3,4) ..  (3,3);	
	\end{tikzpicture} \; = \;
	\begin{tikzpicture}[scale = 1/3, baseline ={(current bounding box.center)},yscale=-1]
	\draw[line width = 1pt, black] (0.5,1) -- (4.5,1);
	\draw[line width = 1pt, black] (0.5,3) -- (2.5,3);
	\draw[line width = 1pt, black] (1,1) .. controls (1,2) and (2,2) .. (2,1);
	\draw[line width = 1pt, black] (3,1) ..controls (3,2) and (1,2) .. (1,3);
	\draw[line width = 1pt, black] (4,1) ..controls (4,2) and (2,2) .. (2,3);
	\end{tikzpicture}\;, \qquad	\qquad
	\begin{tikzpicture}[scale = 1/3, baseline ={(current bounding box.center)},yscale=-1]
	\draw[line width = 1pt, black] (0.5,1) -- (4.5,1);
	\draw[line width = 1pt, black] (0.5,3) -- (4.5,3);
	\draw[line width = 1pt, black] (1,1) .. controls (1,2) and (2,2) .. (2,1);
	\draw[line width = 1pt, black] (1,3) .. controls (1,2) and (2,2) .. (2,3);
	\draw[line width = 1pt, black] (3,1) -- (3,3);
	\draw[line width = 1pt, black] (4,1) -- (4,3);
	\draw[line width = 1pt, black] (0.5,5) -- (2.5,5);
	\draw[line width = 1pt, black] (1,3) -- (1,5);
	\draw[line width = 1pt, black] (2,3) -- (2,5);
	\draw[line width = 1pt, black] (4,3) ..controls (4,4) and (3,4) .. (3,3);	
	\end{tikzpicture} \; = 0.
\end{equation}
This is the way standard modules $\mathsf{S}_{k}(n) $ are defined for the regular Temperley-Lieb algebra $\tl{n}(\q)$, by simply excluding the diagrams with strings crossing the imaginary boundary on each side of the diagrams; while for $\tl{n}(\q)$ such diagrams form a finite dimensional module, it is not true for the affine version $\atl{n}(\q)$, as e.g.\ the translation generators $u^{\pm1}$ produce states with arbitrary winding of through lines. To get a finite dimensional module for $\atl{n}(\q)$, one must also fix the eigenvalues of the two central elements identified in the previous subsection:  $-Y =\q b + \q^{-1} b^{-1} $ and $-\bar{Y}=\q \bar{b} + \q^{-1} \bar{b}^{-1} $. The simplest way to do this is to define the right action of $u$ (the action on through lines) as multiplication by $z$, i.e.
\begin{equation}\label{eq:u-act-standard}
	\begin{tikzpicture}[scale = 1/3, baseline ={(current bounding box.center)},yscale = -1]
	\draw[line width = 1pt, black] (0.5,1) -- (4.5,1);
	\draw[line width = 1pt, black] (0.5,3) -- (2.5,3);
	\draw[line width = 1pt, black] (1,1) .. controls (1,2) and (2,2) .. (2,1);
	\draw[line width = 1pt, black] (3,1) ..controls (3,2) and (1,2) .. (1,3);
	\draw[line width = 1pt, black] (4,1) ..controls (4,2) and (2,2) .. (2,3);
	\draw[line width = 1pt, black] (0.5,5) -- (2.5,5);
	\foreach \s in {1,2,3}{
		\draw[line width = 1pt, black] (\s,3) .. controls (\s, 4) and (\s -1, 4) .. (\s - 1, 5);
	};
	\filldraw[line width = 1pt, white] (-.5,5) -- (0.5,5) -- (0.5,3) -- (-.5,3) -- (-.5,5);
	\filldraw[line width = 1pt, white] (3,5) -- (2.5,5) -- (2.5,3) -- (3,3) -- (3,5);
 	\end{tikzpicture} \; = z \;
	\begin{tikzpicture}[scale = 1/3, baseline ={(current bounding box.center)},yscale=-1]
	\draw[line width = 1pt, black] (0.5,1) -- (4.5,1);
	\draw[line width = 1pt, black] (0.5,3) -- (2.5,3);
	\draw[line width = 1pt, black] (1,1) .. controls (1,2) and (2,2) .. (2,1);
	\draw[line width = 1pt, black] (3,1) ..controls (3,2) and (1,2) .. (1,3);
	\draw[line width = 1pt, black] (4,1) ..controls (4,2) and (2,2) .. (2,3);
 	\end{tikzpicture}, \qquad \qquad
 	\begin{tikzpicture}[scale = 1/3, baseline ={(current bounding box.center)},yscale=-1]
	\draw[line width = 1pt, black] (0.5,1) -- (4.5,1);
	\draw[line width = 1pt, black] (0.5,3) -- (2.5,3);
	\draw[line width = 1pt, black] (1,1) .. controls (1,2) and (2,2) .. (2,1);
	\draw[line width = 1pt, black] (3,1) ..controls (3,2) and (1,2) .. (1,3);
	\draw[line width = 1pt, black] (4,1) ..controls (4,2) and (2,2) .. (2,3);
	\draw[line width = 1pt, black] (0.5,5) -- (2.5,5);
	\foreach \s in {0,1,2}{
		\draw[line width = 1pt, black] (\s,3) .. controls (\s, 4) and (\s + 1, 4) .. (\s + 1, 5);
	};
	\filldraw[line width = 1pt, white] (-0.5,5) -- (0.5,5) -- (0.5,3) -- (-.5,3) -- (-.5,5);
	\filldraw[line width = 1pt, white] (3,5) -- (2.5,5) -- (2.5,3) -- (3,3) -- (3,5);
 	\end{tikzpicture} \; = z^{-1} \;
	\begin{tikzpicture}[scale = 1/3, baseline ={(current bounding box.center)},yscale=-1]
	\draw[line width = 1pt, black] (0.5,1) -- (4.5,1);
	\draw[line width = 1pt, black] (0.5,3) -- (2.5,3);
	\draw[line width = 1pt, black] (1,1) .. controls (1,2) and (2,2) .. (2,1);
	\draw[line width = 1pt, black] (3,1) ..controls (3,2) and (1,2) .. (1,3);
	\draw[line width = 1pt, black] (4,1) ..controls (4,2) and (2,2) .. (2,3);
 	\end{tikzpicture},
\end{equation} 
where the left side of the first equality is the right action of $u$ while the left side of the second equality is the right action of $u^{-1}$. The eigenvalue of the central element $u^n$ is thus $z^n$. It was shown in~\cite{GL} that the endomorphism ring of standard modules is one dimensional, so any central element must act like a multiple of the identity on a standard module; finding the eigenvalue is then simply a matter of choosing a convenient element $x$ such that computing $Y x$ is easy. For example, using $x$ which is filled by non-nested arcs from the right and the rest are the $2k$ through lines, we find that the choice~\eqref{eq:u-act-standard} for the action of $u$ also fixes the eigenvalues of the  central elements $Y$ and~$\bar{Y}$ as follows: 
	\begin{equation}\label{eq:Y-eigenvalues}
	\begin{split}
		Y & = - (\q b + \q^{-1} b^{-1}) =  z (-\q)^{k} + z^{-1} (-\q)^{-k}, \\
		\bar{Y} & = - (\q \bar{b} + \q^{-1} \bar{b}^{-1}) =  z (-\q)^{-k} + z^{-1} (-\q)^{k}.
		\end{split}
	\end{equation}
	To see this, we first recall the diagram presentation for $b$ in~\eqref{eq:b-diag}. Applying then $ -\q b$ to the chosen~$x$ and expanding the braid-crossings according to the rules in Figure~\ref{fig:br-diag}, only one configuration has a non-zero contribution that corresponds to the factor $z^{-1} (-\q)^{-k}$.
As an example of such a calculation for $k=1$, $n=4$, we have 
\begin{align}
	(-\q)b \; 
	\begin{tikzpicture}[scale = 1/3, baseline ={(current bounding box.center)},yscale = -1]
	\draw[line width = 1pt, black] (0.5,1) -- (4.5,1);
	\draw[line width = 1pt, black] (0.5,3) -- (2.5,3);
	\draw[line width = 1pt, black] (3,1) .. controls (3,2) and (4,2) .. (4,1);
	\draw[line width = 1pt, black] (1,1) -- (1,3);
	\draw[line width = 1pt, black] (2,1) -- (2,3);
 	\end{tikzpicture} \;& = (-\q)^{-\half}\;
 	\begin{tikzpicture}[scale = 1/3, baseline ={(current bounding box.center)},yscale = -1]
	\draw[line width = 1pt, black] (3,1) .. controls (3,2) and (4,2) .. (4,1);
	\draw[line width = 1pt, black] (1,1) -- (1,3);
	\draw[line width = 1pt, black] (2,1) -- (2,3);
	\draw[line width = 1pt, black] (1,1) .. controls (1,0)  .. (0,0);
	\foreach \r in {2,3,4}{
		\draw[line width = 3pt, white] (\r,-1) -- (\r,1);
		\draw[line width = 1pt, black] (\r,-1) -- (\r,1);
	}
	\draw[line width = 3pt, white] (1,-1) .. controls (1,0) .. (5,0);
	\draw[line width = 1pt, black] (1,-1) .. controls (1,0) .. (5,0);
	\filldraw[white] (0.5,-1) -- (.5,3) -- (-.5,3) -- (-.5,-1) -- (.5,-1);
	\filldraw[white] (5.5,-1) -- (5.5,3) -- (5-.5,3) -- (5-.5,-1) -- (5.5,-1);
	\draw[line width = 1pt, black] (0.5,-1) -- (4.5,-1);
	\draw[line width = 1pt, black] (0.5,1) -- (4.5,1);
	\draw[line width = 1pt, black] (0.5,3) -- (2.5,3);
 	\end{tikzpicture} \; = (-\q)^{-\half}\;
 	\begin{tikzpicture}[scale = 1/3, baseline ={(current bounding box.center)},yscale = -1]
	\draw[line width = 1pt, black] (3,-1) .. controls (3,0) and (4,0) .. (4,-1);
	\draw[line width = 1pt, black] (1,1) .. controls (1,0)  .. (0,0);
	\foreach \r in {2}{
		\draw[line width = 3pt, white] (\r,-1) -- (\r,1);
		\draw[line width = 1pt, black] (\r,-1) -- (\r,1);
	}
	\draw[line width = 3pt, white] (1,-1) .. controls (1,0) .. (5,0);
	\draw[line width = 1pt, black] (1,-1) .. controls (1,0) .. (5,0);
	\filldraw[white] (0.5,-1) -- (.5,1) -- (-.5,1) -- (-.5,-1) -- (.5,-1);
	\filldraw[white] (5.5,-1) -- (5.5,1) -- (5-.5,1) -- (5-.5,-1) -- (5.5,-1);
	\draw[line width = 1pt, black] (0.5,-1) -- (4.5,-1);
	\draw[line width = 1pt, black] (0.5,1) -- (2.5,1);
 	\end{tikzpicture} \;, \\
 	& = (-\q)^{-1}\;
 	\begin{tikzpicture}[scale = 1/3, baseline ={(current bounding box.center)},yscale = -1]
	\draw[line width = 1pt, black] (3,-1) .. controls (3,0) and (4,0) .. (4,-1);
	\draw[line width = 1pt, black] (1,-1) .. controls (1,0) and (2,0) .. (2,1);
	\draw[line width = 1pt, black] (0,-1) .. controls (0,0) and (1,0) .. (1,1);
	\draw[line width = 1pt, black] (2,-1) .. controls (2,0) .. (5,0);
	\filldraw[white] (0.5,-1) -- (.5,1) -- (-.5,1) -- (-.5,-1) -- (.5,-1);
	\filldraw[white] (5.5,-1) -- (5.5,1) -- (5-.5,1) -- (5-.5,-1) -- (5.5,-1);
	\draw[line width = 1pt, black] (0.5,-1) -- (4.5,-1);
	\draw[line width = 1pt, black] (0.5,1) -- (2.5,1);
 	\end{tikzpicture} + 
	 \begin{tikzpicture}[scale = 1/3, baseline ={(current bounding box.center)},yscale = -1]
	\draw[line width = 1pt, black] (3,-1) .. controls (3,0) and (4,0) .. (4,-1);
	\draw[line width = 1pt, black] (1,-1) .. controls (1,0) and (2,0) .. (2,-1);
	\draw[line width = 1pt, black] (0, 1) .. controls (0,0) and (1,0) .. (1,1);
	\draw[line width = 1pt, black] (2,1) .. controls (2,0) .. (5,0);
	\filldraw[white] (0.5,-1) -- (.5,1) -- (-.5,1) -- (-.5,-1) -- (.5,-1);
	\filldraw[white] (5.5,-1) -- (5.5,1) -- (5-.5,1) -- (5-.5,-1) -- (5.5,-1);
	\draw[line width = 1pt, black] (0.5,-1) -- (4.5,-1);
	\draw[line width = 1pt, black] (0.5,1) -- (2.5,1);
 	\end{tikzpicture} \;, \\
 	& = z^{-1} (-\q)^{-1}
 	\begin{tikzpicture}[scale = 1/3, baseline ={(current bounding box.center)},yscale = -1]
	\draw[line width = 1pt, black] (0.5,1) -- (4.5,1);
	\draw[line width = 1pt, black] (0.5,3) -- (2.5,3);
	\draw[line width = 1pt, black] (3,1) .. controls (3,2) and (4,2) .. (4,1);
	\draw[line width = 1pt, black] (1,1) -- (1,3);
	\draw[line width = 1pt, black] (2,1) -- (2,3);
 	\end{tikzpicture}.
\end{align}
A similar calculation can be done for $\bar{b}^{\pm1}$ confirming the result in~\eqref{eq:Y-eigenvalues}.

It shall be convenient in what follows to use the notation
	\begin{equation}
		\mathsf{W}^{o}_{\pm |k|,\delta}(n) \equiv \mathsf{W}_{|k|,\delta^{\pm 1}(-\q)^{-k}}(n), \qquad \mathsf{W}^{u}_{\pm |k|,\mu}(n) \equiv \mathsf{W}_{|k|,\mu^{\pm 1}(-\q)^{k}}(n),
	\end{equation}
which fixes the eigenvalue of $Y = \delta + \delta^{-1}$, or $ \bar{Y} = \mu + \mu^{-1}$, respectively,  and
the superscript $o/u$ refers here to the central element being fixed: the one with a horizontal line going \textsl{over} ($Y$) or \textsl{under} ($\bar{Y}$) all others.

We conclude this section with a description of the structure of these modules.
 \begin{proposition}\label{prop:homs-Y}
For $\q\in\mathbb{C}^*$, there exists a non-zero homomorphism $f\colon \mathsf{W}_{s,w}(n) \to \mathsf{W}_{r,z}(n)$ if and only if $s \geq r$ and $Y$, $\bar{Y}$ have the same eigenvalues on both modules; furthermore any such morphism is proportional to the identity (for $s=r$) or to a  unique injective map  (for $s>r$). 
 \end{proposition}
The proof is straightforward and follows from the results in~\cite{GL}. Indeed we observe that the condition on equality of the eigenvalues of $Y$ and $\bar{Y}$ is equivalent to the Graham-Lehrer condition~\cite{GL}:
\begin{equation}
	z = \begin{cases}
		w (-\q)^{r-s} & \text{ if } (-\q)^{2(r-s)} = 1 \text{ or } 
		w^{2} = (-\q)^{-2r}, \\
		w^{-1} (-\q)^{r+s} & \text{ if } (-\q)^{2(r+s)} = 1 \text{ or }
		 w^{2} = (-\q)^{2r}.
	\end{cases}
\end{equation}

\medskip

Let us also note that each standard module has a unique simple quotient denoted by $\overline{\mathsf{W}}_{r,z}(n)$, and these form a complete set of irreducible modules~\cite{GL}.

\subsection{The center of  $\atl{n}(\q)$ }\label{sec:proof.center} 

In this section we prove one of our main results about the algebra $\atl{n}(\q)$.
\begin{Thm}\label{thm:main}
For all $\q\in \mathbb{C}^{*} $ not an integer root of 1, the center of $\atl{n}(\q)$ is generated by the two hoop operators, $Y$ and $\bar{Y}$.
\end{Thm}

Before proving this theorem, we establish a few technical results.
\begin{Def}
	Let $z \in \mathbb{C}$, define $\pi_z$ as the canonical projection from $\atl{n}$ onto the quotient algebra $\atl{n}/\mathsf{I}_{z}$, where $\mathsf{I}_{z} \equiv \atl{n}(Y - z \mathbb{I}_{\atl{n}})$.
\end{Def}
\begin{Lem}\label{lem:atl.cy.algebra}
	When seen as a $\mathbb{C}[Y]$-module, $\atl{n}$ is finite dimensional, and admits a basis $\lbrace x_{\lambda} \rbrace_{\lambda \in \Lambda}$ such that $\lbrace \pi_z(x_{\lambda} )\rbrace_{\lambda \in \Lambda}$ is a basis of the $\mathbb{C}$-module $\atl{n}/\mathsf{I}_{z}$ for all $z \in \mathbb{C}$.
\end{Lem}
\begin{proof}
	As a $\mathbb{C}$-algebra, $\atl{n}$ contains all finite sums of finite products of the generators $b, b^{-1}, e_1,$ $ \hdots, e_{n-1}$ subject to the defining relations \eqref{eq:blob-gen-1}. However, recall that by definition $Y = -\q b - \q^{-1}b^{-1}$, so that $b^{-1} = -\q^{2}b - \q Y$. It thus follows that one can remove the generator $b^{-1}$ and replace it, as a generator, with $Y$. In particular, this means that, as a $\mathbb{C}[Y]$-algebra, $\atl{n}$ is generated solely by the generators $b, e_1, \hdots, e_{n-1}$. In other words, as a $\mathbb{C}[Y]$-module, $\atl{n}$ is spanned by a set $\lbrace x_{\lambda} \rbrace_{\lambda \in \Lambda}$, for some yet unknown set $\Lambda$, composed of all elements corresponding to distinct diagrams\footnote{Note that a generic element of $\atl{n}$ can be understood as a sum of diagrams (see section \ref{sec:diag-present}); we are here specifically looking at elements which can be written as a single diagram.} which can be constructed without using the generator $b^{-1}$, relying solely on the generators $b, e_1, e_2, \hdots$. A generic element $a \in \atl{n}$ can thus be written
	\begin{equation}
	  a = \sum_{\lambda \in \Lambda} p_{\lambda}(Y)x_{\lambda}, \qquad p_{\lambda}(Y) \in \mathbb{C}[Y] \; \forall \lambda \in \Lambda.
	\end{equation}

We must now find $|\Lambda|$; note that a generic element $x_{\lambda}$ which contains the generator $b$ can be written in the form
\begin{equation}
	x_{\lambda} = a_1 b a_2 b a_3 \hdots, \qquad \text{ or } \qquad x_{\lambda} = a_1,
\end{equation}
where $a_1, a_2, a_3, \hdots$ can be written using only $e_1,e_2,\hdots, e_{n-1}$. However, each $a_k$ can be put in (reverse) Jones normal form:
\begin{equation}
a_{k} = e_{i_{1}}e_{i_{1}+1}\hdots e_{i_{1} + r_{1}}e_{i_{2}}e_{i_{2}+1}\hdots e_{i_{2} + r_{2}} \hdots ,
\end{equation}
where $i_{1}>i_{2} > i_{3} > \hdots$ and $i_{1} + r_{1} > i_{2} + r_{2} > \hdots $. In particular, $e_1$ appears at most once. Since $b$ commutes with all the $e_{i}$s except $e_1$, it thus follows that one can write
\begin{equation}\label{eq:basiselement.proof.01}
	x_{\lambda} =  \big( \prod_{j = 1}^{r}(e_{i_j} \hdots e_{k_j}) \big)(b e_{1} \hdots e_{k_{r+1}})(b e_{1} \hdots e_{k_{r+2}}) \hdots,
\end{equation}
where $r \geq 0$, $i_1 > i_2 > \hdots > i_{r}$, $k_1 > k_{2} > \hdots $, and we understand that if $r = 0$, the product corresponds to the identity element. In order to count how many elements of the form \eqref{eq:basiselement.proof.01} there are, we encode them as path on a $(n+1)\times (n+1)$ square lattice and a subset of $\mathbb{Z}_{k}$ in the following way; the corresponding path is
\begin{equation}
	 (n,n) \to (i_1,n) \to (i_1,k_1) \to (i_{2},k_1) \to (i_{2},k_{2}) \to \hdots \to (i_{r},k_r) \to (0,k_r),
\end{equation}
while the corresponding set of integers is simply
\begin{equation}
	(k_{r+1},k_{r+2}, \hdots).
\end{equation}
The corresponding path always starts at $(n,n)$, moving along the lattice either left or down, and ends at $(0,k_r)$ arriving from the right, never crossing the diagonal (since $i_{j}< k_{j}$). Furthermore, one can see that every such path corresponds to a unique product of the form  $\prod_{j = 1}^{r}(e_{i_j} \hdots e_{k_j})$ where $(i_{j},k_{j})$ are the corners of the path. Let thus $d_{i,k}[n]$ be the number of paths starting at $(n,n)$ and ending at $(i,k)$, moving along the lattice either left or down without crossing the diagonal; since every such path must pass by $(i+1,k)$ or $(i,k+1)$,
\begin{equation}
	d_{i,k}[n] = d_{i+1,k}[n] + d_{i,k + 1}[n], \qquad d_{n,n} = 1,
\end{equation}
with boundary conditions $d_{i,j} = 0$ whenever $i,j < 0$, $i,j > n$, or $i>j$. Solving this recurrence system, we find that 
\begin{equation*}
 d_{1,k}[n] = \binom{2n - k -1}{n-1} \frac{k}{n}.
\end{equation*}
 Since there are $2^{k}$ distinct subset of $\mathbb{Z}_{k}$, including the empty set, one concludes that
\begin{equation}\label{eq:atl.basis.dimension}
	\mathsf{dim}_{\mathbb{C}[Y]}(\atl{n}) = |\Lambda| = \sum_{k=1}^{n}  \binom{2n - k -1}{n-1} \frac{k}{n} 2^{k}.
\end{equation}

Note now that for any $z \in \mathbb{C}$, $\lbrace \pi_z(x_{\lambda}) \rbrace_{\lambda \in \Lambda}$ must generate $\atl{n}/\mathsf{I}_{z}$; to show that this set is a basis however, one must show that the dimension of this algebra is equal to $|\Lambda|$. Since these algebras all have the same dimension for all values of $z$ (as $\mathbb{C}$-module), it is sufficient to prove the result for generic values of $z$. For $\q,z$ generic, one has
\begin{align}
	\mathsf{dim}_{\mathbb{C}}\big(\atl{n}/\mathsf{I}_{z} \big) & \geq \sum^{n}_{\underset{\text{step }=2}{k = - n}} \binom{n}{\frac{n-k}{2}}^{2} \\
	& = \sum_{p = 0}^{n} \binom{n}{p}\binom{n}{n-p},\\
	& = \binom{2n}{n}.
\end{align}	
To see this, note that the standard modules $\mathsf{W}^{o}_{k,\delta}[n]$ is a non-trivial $\atl{n}/\mathsf{I}_{z}$ module if and only if $\delta + \delta^{-1} = z$, and it's dimension is $\binom{n}{\frac{n-k}{2}}$ \cite{GL}. Since the standard modules are simple for generic values of $\q,z$, any complete set of simple modules for the quotient algebra must include them. A simple combinatorial argument\footnote{The expression on the left counts the number of sequences of $2n+1$ consecutive integers, starting and ending with $0$; the expression on the right counts the same sequences by first grouping them in subsets where $0$ appears $k+1$ times.} then shows that
\begin{equation}
	\binom{2n}{n} = \sum_{k=1}^{n}  \binom{2n - k -1}{n-1} \frac{k}{n} 2^{k}.
\end{equation}
and thus $\lbrace \pi_z(x_{\lambda}) \rbrace_{\lambda \in \Lambda}$ must be a basis of $\atl{n}/\mathsf{I}_{z}$.
\end{proof}
\begin{Lem}\label{lem:infinite.ideals}
Let $\lbrace z_{\delta} \rbrace_{\delta \in \Delta}$ be an infinite set of distinct complex numbers, then 
\begin{equation}
	\bigcap_{\delta \in \Delta} I_{z_{\delta}} = 0.
\end{equation}
\end{Lem}
\begin{proof}
	Let $a \in \bigcap_{\delta \in \Delta} I_{z_{\delta}} $, according to lemma \ref{lem:atl.cy.algebra}, there must exist $\lbrace p_{\lambda}(Y) \rbrace_{\lambda \in \Lambda} \subset \mathbb{C}[Y]$ such that
	\begin{equation}
		a = \sum_{\lambda \in \Lambda} p_{\lambda}(Y) x_{\lambda}.
	\end{equation} 
	Furthermore, since $\lbrace \pi_{z_{\delta}}(x_{\lambda}) \rbrace_{\lambda \in \Lambda}$ is a basis of $\atl{n}/\mathsf{I}_{z_{\delta}}$, it follows that 
	\begin{equation}
		\pi_{z_{\delta}}(p_{\lambda}(Y)) = p_{\lambda}(z_{\delta}) = 0, \qquad \forall \delta\in \Delta, 
	\end{equation}
	so that the polynomial $p_{\lambda}(\; - \;)$ must have infinitely many roots. The only such polynomial is identically zero, and thus $a = 0$.
\end{proof}
\begin{Lem}\label{lem:atl.central.is.central}
 Let $p(\; - \;)$ be some non-zero polynomial in one variable, and $a \in \atl{n}$ be such that $p(Y)a = 0$, then $a = 0$. In particular, $p(Y)a$ is central if and only if $a$ is.
\end{Lem}
\begin{proof}
	Let $\lbrace z_{i} \rbrace_{i \in \mathbb{N}} \subset \mathbb{C}$ be a set of non-zeroes of $p$, i.e. numbers such that $p(z_i) \neq 0$. However, since the canonical projection $\pi_{z_{i}}$ is a homomorphism of algebras,
	\begin{equation}
		0 = \pi_{z_i}(p(Y)a) = \pi_{z_{i}}(p(Y)) \pi_{z_{i}}(a) = p(z_{i})\pi_{z_{i}}(a),
	\end{equation}
	and thus $a \in \mathsf{I}_{z_{i}}$ for all $i \in \mathbb{N}$. It thus follows that
	\begin{equation}
		a \in \bigcap_{i \in \mathbb{N}} \mathsf{I}_{z_{i}}.
	\end{equation}
	Applying lemma \ref{lem:infinite.ideals} then yields $a = 0$. Furthermore, since $Y$ is central, then 
	\begin{equation}
		p(Y)a b - b p(Y) a = p(Y)(a b - ba), \qquad \forall b \in \atl{n}.
	\end{equation}	
	The left side of this equation is zero for all $b$ if and only if $p(Y)a$ is central, while the right one is  if and only if $a$ is central, by the first part of this lemma.
\end{proof}
\begin{Lem}\label{lem:YYbar.poly.rel}
For all $n \in \mathbb{N}$,
\begin{equation}
	\prod_{\underset{\text{step } =2}{k = n \; \mathsf{mod} \;2}}^{n} Q_k = 0,
\end{equation} 
where
\begin{equation}
Q_k \equiv \left. \begin{cases} 
	Y - \bar{Y} & k =0 \\
 Y^2 + \bar{Y}^2  - ((-\q)^k + (-\q)^{-k} ) Y \bar{Y}+ (\q^k  - \q^{-k})^2 &  k \neq 0
 \end{cases}\right\rbrace \in \atl{n}.
\end{equation}
\end{Lem}
\begin{proof}
	Using \eqref{eq:Y-eigenvalues}, one verifies that $Q_k$ takes the eigenvalue $0$ on $\mathsf{W}_{k,w}(n)$ for all $n \in \mathbb{N}$ and all $w \in \mathbb{C}^{*}$. In particular, it follows that $Q = \prod_{\underset{\text{step } =2}{k = n \; \mathsf{mod} \;2}}^{n} Q_k$ takes the eigenvalue zero on all simple $\atl{n}$-modules. Since there are infinitely many values of $z$ for which $\atl{n}/\mathsf{I}_{z}$ is semisimple, it follows that $Q$ is in the intersection of infinitely many $\mathsf{I}_{z}$, and must therefore be zero by lemma \ref{lem:infinite.ideals}
\end{proof}

\begin{proof}[Proof of theorem \ref{thm:main}]

Let $\lbrace c_{i} \rbrace^{N}_{i = 1}$ be a basis of the center of $\atl{n}$, seen as a $\mathbb{C}[Y]$-module; such a basis exists by lemma \ref{lem:atl.cy.algebra}. Let then $z \in \mathbb{C}$ be such that 
 $\lbrace \pi_z(c_{i}) \rbrace^{N}_{i = 1}$ is not linearly independent over $\mathbb{C} $, i.e. there exists $\lbrace y^{(z)}_{i} \rbrace_{i=1}^{N} \subset \mathbb{C}$ such that 
 \begin{equation}
 	\sum_{i=1}^{N} y^{(z)}_{i} \pi_z(c_{i}) = 0.
 \end{equation}
 By definition of the canonical projection, it thus follows that there exist some $a \in \atl{n}$ such that
 \begin{equation}
 	\sum_{i=1}^{N} y^{(z)}_{i}c_{i} = a (Y - z).
 \end{equation}
However, by lemma \ref{lem:atl.central.is.central}, $a$ must be central, so there exist a set $\lbrace p^{(z)}_{i}(Y) \rbrace_{i=1}^{N} \subset \mathbb{C}[Y]$ such that
\begin{equation}
	 (Y -z)\sum_{i = 1}^{N} p^{(z)}_{i}(Y)c_{i} =  \sum_{i=1}^{N} y^{(z)}_{i}c_{i},
\end{equation}  
 and since by assumption $\lbrace c_{i} \rbrace^{N}_{i = 1}$ is a basis, one must have
 \begin{equation}
 	(Y-z)p^{(z)}_{i}(Y) = y^{(z)}_{i}.
 \end{equation}
 Since $(Y-z)$ does not have an inverse in $\mathbb{C}[Y]$, it thus follows that $y^{(z)}_{i} = 0$ for all $i = 1, \hdots, N$. We have thus proven that the dimension of the center of $\atl{n}/\mathsf{I}_{z}$, seen as a $\mathbb{C}$-module, must be greater than the dimension of the center of $\atl{n}$ seen as a $\mathbb{C}[Y]$-module.
 
 However, if $\q$ is not a root of $1$, for generic values of $z$, i.e. not of the form $z_i = - (-\q)^{t} - (-\q)^{-t}$ for some $t \in \mathbb{Z}$, the quotient algebra $\atl{n}/\mathsf{I}_{z}$ is isomorphic to the finite dimensional blob algebra, which is then semisimple (see the discussion below~\eqref{eq:blob-gen-1} as well as \cite{MartinWoodcock} for the structure of this algebra). In particular, it has $n+1$ distinct simple modules, so its center has dimension $n+1$. It then follows that the center of $\atl{n}$, seen as a $\mathbb{C}[Y]$-algebra, is at most of dimension $n+1$. 
 
 Finally, note that for generic values of $z$, $\pi_z(\bar{Y})$ has different eigenvalues on all simple $\atl{n}/\mathsf{I}_{z}$-modules, so the polynomials in $\bar{Y}$ must generate an algebra of dimension at least $n+1$. One then concludes that the center of $\atl{n}$, seen as a $\mathbb{C}[Y]$-module, is of dimension $n+1$, and is thus spanned by the integer powers of $\bar{Y}$.
\end{proof}

Note that if $\q$ is a root of $1$, in general, the center of $\atl{n}/\mathsf{I}_{z}$ will not be generated by $\pi_z(\bar{Y})$; however, one could extend this result to the root of unity case by finding the center of the Blob algebra, and then finding it's preimage in $\atl{n}$.

Also of note is that, by lemma \ref{lem:YYbar.poly.rel}, $\bar{Y}^{n+1}$ is a polynomial of order $n+1$ in $Y$ and of order $n$ in $\bar{Y}$; since the center has to have dimension $n+1$ as a $\mathbb{C}[Y]$-module, it follows that there cannot be any identity of lower order linking powers of $Y$ and $\bar{Y}$.

Note also that while the theorem states that central elements can be expressed as a polynomial in $Y$ and $\bar{Y}$, finding such an expression for a given central element can be very challenging. For instance, two well known central elements are $u[n]^n$ and its inverse $u[n]^{-n}$, which can be expressed in terms of Jucys-Murphy elements\footnote{See section \ref{sec:JM-elements}} as
\begin{equation}
 u[n]^{n} = (-\q)^{3 n/2} J_n J_{n-1} \hdots J_{1}\;, \qquad u[n]^{-n} = (-\q)^{3 n/2} M_1M_2 \hdots M_{n},
\end{equation}
which is readily verified by simply drawing the corresponding diagrams; for instance if $n = 3$,
\begin{equation}
 (-\q)^{9/2} J_3J_2J_1 = \;
 \begin{tikzpicture}[scale = 1/3, baseline ={(current bounding box.center)}]
	\clip (.5,0.5) -- (.5,7.5) -- (3.5,7.5) -- (3.5,0.5) -- (.5,0.5);
	\draw[black, line width = 1pt] (1,7) .. controls (1,6) and (0,6) .. (0,5);
	\draw[black, line width = 1pt] (1,1) -- (1,5) .. controls (1,6) and (4,6) .. (4,7);
	\draw[white, line width = 3pt] (2,7) -- (2,5) .. controls (2,4) and (0,4) .. (0,3);
	\draw[black, line width = 1pt] (2,7) -- (2,5) .. controls (2,4) and (0,4) .. (0,3);
	\draw[white, line width = 3pt] (2,1) -- (2,3) .. controls (2,4) and (4,4) .. (4,5);
	\draw[black, line width = 1pt] (2,1) -- (2,3) .. controls (2,4) and (4,4) .. (4,5);
	\draw[white, line width = 3pt] (3,7) -- (3,3) .. controls (3,2) and (0,2) .. (0,1);
	\draw[black, line width = 1pt] (3,7) -- (3,3) .. controls (3,2) and (0,2) .. (0,1);
	\draw[black, line width = 1pt] (3,1) .. controls (3,2) and (4,2) .. (4,3);
	\draw[black, line width = 2pt] (.5,1) -- (3.5,1);
	\draw[black, line width = 2pt] (.5,3) -- (3.5,3);
	\draw[black, line width = 2pt] (.5,5) -- (3.5,5);
	\draw[black, line width = 2pt] (.5,7) -- (3.5,7);
 	\end{tikzpicture} \qquad , \text{ and } 
 	(-\q)^{9/2} M_1M_2M_3 = \;
\begin{tikzpicture}[scale = 1/3, baseline ={(current bounding box.center)}]
	\clip (.5,0.5) -- (.5,7.5) -- (3.5,7.5) -- (3.5,0.5) -- (.5,0.5);
	\draw[black, line width = 1pt] (1,1) .. controls (1,2) and (0,2) .. (0,3);
	\draw[black, line width = 1 pt] (2,1) -- (2,3) .. controls (2,4) and (0,4) .. (0,5);
	\draw[black, line width = 1 pt] (3,1) -- (3,5) .. controls (3,6) and (0,6) .. (0,7);
	\draw[white, line width = 3pt] (1,7) -- (1,3) .. controls (1,2) and (4,2) .. (4,1);
	\draw[black, line width = 1pt] (1,7) -- (1,3) .. controls (1,2) and (4,2) .. (4,1);
	\draw[white, line width = 3pt] (2,7) -- (2,5) .. controls (2,4) and (4,4) .. (4,3);
	\draw[black, line width = 1pt] (2,7) -- (2,5) .. controls (2,4) and (4,4) .. (4,3);
	\draw[black, line width = 1pt] (3,7) .. controls (3,6) and (4,6) .. (4,5);
	\draw[black, line width = 2pt] (.5,1) -- (3.5,1);
	\draw[black, line width = 2pt] (.5,3) -- (3.5,3);
	\draw[black, line width = 2pt] (.5,5) -- (3.5,5);
	\draw[black, line width = 2pt] (.5,7) -- (3.5,7);
 	\end{tikzpicture} \qquad.
\end{equation}
Using the formulas in section \ref{sec:JM-elements} to express the Jucys-Murphy elements in terms of the hoop operators, we find
\begin{equation}
	u[2]^{2} = \mathbb{I}_{\atl{2}} + \frac{Z_{2}Z_{0}}{(\q^2 - \q^{-2})(\q - \q^{-1})},
\end{equation}
\begin{equation}
	u[2]^{-2} = \mathbb{I}_{\atl{2}} + \frac{Z_{-2}Z_{0}}{(\q^2 - \q^{-2})(\q - \q^{-1})},
\end{equation}
\begin{align}
	u[3]^{3} = \frac{\left( Z_3 Z_1 Z_{-1} - (\q^2 - \q^{-2})^2 (\q + \q^{-1}) Z_{1} - (\q- \q^{-1})^2 Z_{-1}\right) }{(\q^3 - \q^{-3})(\q^2 - \q^{-2})(\q - \q^{-1})},
\end{align}
\begin{equation}
	u[3]^{-3} = -\frac{\left( Z_{-3} Z_1 Z_{-1} - (\q^2 - \q^{-2})^2 (\q + \q^{-1}) Z_{-1} - (\q- \q^{-1})^2 Z_{1}\right) }{(\q^3 - \q^{-3})(\q^2 - \q^{-2})(\q - \q^{-1})},
\end{equation}
where 
\begin{equation}
	Z_{k} = (-\q)^{k/2}Y - (-\q)^{-k/2}\bar{Y}, \qquad k \in \mathbb{Z}.
\end{equation}
Note that these expressions may not be well-defined if $\q$ is a root of $1$, so the condition that $\q$ be generic is necessary. As an example, one can verify directly that in the twisted XXZ spin chain (see section \ref{eq:sec4-ex}) at $\q = 1$, $u[n]^{n}$ is algebraicaly independent of the two hoop operators.

Let us also mention here that the center of  $\atl{n}(\q)$ has two interesting properties, which will be proven in Sections \ref{sec:JM-elements}, and \ref{sec:higherspin}, respectively:
\begin{enumerate}
\item  The center is the image of the center of the affine Hecke algebra under the standard covering map: $\widehat{H}_n(\q)\to \atl{n}(\q)$, $T_{i} \mapsto g_{i}$, $J_{i} \mapsto J_{i}$, see Section~\ref{sec:JM-elements} for definitions. 
Recall that the center of $\widehat{H}_n(\q)$ is spanned by symmetric polynomials in the Jucys-Murphy elements $J_i$, for $1\leq i\leq n$
(see more about the affine Hecke algebra in~\cite[Sec.\,3]{GLDiagramAlgebras} where the Jucys-Murphy  elements are denoted by $X_i$).
\item There is a special ``canonical" basis (made of Chebyshev polynomials) such that the structure constants in the center are non-negative integers, i.e.\ the center of  $\atl{n}(\q)$ endowed with this basis is a Verlinde algebra.
\end{enumerate}
The second point is very important for our defect operators construction, and we will  see below in Section~\ref{sec:topdef.crossedchannel} that the central elements in the canonical basis provide (on certain $\aTL$ representations)  operators that represent topological defects in the crossed channel.

Finally one should point out that for a decomposable module, the centralizer is often very different from the center of the algebra. For instance, for any positive integer $N$, the centralizer of $M_{N} = \bigoplus_{i = 1}^{N} \mathsf{W}_{2,z}(2)$ is isomorphic to the algebra of $N \times N$ matrices with complex coefficients, but any central element will have the same eigenvalue on each copy of $\mathsf{W}_{2,z}(2)$.

\subsection{Tower structure}\label{sec:def.tower}
The family of affine Temperley-Lieb  algebras admits inclusions of the form (we will often abbreviate $\atl{n}\equiv \atl{n}(\q)$)
$$
\atl{n} \subset \atl{n+1} \subset \atl{n+2} \subset \hdots\ , \qquad n\geq 1\ , 
$$
 giving the structure of a tower of algebras~\cite[Sec.\,3.3]{GS}. Some of these inclusions will play a role in our construction of topological defects so we describe them here. We now assume that $k$ is a positive integer, and define a morphism\footnote{For brevity, we will  use the term ``morphism" instead of the more standard ``homomorphism".} of algebras 
 \be
 \phi^{u}_{n,k}\colon\; \atl{n} \to \atl{n+k},
 \ee
  by its action on the various sets of generators of the algebra. For clarity, we add a superscript to the generators to indicate  which algebra they belong to; for instance $u^{(n)}$ is the shift generator in $\atl{n}$, while $u^{(n+2)}$ is the shift generator in $\atl{n+2}$, etc. With this notation, the map  $\phi^{u}_{n,k}$ on the blobbed set of  generators is
\begin{align}
\phi^{u}_{n,k}\colon \quad  	\big(b^{(n)} \big)^{\pm 1} &\mapsto  \big(b^{(n+k)}\big)^{\pm 1},\\
 e^{(n)}_{i}  &\mapsto e^{(n+k)}_{i}.
\end{align}
It is straightforward to verify that $\phi^{u}_{n,k}$ defines an inclusion of algebras. We note that this definition is parallel to what was done in affine Hecke algebra terms in~\cite[Sec.\,4.4.2]{GS}. While the map is very simple with the blobbed generators, it is more complicated when expressed on the periodic set of generators, for instance
\begin{equation}
\phi^{u}_{n,k} \colon \quad 	u^{(n)}  \mapsto  u^{(n+k)}g^{(n+k)}_{n+k-1}g^{(n+k)}_{n+k-2}\hdots g^{(n+k)}_{n} = \;
	\begin{tikzpicture}[scale = 1/3,baseline={(current bounding box.center)}, yscale = -1] 
	\foreach \r in {1,2,3,5}{
		\draw[black, line width = 1pt] (\r, 1) .. controls (\r, 2) and (\r -1, 2 ) .. (\r - 1,3);
	};
	\draw[black, line width = 1pt] (6,1) -- (6,3);
	\draw[black, line width = 1pt] (8,1) -- (8,3);
	\draw[white, line width = 3pt] (9,2) .. controls (5,2) .. (5,3);
	\draw[black, line width = 1pt] (9,2) .. controls (5,2) .. (5,3);
	\draw[black, line width = 2pt] (.5,1) -- (8.5,1);
	\draw[black, line width = 2pt] (.5,3) -- (8.5,3);
	\node[anchor = north] at (3.30,1.75) {$\hdots$};
	\node[anchor = north] at (7,1) {$\hdots$};
	\node[anchor = south] at (7,3) {$\hdots$};
	\filldraw[white] (-.5,1) -- (.5,1) -- (.5,3) -- (-.5,3) -- (-.5,1);
	\filldraw[white] (8.5,1) -- (9.5,1) -- (9.5,3) -- (8.5,3) -- (8.5,1);
	\draw[decorate, decoration = {brace, mirror, amplitude = 2 pt}, yshift = 3pt] (.5,3.5) -- (5.5,3.5) node [midway,yshift = -7pt] {\footnotesize{n}};
	\draw[decorate, decoration = {brace, mirror, amplitude = 2 pt}, yshift = 3pt] (5.5,3.5) -- (8.5,3.5) node [midway,yshift = -7pt] {\footnotesize{k}};
	\end{tikzpicture}
\end{equation}
which  agrees with~\cite[Eq.\,(3.9)]{GS}. 
One therefore sees that, in terms of diagrams, the morphism $\phi^{u}_{n,k}$  consists in adding $k$ through lines on the right side of each diagrams, going \emph{under} every lines that wraps around the cylinder (hence the superscript $u$ on the morphism). 

Similarly, one defines another morphism of algebras 
\be
\phi^{o}_{n,k}\colon \;  \atl{n} \to \atl{n+k}
\ee
 by adding the $k$ lines \emph{over} the lines that wrap around the cylinder, i.e.\ 
  on the periodic set of generators the map is
\begin{equation}
\phi^{o}_{n,k}\colon \quad u^{(n)} \mapsto u^{(n+k)}\big(g^{(n+k)}_{n+k-1}\big)^{-1}\big(g^{(n+k)}_{n+k-2}\big)^{-1}\hdots \big(g^{(n+k)}_{n}\big)^{-1} = \;
	\begin{tikzpicture}[scale = 1/3,baseline={(current bounding box.center)},yscale=-1] 
	\foreach \r in {1,2,3,5}{
		\draw[black, line width = 1pt] (\r, 1) .. controls (\r, 2) and (\r -1, 2 ) .. (\r - 1,3);
	};
	\draw[black, line width = 1pt] (9,2) .. controls (5,2) .. (5,3);
	\draw[white, line width = 3pt] (6,1) -- (6,3);
	\draw[white, line width = 3pt] (8,1) -- (8,3);
	\draw[black, line width = 1pt] (6,1) -- (6,3);
	\draw[black, line width = 1pt] (8,1) -- (8,3);
	\draw[black, line width = 2pt] (.5,1) -- (8.5,1);
	\draw[black, line width = 2pt] (.5,3) -- (8.5,3);
	\node[anchor = north] at (3.30,1.75) {$\hdots$};
	\node[anchor = north] at (7,1) {$\hdots$};
	\node[anchor = south] at (7,3) {$\hdots$};
	\filldraw[white] (-.5,1) -- (.5,1) -- (.5,3) -- (-.5,3) -- (-.5,1);
	\filldraw[white] (8.5,1) -- (9.5,1) -- (9.5,3) -- (8.5,3) -- (8.5,1);
	\draw[decorate, decoration = {brace, mirror, amplitude = 2 pt}, yshift = 3pt] (.5,3.5) -- (5.5,3.5) node [midway,yshift = -7pt] {\footnotesize{n}};
	\draw[decorate, decoration = {brace, mirror, amplitude = 2 pt}, yshift = 3pt] (5.5,3.5) -- (8.5,3.5) node [midway,yshift = -7pt] {\footnotesize{k}};
	\end{tikzpicture}.
\end{equation}
On the blobbed set of blob type generators, the map is simply
\begin{align}
\phi^{o}_{n,k}\colon \quad	\big(\bar{b}^{(n)}\big)^{\pm 1} & \mapsto  \big(\bar{b}^{(n+k)}\big)^{\pm 1}, \\
 e^{(n)}_{i} &\mapsto e^{(n+k)}_{i}.
\end{align}

Furthermore, while we placed the extra lines on the right side of the diagram, we could have put them on the left side instead; we name the resulting morphisms 
\be\label{eq:psi-def}
\psi^{u/o}_{n,k}\colon \;  \atl{n} \to \atl{n+k},
\ee
for the corresponding \textit{under} and \textit{over} versions.
We then notice that the two subalgebras $\phi^{u}_{n,k}(\atl{n})$ and $\psi^{o}_{k,n}(\atl{k})$ commute with each others; this can be seen by a direct calculation as in~\cite{GS} or showing that 
$$
\phi^{u}_{n,k}\big(b^{(n)}\big) \propto J^{(n+k)}_{1} , \qquad \psi^{o}_{k,n}\big(b^{(k)}\big) \propto J^{(n+k)}_{n+1} ,
$$
 where $J_{i}$ is the Jucys-Murphy element\footnote{The Jucys-Murphy elements form a commutative subalgebra.} of the affine Temperley-Lieb algebra (see Section~\ref{sec:JM-elements}). This fact can be exploited to define a monoidal structure on the affine Temperley-Lieb category~\cite{GS}, see also~\cite{GJS} for the corresponding fusion calculation. We note that $ \phi^{o}_{n,k}(\atl{n})$ and $\psi^{u}_{k,n}(\atl{k})$ also commute.

\subsection{Tile formalism and the transfer matrix}\label{sec:transf}

While we formulate most of our results in terms of diagrams with strings and arcs on a cylinder, a very significant body of work on this subject is written in terms of \emph{planar tiles} (see for instance~\cite{PeRasmPolymers,PeMDFusionHierarchy}); we present here a brief translation between the two formalisms and use it to introduce the usual transfer matrix.

The planar tile with spectral parameter $x$ is defined by (where the corner mark allows to keep track of the orientation of the tile)\footnote{This tile is often divided by $(\q - \q^{-1})$ to normalize it, but then the natural defect operator would be $(\q - \q^{-1})^{-n}Y$ instead of $Y$.}
\begin{equation}
	\begin{tikzpicture}
		\foreach \r in {0,1}{
			\draw[line width = 1] (-3/4,1/4 -\r /2) -- (3/4,1/4 -\r/2);
			}
		\node at (0,0) {\utiles{x}};
		\node at (1,0) {$=$};
	\end{tikzpicture}
	\begin{tikzpicture}
		\foreach \r in {0,1}{
			\draw[line width = 1] (-1/2,1/4 -\r /2) -- (1/2,1/4 -\r/2);
			}
		\node at (0,0) {\idtiles};
		\node[anchor = east] at (- 1/2,0) {$\left( \frac{q}{x} - \frac{x}{q}\right)$};
	\end{tikzpicture}
	\begin{tikzpicture}
		\foreach \r in {0,1}{
			\draw[line width = 1] (-1/2,1/4 -\r /2) -- (1/2,1/4 -\r/2);
			}
		\node at (0,0) {\etiles};
		\node[anchor = east] at (- 1/2,0) {$+ \left( x - x^{-1}\right)$};
	\end{tikzpicture}.
\end{equation}
These satisfy three particular relations:
\begin{align}
	\begin{tikzpicture}[baseline = {(current bounding box.center)}]
		\foreach \r in {0,1}{
			\draw[line width = 1] (-3/4,1/4 -\r /2) -- (6/4,1/4 -\r/2);
			}
		\node at (0,0) {\utiles{x}};
		\node at (1,0) {\utiles{x^{-1}}};
	\end{tikzpicture} \; = & \;\; (\q^{2} + \q^{-2} - x^{2} - x^{-2})\;
	\begin{tikzpicture}[baseline = {(current bounding box.center)}]
		\foreach \r in {0,1}{
			\draw[line width = 1] (-1/2,1/4 -\r /2) -- (1/2,1/4 -\r/2);
			}
		\node at (0,0) {\idtiles};
	\end{tikzpicture}\ , \\
	\begin{tikzpicture}[baseline = {(current bounding box.center)}]
		\foreach \r in {0,1,2}{
			\draw[line width = 1] (-3/4,1/4 -\r /2) -- (6/4,1/4 -\r/2);
			}
		\node at (0,0) {\utiles{x}};
		\node at (1,0) {\utiles{y}};
		\node at (.5,-.5) {\utiles{ x y}};
	\end{tikzpicture} \; = & \;\; 
	\begin{tikzpicture}[baseline = {(current bounding box.center)}]
		\foreach \r in {-1,0,1}{
			\draw[line width = 1] (-3/4,1/4 -\r /2) -- (6/4,1/4 -\r/2);
			}
		\node at (0,0) {\utiles{y}};
		\node at (1,0) {\utiles{x}};
		\node at (.5,.5) {\utiles{ x y}};
	\end{tikzpicture}\ ,\\
	\begin{tikzpicture}[baseline = {(current bounding box.center)}]
		\foreach \r in {0,1}{
			\draw[line width = 1] (-3/4,1/4 -\r /2) -- (3/4,1/4 -\r/2);
			}
		\node at (0,0) {\utiles{x}};
	\end{tikzpicture} \; = &\; \;
	\begin{tikzpicture}[baseline = {(current bounding box.center)}]
		\foreach \r in {0,1}{
			\draw[line width = 1] (-3/4,1/4 -\r /2) -- (3/4,1/4 -\r/2);
			}
		\node at (0,0) {\rutiles{\q x^{-1}}{90}};
	\end{tikzpicture} \;.
\end{align}
These are respectively called the inversion and the Yang-Baxter identities, and the crossing symmetry.

The transfer matrix $T_{n}(\vec{x})$ can then be defined as
\begin{equation}
	T_{n}(\vec{x}) = \;
	\begin{tikzpicture}[baseline={(current bounding box.center)},yscale = -1]
				\draw[line width = 2] (0,-0.353553) -- (4*.707,-0.353553);
				\draw[line width = 2] (-.354553,0.353553) -- (13*.353553,0.353553);
				\node at (0 + 0.707*3,0) {$\hdots$};
				\node at (0 + 0.707*0 ,0) {\rutiles{x_{1}}{-135}};
				\node at (0 + 0.707*1 ,0) {\rutiles{x_{2}}{-135}};
				\node at (0 + 0.707*2 ,0) {\rutiles{x_{3}}{-135}};
				\node at (0 + 0.707*4 ,0) {\rutiles{x_{n-2}}{-135}};
				\node at (0 + 0.707*5 ,0) {\rutiles{x_{n-1}}{-135}};
				\node at (0 + 0.707*6 ,0) {\rutiles{x_{n}}{-135}};
	\end{tikzpicture}, \qquad \vec{x} = \lbrace x_{1}, x_{2}, \hdots, x_{n} \rbrace \;,
\end{equation}
where there are $n$ tiles and the opposing vertical sides are identified so that this defines an element of $\atl{n}(\q)$ for each $n$-dimensional vector $\vec{x}$. If $x_{1} = x_{2} = \hdots = x_{n} $ the transfer matrix is said to be \emph{homogeneous} and is simply written $T_{n}(x_{1})$. Using the three previous identities, one readily shows that homogeneous transfer matrices commute with each others, i.e.\ $[T_{n}(x),T_{n}(y)]=0$, and thus define families of integrable lattice models.

We note four specific cases of the homogeneous transfer matrix. Setting the spectral parameter $x$ to $1$ or $\q$ gives the translation operators $u^{\mp1}$:
\begin{equation}
	T_{n}(1) = (\q - \q^{-1})^{n} \;
	\begin{tikzpicture}[baseline = {(current bounding box.center)}]
		\foreach \r in {1,...,6}{
		\node at (0 + 0.707*\r,0) {
			\begin{tikzpicture}[scale=1/2,rotate= 45]
				\filldraw[white] (0,0) -- (1,1) -- (2 ,0) -- (1 , -1 ) -- (0, 0);
				\draw[line width = 2] (0,0) -- (1,1) -- (2 ,0) -- (1 , -1 ) -- (0, 0);
				\draw[line width = 1] (.5,-.5) arc (-45:45:.707);
				\draw[line width = 1] (1.5,-.5) arc (225:135:.707);
			\end{tikzpicture}
			};
		};
	\end{tikzpicture} \; =  (\q - \q^{-1})^{n} u^{-1},
\end{equation}
\begin{equation}
	T_{n}(\q) = (\q - \q^{-1})^{n} \;
	\begin{tikzpicture}[baseline = {(current bounding box.center)}]
		\foreach \r in {1,...,6}{
		\node at (0 + 0.707*\r,0) {
			\begin{tikzpicture}[scale=1/2,rotate= -45]
				\filldraw[white] (0,0) -- (1,1) -- (2 ,0) -- (1 , -1 ) -- (0, 0);
				\draw[line width = 2] (0,0) -- (1,1) -- (2 ,0) -- (1 , -1 ) -- (0, 0);
				\draw[line width = 1] (.5,-.5) arc (-45:45:.707);
				\draw[line width = 1] (1.5,-.5) arc (225:135:.707);
			\end{tikzpicture}
			};
		};
	\end{tikzpicture} \; =  (\q - \q^{-1})^{n} u,
\end{equation}
while taking the limits in the spectral parameter to zero or infinity  produces the two hoop operators $Y$ and $\bar{Y}$:
\begin{equation}
	\lim_{x \to 0}((-(-\q)^{-\half}x)^{n} T_{n}(x)) = \;
	\begin{tikzpicture}[baseline = {(current bounding box.center)}]
		\foreach \r in {1,...,6}{
		\node at (0 + 0.707*\r,0) {
			\begin{tikzpicture}[scale=1/2,rotate= 45]
				\filldraw[white] (0,0) -- (1,1) -- (2 ,0) -- (1 , -1 ) -- (0, 0);
				\draw[line width = 1] (.5,.5) -- (1.5,-.5);
				\draw[white, line width = 3pt] (.5,-.5) -- (1.5,.5);
				\draw[black, line width = 1pt] (.5,-.5) -- (1.5,.5);
				\draw[line width = 2] (0,0) -- (1,1) -- (2 ,0) -- (1 , -1 ) -- (0, 0);
			\end{tikzpicture}
			};
		};
	\end{tikzpicture} \; = \bar{Y},
\end{equation}
\begin{equation}\label{eq:T-Y}
	\lim_{x \to \infty}(((-\q)^{-\half}x)^{-n} T_{n}(x)) = \;
	\begin{tikzpicture}[baseline = {(current bounding box.center)}]
		\foreach \r in {1,...,6}{
		\node at (0 + 0.707*\r,0) {
			\begin{tikzpicture}[scale=1/2,rotate= -45]
				\filldraw[white] (0,0) -- (1,1) -- (2 ,0) -- (1 , -1 ) -- (0, 0);
				\draw[line width = 1] (.5,.5) -- (1.5,-.5);
				\draw[white, line width = 3pt] (.5,-.5) -- (1.5,.5);
				\draw[black, line width = 1pt] (.5,-.5) -- (1.5,.5);
				\draw[line width = 2] (0,0) -- (1,1) -- (2 ,0) -- (1 , -1 ) -- (0, 0);
			\end{tikzpicture}
			};
		};
	\end{tikzpicture} \; = Y.
\end{equation}
Finally, the standard periodic Temperley-Lieb Hamiltonian $H_n = \sum_{j=1}^n e_j$ can be obtained as the logarithmic derivative evaluated at $x = 1$ of the homogeneous transfer matrix $T_{n}(x)$, see e.g.~\cite{Baxter}.

\section{Lattice topological defects: crossed channel}\label{sec:topdef.crossedchannel}\label{sec:3}
In this section, we formulate our lattice topological defects in terms of the affine TL algebra using the hoop operators --  the central elements $Y$ and $\bar{Y}$ introduced in the previous section -- and describe their fusion rules.
 In more mathematical terms, we prove that the center $\Z$ of $\atl{n}$ admits a certain basis with \textsl{non-negative integer} structure constants. Interestingly, at least for generic values of $\q$, the structure constants do not depend on~$n$ or~$\q$. Then we also show that $\Z$  agrees with the algebra of symmetric Laurent polynomials in the famous Jucys-Murphy elements, and give a precise relations in terms of Chebyshev polynomials of $Y$ and $\bar{Y}$. 

\subsection{The algebra of defects $Y$ and $\bar{Y}$}
Recall that the hoop operators defined in Section~\ref{sec:diag-present} can be represented by diagrams with a single closed string wrapping   over or under all the other strings:
	\begin{equation}
		Y = -(\q b + \q^{-1} b^{-1}) =  \;
	\begin{tikzpicture}[scale = 1/3, baseline ={(current bounding box.center)}]
	\foreach \s in {1,2,4,5}{
		\draw[line width = 1pt, black ] (\s, 1) -- (\s, 3);
	};
	\draw[line width = 3pt, white] (0.5,2) -- (5.5,2);
	\draw[line width = 1pt, black] (0.5,2) -- (5.5,2);
	\draw[line width = 1pt, black] (0.5,1) -- (5.5,1);
	\draw[line width = 1pt, black] (0.5,3) -- (5.5,3);
	\filldraw[line width = 1pt, white] (0,1) -- (.5,1) -- (.5,3) -- (0,3) -- (0,1); 
	\filldraw[line width = 1pt, white] (6,1) -- (5.5,1) -- (5.5,3) -- (6,3) -- (6,1);
	\node[anchor = north] at (3,2) {$\hdots$};
	\node[anchor = south] at (3,2) {$\hdots$};
	\draw[decorate, decoration = {brace, mirror, amplitude = 2 pt}, yshift = -3pt] (1,0.5) -- (5,0.5) node [midway,yshift = -7pt] {\footnotesize{n}};
	\end{tikzpicture},\; \qquad
		\bar{Y} = -(\q \bar{b} + \q^{-1} \bar{b}^{-1}) = \;
	\begin{tikzpicture}[scale = 1/3, baseline ={(current bounding box.center)}]
	\draw[line width = 1pt, black] (0.5,2) -- (5.5,2);
	\foreach \s in {1,2,4,5}{
		\draw[line width = 3pt, white ] (\s, 1) -- (\s, 3);
		\draw[line width = 1pt, black ] (\s, 1) -- (\s, 3);
	};
	\draw[line width = 1pt, black] (0.5,1) -- (5.5,1);
	\draw[line width = 1pt, black] (0.5,3) -- (5.5,3);
	\filldraw[line width = 1pt, white] (0,1) -- (.5,1) -- (.5,3) -- (0,3) -- (0,1); 
	\filldraw[line width = 1pt, white] (6,1) -- (5.5,1) -- (5.5,3) -- (6,3) -- (6,1);
	\node[anchor = north] at (3,2) {$\hdots$};
	\node[anchor = south] at (3,2) {$\hdots$};
	\draw[decorate, decoration = {brace, mirror, amplitude = 2 pt}, yshift = -3pt] (1,0.5) -- (5,0.5) node [midway,yshift = -7pt] {\footnotesize{n}};
	\end{tikzpicture},
	\end{equation}
and these are central elements in $\atl{n}(\q)$. 
This wrapping string can be isotopically deformed at will without changing the spectrum of the transfer matrix from Section~\ref{sec:transf}, and it thus can be thought of as a defect line (in the crossed channel). We are interested in the algebra   generated by these hoop operators,
and first study their powers.

\begin{figure}
\begin{center}
\begin{tikzpicture}[scale = 1/3, baseline ={(current bounding box.center)}]
	\foreach \s in {1,2,4,5}{
		\draw[line width = 1pt, black] (\s,1) -- (\s,3);
	}
	\filldraw[white] (.5,1.25) -- (5.5,1.25) -- (5.5,2.75) -- (.5,2.75) -- (.5,1.25);
	\draw[line width = 1pt, black] (.5,1.25) -- (5.5,1.25) -- (5.5,2.75) -- (.5,2.75) -- (.5,1.25);
	\draw[line width = 1pt, black, ->] (1,1.5) -- (1,2.5);
	\draw[line width = 1pt, black, ->] (5,1.5) -- (5,2.5);
	\node[anchor = south] at (3,1.25) {D};
	\draw[decorate, decoration = {brace, mirror, amplitude = 2 pt}, yshift = -3pt] (1,0.5) -- (5,0.5) node [midway,yshift = -7pt] {\footnotesize{m}};
\end{tikzpicture}	$ \qquad \overset{Y^{m}}{\longrightarrow} $
\begin{tikzpicture}[scale = 1/3, baseline ={(current bounding box.center)}]
	\foreach \s in {1,2,6,7}{
		\draw[line width = 1pt, black] (\s, 1) -- (\s, 7);
	};
	\foreach \s in {2,3,5,6}{
		\draw[line width = 3pt, white] (.5,\s) -- (7.5,\s);
		\draw[line width = 1pt, black] (.5,\s) -- (7.5,\s);
	}
	\filldraw[white] (3.25,1.5) -- (4.75,1.5) -- (4.75,6.5) -- (3.25,6.5) -- (3.25,1.5);
	\draw[line width = 1pt, black] (3.25,1.5) -- (4.75,1.5) -- (4.75,6.5) -- (3.25,6.5) -- (3.25,1.5);
	\draw[line width = 1pt, black, ->] (3.5,2) -- (4.5,2);
	\draw[line width = 1pt, black, ->] (3.5,6) -- (4.5,6);
	\node[anchor = west] at (3.25,4) {D};
	\draw[decorate, decoration = {brace, mirror, amplitude = 2 pt}, yshift = -3pt] (1,0.5) -- (7,0.5) node [midway,yshift = -7pt] {\footnotesize{n}};
	\draw[decorate, decoration = {brace, mirror, amplitude = 2 pt}, xshift = -3pt] (.5,6.5) -- (.5,1.5) node [midway,xshift = -7pt] {\footnotesize{m}};
	\draw[line width = 2pt, black] (.5,1) -- (7.5,1);
	\draw[line width = 2pt, black] (.5,7) -- (7.5,7);
\end{tikzpicture}
\end{center}
\caption{An illustration of the action of the map $Y^{m}$; the D box represent some diagram in $\tl{m}$ and the arrows illustrate its orientation. The map then rotates the diagram $90$ degrees clockwise, and insert it on the defect. The result is a central element of $\atl{n}$.}\label{fig:defectmap}
\end{figure}
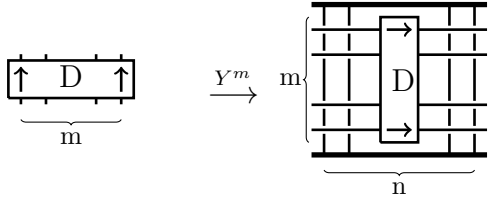

 Taking powers of the hoop operators will increase the width of the defects by increasing the number of lines going across the system; one can then imagine Temperley-Lieb operators acting horizontally on the defect. For instance, we can introduce the operators
	\begin{align}
		Y^{2}(e_{1}) = \;&
	\begin{tikzpicture}[scale = 1/3, baseline ={(current bounding box.center)}]
	\draw[line width = 1pt, black] (0.5,1) -- (7.5,1);
	\draw[line width = 1pt, black] (0.5,4) -- (7.5,4);
	\foreach \s in {1,2,6,7}{
		\draw[line width = 1pt, black] (\s, 1) -- (\s,4);
	}
	\draw[line width = 3pt, white] (0,2) -- (8,2);
	\draw[line width = 3pt, white] (0,3) -- (8,3);
	\draw[line width = 1pt, black] (0.5,2) -- (7.5,2);
	\draw[line width = 1pt, black] (0.5,3) -- (7.5,3);
	\filldraw[line width = 1pt, white] (3,2) -- (5,2) -- (5,3) -- (3,3) -- (3,2);
	\draw[line width = 1pt, black] (3,2) .. controls (4,2) and (4,3) .. (3,3);
	\draw[line width = 1pt, black] (5,2) .. controls (4,2) and (4,3) .. (5,3); 
 	\end{tikzpicture} \; = (\q + \q^{-1}) 1_{\atl{n}},\label{eq:defectmap1}\\
	Y^{3}(e_{1}e_{2}) = \;&
	\begin{tikzpicture}[scale = 1/3, baseline ={(current bounding box.center)}]
	\draw[line width = 1pt, black] (0.5,1) -- (7.5,1);
	\draw[line width = 1pt, black] (0.5,5) -- (7.5,5);
	\foreach \s in {1,2,6,7}{
		\draw[line width = 1pt, black] (\s, 1) -- (\s,5);
	}
	\draw[line width = 3pt, white] (0,2) -- (7.5,2);
	\draw[line width = 3pt, white] (0,3) -- (7.5,3);
	\draw[line width = 3pt, white] (0,4) -- (7.5,4);
	\draw[line width = 1pt, black] (0.5,2) -- (7.5,2);
	\draw[line width = 1pt, black] (0.5,3) -- (7.5,3);
	\draw[line width = 1pt, black] (0.5,4) -- (7.5,4);
	\filldraw[line width = 1pt, white] (3,2) -- (5,2) -- (5,4) -- (3,4) -- (3,2);
	\draw[line width = 1pt, black] (3,3) .. controls (4,3) and (4,4) .. (3,4);
	\draw[line width = 1pt, black] (3,2) .. controls (4,2) and (4,4) .. (5,4);
	\draw[line width = 1pt, black] (5,2) .. controls (4,2) and (4,3) .. (5,3); 
 	\end{tikzpicture} \; = Y.\label{eq:defectmap2}
	\end{align}
One recognize that this corresponds to taking a Markov trace in the horizontal direction; in particular, the operator $Y^{m}$ can be seen as a map from $\tl{m}$ to the ring of endomorphisms of~$\atl{n}$: 
\be
Y^m \colon \; \tl{m} \to \mathrm{End}_{\atl{n}} \big(\atl{n}\big)\ ,
\ee 
where a given element in $\tl{m}$ considered as a diagram is just placed on the $m$ horizontal strands, as in Fig.~\ref{fig:defectmap}. It is easy to see that the image of this map lives in the center of $\atl{n}$, and the central elements provide an endomorphism via the multiplication. We similarly define the mapping 
\be\label{eq:bYm}
\bar{Y}^m \colon \tl{m} \to \mathrm{End}_{\atl{n}} \big(\atl{n}\big)
\ee
 whose image is also in the center of~$\atl{n}$, and that can be represented graphically similarly to Fig.~\ref{fig:defectmap}, however with horizontal lines going under the vertical ones.

\subsection{Higher-spin operators $Y_{j}$ and $\bar{Y}_{j}$}\label{sec:higherspin}
Instead of applying the defect operators $Y^m$ on individual elements of the Temperley-Lieb algebra, we can  have them act on an entire ideal, sending each to a sub-ring of the ring of endomorphisms of $\atl{n}$. If $\q$ is generic, every indecomposable left-ideal of $\tl{m}$ is isomorphic to one of the form $\mathsf{S}_{j}(m) = \tl{m}P_{j} $, where $P_{j}$
 is an idempotent of spin $j$
 (a primitive idempotent in $\tl{m}$ such that $\tl{m} P_j$ is isomorphic to the standard representation with $2j$ through lines).
 When  $j= m/2$ one can use the Jones-Wenzl projectors 
\be\label{eq:JW-P}
P_{m/2} = W^{m+1}_{1}\; ,
\ee
 defined recursively through the following formula: 
\begin{align}\label{eq:JW-recursive}
	W^{1}_{i}(n) & \equiv W^{2}_{i}(n) \equiv 1_{\tl{n}},  \notag \\
	W^{m}_{i}(n) & \equiv W^{m-1}_{i+1}(n)\left( 1_{\tl{n}} - \frac{\q^{m-2} - \q^{2-m}}{\q^{m-1} - \q^{1-m}}e_{i} \right)W^{m-1}_{i+1}(n),
\end{align}
where the index $m$ is related to the spin as above, and $i$ is just the lattice position.

Recall that $P_{j}$ is an idempotent, i.e.\ $P_j P_j = P_j$,  and the map $Y^{m}$ has the property of a trace, we then have 
$$
Y^{m}(x P_{j}) = Y^{m}(P_{j} x P_{j})
$$
 for all $x\in \tl{m}$. By construction, $P_{j} x P_{j}$ is an endomorphism of the ideal $\mathsf{S}_{j}(m)$ (by multiplication on the right), which is simple whenever $\q$ is generic; it follows that $ P_{j} x P_{j} = \lambda_{x}P_{j}$ for some $\lambda_{x} \in \mathbb{C}$, and thus that  
	\begin{equation} \label{eq:Ym-Sj}
		Y^{m}(\mathsf{S}_{j}(m)) =  \mathbb{C}Y_{j}, 
	\end{equation}
where we introduced a special central element
\be\label{eq:Yj-def}
 Y_{j} := Y^{2j}(W^{2j+1}_{1}) .
\ee
Here, we used the fact that the trace of $P_{j}$ is independent both of $m$,  and of the particular choice of $P_{j}$ we made (see Appendix~\ref{sec:rigor.hspindefect} for details of the proof). In particular, the identity~\eqref{eq:Ym-Sj} makes sense and is true for any valid value of $m$ when the ideal $\mathsf{S}_{j}(m)$ is non-zero. 

\medskip

Using the recurrence relation~\eqref{eq:JW-recursive} for the Jones-Wenzl projectors, we find in Appendix~\ref{sec:rigor.hspindefect} that
\be\label{eq:Yj-U}
Y_{j} = \mathsf{U}_{2j}\big(Y/2\big),
\ee
 where $\mathsf{U}_{k}(x)$ is the Chebyshev polynomial of the second kind, of order $k$. For instance, we have
\begin{eqnarray}
 Y_{1/2} &=& Y  \,, \nonumber \\
 Y_1 &=& (Y_{1/2})^2 - 1 \,, \nonumber \\
 Y_{3/2} &=& (Y_{1/2})^3 - 2 Y_{1/2} \,,\nonumber \\
 Y_2 &=& (Y_{1/2})^4 - 3 (Y_{1/2})^2 + 1 \,.\nonumber 
\end{eqnarray}
Recall that $Y$ acts on $\mathsf{W}^{o}_{k,\delta}$ as $(\delta + \delta^{-1})$; writing $\delta = e^{i \theta}$, the higher-spin operator eigevalues are thus
\begin{align*}
Y_{j} = \frac{\sin((2j+1)\theta)}{\sin \theta}.
\end{align*}

The important observation is that the properties of the Chebyshev polynomials allow us to decompose products of $Y_{j} $s:
\begin{equation}\label{eq:Y-fusion}
	Y_{j}\cdot Y_{k} = \sum_{r = |j-k|}^{j+k} Y_{r}.
\end{equation}

We finally note that the whole construction of this section would work equally well if the defect had been going under the strings instead of over them, by simply replacing $Y$ with $\bar{Y}$ everywhere it appears. We begin with the map $\bar{Y}^m$ defined in~\eqref{eq:bYm}. Its properties are identical to those of the map $Y^m$ in every way; applying it to the ideals $\mathsf{S}_{j}(m)$  yields higher-spin defect operators $\bar{Y}_{j}$ whose eigenvalues on $\mathsf{W}^{u}_{k,\delta}$ are
\begin{align*}
\bar{Y}_{j} = \frac{\sin((2j+1)(\phi))}{\sin \phi},
\end{align*}
where $\delta \equiv e^{i \phi}$.
And they have similarly the fusion
\begin{equation}\label{eq:bY-fusion}
	\bar{Y}_{j}\cdot \bar{Y}_{k} = \sum_{r = |j-k|}^{j+k} \bar{Y}_{r}.
\end{equation}

Using Theorem~\ref{thm:main}, we see that the algebra generated by $Y_j$ and $\bar{Y}_j$, for all  non-negative half-integer $j$, is identified with the whole center $\Z$. 


Finally, we note that by theorem \ref{thm:main}, the center of the algebra is spanned by polynomials in $Y$ and $\bar{Y}$; since the Chebyshev polynomials form a basis of the space of polynomials in one variable, one can then conclude that any element in the center can be  written as a linear combination of ${Y}_{j}$, $ \bar{Y}_{k}$, and ${Y}_{j}\cdot \bar{Y}_{k}$ using~\eqref{eq:Y-fusion} and~\eqref{eq:bY-fusion}.
We finally note that for the ``mixed" fusion ${Y}_{j}\cdot \bar{Y}_{k}$ there is, to our knowledge, no interesting decomposition: the relation in lemma \ref{lem:YYbar.poly.rel} between $Y$ and $\bar{Y}$ is polynomial of order $n+1$, so any decomposition of ${Y}_{j}\cdot \bar{Y}_{k}$ would heavily depend on the size of the lattice.  In particular, in the continuum limit of the model, one would then expect the two hoop operators to become algebraically independent. 
\subsection{Relation to symmetric polynomials.}\label{sec:JM-elements}
While identifying the topological defect operators with the hoop operator is an intuitive choice, there are many other known central elements, which could also lead to topological defects. These are built from the so-called Jucys-Murphy elements; let 
\begin{equation}\label{eq:def-J}
	J_{1} \equiv  \bar{b}, \qquad J_{i} \equiv g_{i-1}J_{i-1}g_{i-1} = (-\q)^{-3/2} \;
	\begin{tikzpicture}[scale = 1/3,baseline={(current bounding box.center)},yscale= -1] 
	\draw[black, line width = 1pt] (3,2) .. controls (3,1) ..  (7,1);
	\foreach \r in {0,2,4,6}{
		\draw[white, line width = 3pt] (\r,0) -- (\r,2);
		\draw[black, line width = 1pt] (\r,0) -- (\r,2);
	}
	\draw[white, line width = 3pt] (-1,1) .. controls (3,1) .. (3,0);
	\draw[black, line width = 1pt] (-1,1) .. controls (3,1) .. (3,0);
	\filldraw[white] (-3/2,0)-- (-1/2,0) -- (-1/2,2) -- (-3/2,2) -- (-3/2,0);
	\filldraw[white] (15/2,0)-- (13/2,0) -- (13/2,2) -- (15/2,2) -- (15/2,0);
	\draw[black, line width = 2pt] (-1/2,0) -- (13/2,0);
	\draw[black, line width = 2pt] (-1/2,2) -- (13/2,2);
	\foreach \r in {1,5}{
		\node[anchor = north] at (\r, 0) {$\hdots$};
		\node[anchor = south] at (\r, 2) {$\hdots$};
	}
	\draw[decorate, decoration = {brace, mirror, amplitude = 4 pt}, yshift = 3pt] (-.5,2.5) -- (2.5,2.5) node [midway,yshift = -7pt] {\footnotesize{i-1}};
	\draw[decorate, decoration = {brace, mirror, amplitude = 4 pt}, yshift = 3pt] (3.5,2.5) -- (6.5,2.5) node [midway,yshift = -7pt] {\footnotesize{n-i}};
	\end{tikzpicture} \;, \qquad i = 2,\hdots n,
\end{equation}
\begin{equation}\label{eq:def-M}
	M_{1} \equiv  b, \qquad M_{i} \equiv g_{i-1}M_{i-1}g_{i-1} = (-\q)^{-3/2} \;
	\begin{tikzpicture}[scale = 1/3,baseline={(current bounding box.center)},yscale = -1] 
	\draw[black, line width = 1pt] (3,2) .. controls (3,1) ..  (-1,1);
	\foreach \r in {0,2,4,6}{
		\draw[white, line width = 3pt] (\r,0) -- (\r,2);
		\draw[black, line width = 1pt] (\r,0) -- (\r,2);
	}
	\draw[white, line width = 3pt] (7,1) .. controls (3,1) .. (3,0);
	\draw[black, line width = 1pt] (7,1) .. controls (3,1) .. (3,0);
	\filldraw[white] (-3/2,0)-- (-1/2,0) -- (-1/2,2) -- (-3/2,2) -- (-3/2,0);
	\filldraw[white] (15/2,0)-- (13/2,0) -- (13/2,2) -- (15/2,2) -- (15/2,0);
	\draw[black, line width = 2pt] (-1/2,0) -- (13/2,0);
	\draw[black, line width = 2pt] (-1/2,2) -- (13/2,2);
	\foreach \r in {1,5}{
		\node[anchor = north] at (\r, 0) {$\hdots$};
		\node[anchor = south] at (\r, 2) {$\hdots$};
	}
	\draw[decorate, decoration = {brace, mirror, amplitude = 4 pt}, yshift = 3pt] (-.5,2.5) -- (2.5,2.5) node [midway,yshift = -7pt] {\footnotesize{i-1}};
	\draw[decorate, decoration = {brace, mirror, amplitude = 4 pt}, yshift = 3pt] (3.5,2.5) -- (6.5,2.5) node [midway,yshift = -7pt] {\footnotesize{n-i}};
	\end{tikzpicture} \;, \qquad i = 2,\hdots n.
\end{equation}
It is straightforward, though tedious, to prove that all $J_i$, for $1\leq i\leq n$, commute with each others and so do the~$M_i$, see~\cite{GLDiagramAlgebras,Halverson}. Furthermore, $J_i$ and $M_i$ are invertible and if $P(x_1, \hdots, x_{n})$ is a symmetric Laurent polynomial, then $P((-\q)J_{1}, \hdots, (-\q)^{n}J_{n})$ and  $P((-\q)^{1}M_{1}, \hdots, (-\q)^{n}M_{n})$ are central in $\atl{n}$. All of these can be generated from the power-sum symmetric polynomials
\begin{equation}\label{def:Ck}
	C_{k}(n) = \sum_{i=1}^{n} ((-\q)^{i+1} M_{i})^{k}, \qquad \bar{C}_{k}(n) = \sum_{i=1}^{n} ((-\q)^{i+1} J_{i})^{k},
\qquad k\in \mathbb{Z}.	
\end{equation}
However, it turns out that these are related to the hoop operators through the following relations:
\begin{equation}\label{eq:JMandTopD1}
	C_{k}(n) + C_{-k}(n) = (-\q)^{- n k}\bar{C}_{k}(n) + (-\q)^{n k}\bar{C}_{-k}(n)=  2 [n]_{k} \mathsf{T}_{k}(\bar{Y}/2),
\end{equation}
\begin{equation}\label{eq:JMandTopD2}
	\bar{C}_{k}(n) + \bar{C}_{-k}(n) = (-\q)^{- n k}C_{k}(n) + (-\q)^{ n k}C_{-k}(n) = 2 [n]_{k} \mathsf{T}_{k}(Y/2),
\end{equation}
where we defined 
$$
[n]_{k} \equiv \frac{ (-\q)^{k n} - (-\q)^{-k n}}{(-\q)^{k} - (-\q)^{-k}},
$$
and here it is understood that $[n]_{0} \equiv n $, and $\mathsf{T}_{k}(x)$ is the $k$th Chebyshev polynomial of the first kind. The proof of these identities can be found in Appendix~\ref{sec:rigor.JM} and it uses the important relations that hold in $\atl{n}$\footnote{However note that there relations do not hold in the corresponding affine Hecke algebra that covers $\atl{n}$.}
 $$
  (-\q)^{2} J_{i}  + (-\q)^{-2} J_{i}^{-1}  = (-\q) J_{i+1}  + (-\q)^{-1} J_{i+1}^{-1}.
 $$
  If $(-\q)^{n k} \neq 1$, the relations~\eqref{eq:JMandTopD1}-\eqref{eq:JMandTopD2} can be combined to find
\begin{equation}
	((-\q)^{k} - (-\q)^{-k})C_{k}(n) = 2\left((-\q)^{k n} \mathsf{T}_{|k|}( \bar{Y}/2 ) - \mathsf{T}_{|k|}( Y/2 )\right),
\end{equation}
\begin{equation}
	((-\q)^{k} - (-\q)^{-k})\bar{C}_{k}(n) = 2 \left((-\q)^{k n} \mathsf{T}_{|k|}( Y/2 ) - \mathsf{T}_{|k|}( \bar{Y}/2 )\right).
\end{equation}
Finally, using the properties of the Chebyshev polynomials it follows that
\begin{equation}
	Y_{k/2} = \sum_{\underset{\text{step } = 2}{j = -k}}^{k} \frac{1}{[n]_{j}}\bar{C}_{j}(n), \qquad \bar{Y}_{k/2} = \sum_{\underset{\text{step } = 2}{j = -k}}^{k} \frac{1}{[n]_{j}}C_{j}(n).
\end{equation}

\section{Lattice topological defects: direct channel}\label{sec:4}
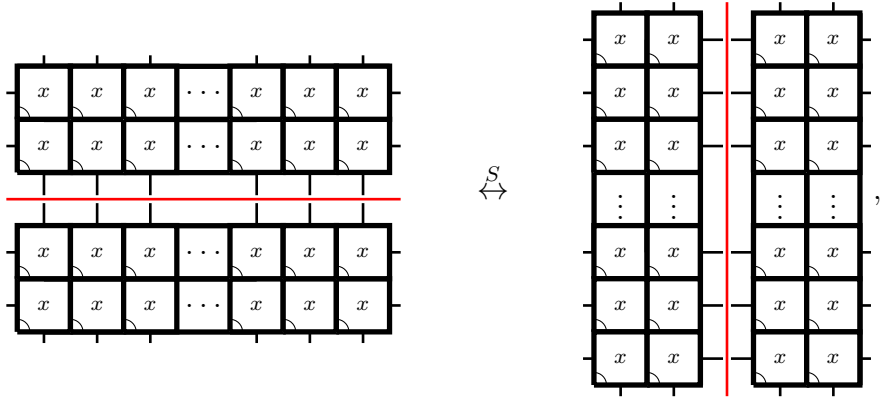
\begin{figure}
\begin{center}
	\begin{tikzpicture}[baseline={(current bounding box.center)}]
	\foreach \r in {0,1,2,4,5,6}{
		\draw[black, line width = 1pt] (.707*\r, -.5) -- (.707*\r, .5 + .707*4);
	};
	\foreach \s in {0,1,3,4}{
		\draw[black, line width = 1pt] (-.5, .707*\s) -- (.5 + .707*1, .707*\s);
		\draw[black, line width = 1pt] (.5+.707*4, .707*\s) -- (.5 + .707*6, .707*\s);
	};
	\draw[white, line width = 3pt] (-.5, .707*2) -- (.5 + .707*6, .707*2);
	\draw[red, line width = 1pt] (-.5, .707*2) -- (.5 + .707*6, .707*2);
	\foreach \s in {0,1,3,4}{
	\node at (0 + 0.707*3,0.707*\s) {$\hdots$};
	\draw[line width = 2] (0,0.707*\s -0.353553) -- (4*.707,0.707*\s -0.353553);
	\draw[line width = 2] (-.354553,0.707*\s + 0.353553) -- (13*.353553,0.707*\s + 0.353553);
		\foreach \r in {0,1,2,4,5,6}{
	\node at (0 + 0.707*\r ,0.707*\s) {\rutiles{x}{-135}};
	};
	};
\end{tikzpicture} $\qquad \overset{S}{\leftrightarrow} \qquad $
\begin{tikzpicture}[rotate = -90,baseline={(current bounding box.center)}]
	\foreach \r in {0,1,2,4,5,6}{
		\draw[black, line width = 1pt] (.707*\r, -0.5) -- (.707*\r, .5 + .707*4);
	};
	\foreach \s in {0,1,3,4}{
		\draw[black, line width = 1pt] (-.5, .707*\s) -- (.5 + .707*1, .707*\s);
		\draw[black, line width = 1pt] (.5+.707*4, .707*\s) -- (.5 + .707*6, .707*\s);
	};
	\draw[white, line width = 3pt] (-.5, .707*2) -- (.5 + .707*6, .707*2);
	\draw[red, line width = 1pt] (-.5, .707*2) -- (.5 + .707*6, .707*2);
	\foreach \s in {0,1,3,4}{
	\node at (0 + 0.707*3,0.707*\s) {$\vdots$};
	\draw[line width = 2] (0,0.707*\s -0.353553) -- (4*.707,0.707*\s -0.353553);
	\draw[line width = 2] (-.354553,0.707*\s + 0.353553) -- (13*.353553,0.707*\s + 0.353553);
		\foreach \r in {0,1,2,4,5,6}{
	\node at (0 + 0.707*\r ,0.707*\s) {\rutiles{x}{-135}};
	};
	};
\end{tikzpicture},

     \caption{The modular $S$-transformation, which is the lattice rotation by $90^o$,  sends a defect $Y$ (in red) in the crossed channel to a defect in the direct channel, and vice versa.}\label{DMfig3}
\end{center}
\end{figure}
In this section, we are interested in interpretation of previously introduced defects $Y_j$ and $\bar{Y}_j$ in the direct channel, or in their Hamiltonian realization. We will  consider only the case of generic $\q$.
The action of the defect $Y_{1/2}$ in the direct channel can be inferred by a simple modular transformation -- that is, a rotation by $90^o$ as in Fig.~\ref{DMfig3}. What this means microscopically is that we should have a system where, on top of the usual TL interaction terms, we have an extra line that simply goes over/under the others, and this contributes to defect terms in the Hamiltonian that we also call ``impurities". Below, we calculate explicitly the Hamiltonians with impurities that correspond to the defect operators  $Y_j$ and $\bar{Y}_j$, and study their spectral problem.

The Hamiltonian with $m$ defect lines can be obtained as a logarithmic derivative evaluated at $x = 1$ of the transfer matrix $T_{n}(x;m)$ given in Fig.~\ref{fig:transfermatrix.defect}.
In this case, we obtain the Hamiltonian on $n+m$ sites,  as an  element in $\atl{n+m}$,
\begin{equation*}
		H^{u}_{n,m}  = \sum_{j = 1}^{n-1} e^{(n+m)}_{j}   +  \mu_{n,m}^{-1} e^{(n+m)}_{n}  \mu_{n,m},
	\end{equation*}
	where $\mu_{n,m} = g_{n}g_{n+1}\hdots g_{n+m}$, with $g_i$ defined in Fig.~\ref{fig:br-diag}. 
This derivation is similar to the standard calculation with the homogeneous transfer matrix $T_n(x)$, recall the remark below~\eqref{eq:T-Y}.	We are interested in the  spectral problem of $H^{u}_{n,m}$ acting on some module $M$ over $\atl{n+m}$.
 It is important to note that this Hamiltonian can be written as\footnote{This is even clearer on the level of transfer matrices: $T_{n}(x;m)=  \phi^{u}_{n,m}(T_n(x))$, by definition of  $\phi^{u}_{n,m}$.} 
	\begin{equation}\label{eq:Hnm-phi}
	H^{u}_{n,m}  = \phi^{u}_{n,m}\big(\sum_{j=1}^{n}e^{(n)}_{j}\big),
	\end{equation}
 i.e.\ as the image of the standard periodic TL Hamiltonian
 	\begin{equation}
 	H_{n}= \sum_{j=1}^{n} e_{j}
	\end{equation}
on $n$ sites  under the embedding map  $\phi^{u}_{n,m}$ defined in Section~\ref{sec:def.tower}. In order to simplify the solution of this spectral problem as much as possible, it is customary to start by searching for elements of $\mathsf{End}_{\mathbb{C}}(M)$ which commute with $H^{u}_{n,m}$. With the property of the embeddings $\phi^u$ and $\psi^o$ discussed after~\eqref{eq:psi-def} and using~\eqref{eq:Hnm-phi}, it is clear that for any $a \in \tl{m}$, $ \psi^o(a)$  commutes with $H^{u}_{n,m}$.
 It follows in particular that, as matrices\footnote{We can introduce similarly the versions $H^{o}_{n,m}$ and $H^{o}_{n,m}[\lambda]$ using the over lines  map $\phi^o$.},
\begin{equation}
	 H^{u}_{n,m} = \sum_{\lambda\in \Lambda_m} H^{u}_{n,m}[\lambda],
\end{equation}
where $\Lambda_m$ is a complete set of primitive orthogonal idempotents of $\tl{m}$, and
\begin{equation}\label{eq:Hnm-idemp}
	H^{u}_{n,m}[\lambda] \equiv \psi^o_{m,n}(\lambda) H^{u}_{n,m} =  H^{u}_{n,m} \psi^o_{m,n}(\lambda).
\end{equation}
Note that because the $\lambda$s are orthogonal, these matrices are themselves orthogonal to each others.
The spectral problem of $H^{u}_{n,m}$ is thus reduced to that of $H^{u}_{n,m}[\lambda]$.
To solve this reduced spectral problem,  we present an algebraic construction linking  the spectrum of $H^{u}_{n,m}[\lambda]$ with the spectrum of  the standard Hamiltonian $H_{n}$   acting on a certain ``fusion quotient" module of $\atl{n}(\q)$. This requires certain preparation and an algebraic discussion below. We then come back  to the spectral problem in Section~\ref{sec:4.4} with the final result formulated in Theorem~\ref{eq:thm-H}, and then provide an explicit example based on the twisted XXZ chains in Section~\ref{eq:sec4-ex}.

\begin{figure}
	\begin{center}
	$T_{n}(x;m) = \;\; $ 
		\begin{tikzpicture}[scale = 1, baseline={(current bounding box.center)}]	
	\draw[black, line width = 2pt] (1,.5) -- (10,.5);
	\draw[black, line width = 2pt] (1,1.5) -- (10,1.5);
	\foreach \r in {3,9}{
		\node at (\r, 1) {$\hdots $};
	}
	\node at (6, .75) {$\hdots $};
	\node at (6, 1.25) {$\hdots $};
	\draw[black, line width = 1pt] (5,.5) -- (5,1.5);
	\draw[black, line width = 1pt] (5+.5,.5) -- (5+.5,1.5);
	\draw[black, line width = 1pt] (7-.5,.5) -- (7-.5,1.5);
	\draw[black, line width = 1pt] (7,.5) -- (7,1.5);
	\draw[white, line width = 3pt] (4,1) -- (8,1);
	\draw[black, line width = 1pt] (4,1) -- (8,1);	
	\foreach \r in {1,2,4,8,10}{
		\node at (\r,1) {\correcttiles{x}{-135}{.707}};
	}
	\draw[decorate, decoration = {brace, mirror, amplitude = 4 pt}, yshift = -5pt] (5,.5) -- (7,.5) node [midway,yshift = -7pt] {\footnotesize{m}};
\end{tikzpicture}
\caption{$T_{n}(x;m)$ is a transfer matrix carrying a defect of width $m$ going under the other lines; taking its logarithmic derivative evaluated at $x = 1$ yields the Hamiltonian $H^{u}_{n,m}$
 (up to a normalization factor).}\label{fig:transfermatrix.defect}
	\end{center}
\end{figure}

\medskip

Let us begin with the idea that stays behind the two algebraic constructions formulated below. Adding the extra lines/defects in the direct channel can be realised as a functor that combine a module of $\atl{}$ (the bulk model) with a module of $\tl{}$ (the defect) into a new module of $\atl{}$ (the bulk model with a defect); it turns out that there are (at least) two natural ways of doing this: one can \textsl{add} new strands carrying the defect to the module, a process we  call the \emph{fusion product}, or one can \textsl{impose} the defect on an existing part of the module, a process we  call the \emph{fusion quotient}. 

\subsection{The fusion product}\label{sec:the_fusion_product}
{\center \emph{This section uses the notation introduced in Section \ref{sec:def.tower}}.}

Let $m,k$ both be positive integers, we give $\atl{m+k}$ the structure of a $(\atl{m+k}, \atl{m} \otimes_{\mathbb{C}} \tl{k})$ bimodule by letting $\atl{m+k}$ act on the left through the natural representation, and $\atl{m} \otimes_{\mathbb{C}} \tl{k}$ acts on the right by the morphism $\phi^{u/o}_{m,k} \otimes_{\mathbb{C}} \psi^{o/u}_{k,m}$, where we identified $\tl{k}$ with its image in $\atl{k}$. For $M$ an $\atl{m}$ module, and $V$ a $\tl{k} $-module, our definition of the fusion product can then be written 
\begin{equation}
	M \times^{u/o}_{f} V \equiv \atl{m+k} \otimes_{\atl{m} \otimes_{\mathbb{C}} \tl{k}} \left(M \otimes_{\mathbb{C}} V \right),
\end{equation}
where the superscript $u/o$ denotes which one of $\phi^{u/o}_{m,k}$ we used to define the bimodule structure of $\atl{m+k} $. From a more physical point of view, this corresponds to having a bulk model described by $M$ which contains an isolated sub-system $V$, such that they are both entirely blind to each others so that the Hilbert space of the system is simply the tensor product of the Hilbert spaces of $M \otimes_{\mathbb{C}} V$; at some point one then remove the barrier between the two sub-systems and thus letting $V$ \emph{propagate} freely inside $M$. Note also that this fusion is related, though different, to ones introduced
 previously in the literature~\cite{GS,BSA}. We discuss this more in Appendix~\ref{sec:previousfusions} where also important properties of the hoop operators are studied in relation to the tower homomorphisms  $\phi^{u/o}$ and  $\psi^{o/u}$.

Before giving the general result for the fusion we give a small example and compute the fusion product of two standard modules $\mathsf{W}_{1/2,z}(3) \times^{o}_{f} \mathsf{S}_{1/2}(1) $. Since the standard modules are cyclic, their fusion is also, and thus $\mathsf{W}_{1/2,z}(3) \times^{o}_{f} \mathsf{S}_{1/2}(1) = \atl{4}y$, with
\begin{equation}
	y =\; 
	\begin{tikzpicture}[scale = 1/3,baseline={(current bounding box.center)},yscale = -1] 
	\draw[black, line width = 1pt] (3,3) --  (3,1);
	\draw[black, line width = 1pt] (1,3) -- (1,1);
	\draw[black, line width = 1pt] (2,3) -- (2,1);
	\draw[black, line width = 1pt] (4,1) -- (4,3);
	\draw[black, line width = 2pt] (.5,1) -- (4.5,1);
	\draw[black, line width = 2pt] (.5,3) -- (4.5,3);
	\draw[black, line width = 1pt] (.5,4) -- (.5,6);
	\draw[black, line width = 1pt] (1.5,4) .. controls (1.5,5) and (2.5,5) .. (2.5,4);
	\node[anchor = north] at (3.5,4) {$\otimes$};
	\draw[black, line width = 1pt] (4.5,4) -- (4.5,6);
	\draw[black, line width = 2pt] (0,4) -- (3,4);
	\draw[black, line width = 2pt] (4,4) -- (5,4);
	\end{tikzpicture} \;.
\end{equation}
where we also introduced our diagram notation for the fusion product: the diagram at the top is the element of $\atl{4}$, the one on the bottom left corner is the element of 
$\mathsf{W}_{1/2,z}(3)$, 
and the one on the bottom right corner is the element of $\mathsf{S}_{1/2}(1)$. Since this module is cyclic, we can choose a basis of the form $\lbrace a_{i} y | i= 1, ... \rbrace $ for some subset $\lbrace a_{i} \rbrace \subset \atl{4}$; in the case at hand the simplest choice is\footnote{See appendix \ref{app:fusion.basis} to see how this particular choice was obtained.}
\begin{equation}\label{eq:fusion.base1}
	a_{1} = \;
	\begin{tikzpicture}[scale = 1/3,baseline={(current bounding box.center)},yscale = -1] 
	\draw[black, line width = 2pt] (.5,1) -- (4.5,1);
	\draw[black, line width = 2pt] (.5,3) -- (4.5,3);
	\draw[black, line width = 1pt] (1,1) .. controls (1,2) and (2,2) .. (2,1);
	\draw[black, line width = 1pt] (1,3) .. controls (1,2) and (2,2) .. (2,3);
	\draw[black, line width = 1pt] (3,3) -- (3,1);
	\draw[black, line width = 1pt] (4,3) -- (4,1);
	\end{tikzpicture} \qquad
	a_{2} = \; 
	\begin{tikzpicture}[scale = 1/3,baseline={(current bounding box.center)},yscale = -1] 
	\draw[black, line width = 2pt] (.5,1) -- (4.5,1);
	\draw[black, line width = 2pt] (.5,3) -- (4.5,3);
	\draw[black, line width = 1pt] (3,1) .. controls (3,2) and (2,2) .. (2,1);
	\draw[black, line width = 1pt] (1,3) .. controls (1,2) and (2,2) .. (2,3);
	\draw[black, line width = 1pt] (3,3) .. controls (3,2) and (1,2) .. (1,1);
	\draw[black, line width = 1pt] (4,3) -- (4,1);
	\end{tikzpicture}  \qquad
	a_{3} =\; 
	\begin{tikzpicture}[scale = 1/3,baseline={(current bounding box.center)},yscale = -1] 
	\draw[black, line width = 2pt] (.5,1) -- (4.5,1);
	\draw[black, line width = 2pt] (.5,3) -- (4.5,3);
	\draw[black, line width = 1pt] (3,1) .. controls (3,2) and (4,2) .. (4,1);
	\draw[black, line width = 1pt] (1,3) .. controls (1,2) and (2,2) .. (2,3);
	\draw[black, line width = 1pt] (3,3) .. controls (3,2) and (1,2) .. (1,1);
	\draw[black, line width = 1pt] (4,3) .. controls (4,2) and (2,2) .. (2,1);
	\end{tikzpicture}  \qquad
	a_{4} = \;
	\begin{tikzpicture}[scale = 1/3,baseline={(current bounding box.center)},yscale = -1] 
	\draw[black, line width = 1pt] (5,1) .. controls (5,2) and (4,2) .. (4,1);
	\draw[black, line width = 1pt] (0,1) .. controls (0,2) and (1,2) .. (1,1);
	\draw[black, line width = 1pt] (1,3) .. controls (1,2) and (2,2) .. (2,3);
	\draw[black, line width = 1pt] (3,3) .. controls (3,2) and (2,2) .. (2,1);
	\draw[black, line width = 1pt] (4,3) .. controls (4,2) and (3,2) .. (3,1);
	\filldraw[white] (-.5,1) -- (.5,1) -- (.5,3) -- (-.5,3) -- (-.5,1);
	\filldraw[white] (4.5,1) -- (5.5,1) -- (5.5,3) -- (4.5,3) -- (4.5,1);
	\draw[black, line width = 2pt] (.5,1) -- (4.5,1);
	\draw[black, line width = 2pt] (.5,3) -- (4.5,3);
\end{tikzpicture}
\end{equation}
\begin{equation}\label{eq:fusion.base2}
	a_5 = 
	\begin{tikzpicture}[scale = 1/3,baseline={(current bounding box.center)}] 
	\draw[black, line width = 1pt] (1,1) .. controls (1,2) and (2,2) .. (2,1);
	\draw[black, line width = 1pt] (3,1) .. controls (3,2) and (4,2) .. (4,1);
	\draw[black, line width = 1pt] (1,3) .. controls (1,2) and (2,2) .. (2,3);
	\draw[black, line width = 1pt] (3,3) .. controls (3,2) and (4,2) .. (4,3);
	\draw[black, line width = 2pt] (.5,1) -- (4.5,1);
	\draw[black, line width = 2pt] (.5,3) -- (4.5,3);
	\filldraw[white] (-.5,.5) -- (.5,.5) -- (.5,3.5) -- (-.5,3.5) -- (-.5,.5);
	\filldraw[white] (5-.5,.5) -- (5.5,.5) -- (5.5,3.5) -- (5-.5,3.5) -- (5-.5,.5);
	\end{tikzpicture}, 
	\; a_6 = 
	\begin{tikzpicture}[scale = 1/3,baseline={(current bounding box.center)}] 
	\draw[black, line width = 1pt] (1,1) .. controls (1,2) and (2,2) .. (2,1);
	\draw[black, line width = 1pt] (3,1) .. controls (3,2) and (4,2) .. (4,1);
	\draw[black, line width = 1pt] (1,3) .. controls (1,2) and (0,2) .. (0,3);
	\draw[black, line width = 1pt] (5,3) .. controls (5,2) and (4,2) .. (4,3);
	\draw[black, line width = 1pt] (3,3) .. controls (3,2) and (2,2) .. (2,3);
	\draw[black, line width = 2pt] (.5,1) -- (4.5,1);
	\draw[black, line width = 2pt] (.5,3) -- (4.5,3);
	\filldraw[white] (-.5,.5) -- (.5,.5) -- (.5,3.5) -- (-.5,3.5) -- (-.5,.5);
	\filldraw[white] (5-.5,.5) -- (5.5,.5) -- (5.5,3.5) -- (5-.5,3.5) -- (5-.5,.5);
	\end{tikzpicture},
	\; a_7 =
	\begin{tikzpicture}[scale = 1/3,baseline={(current bounding box.center)}] 
	\draw[black, line width = 1pt] (1,1) .. controls (1,2) and (2,2) .. (2,1);
	\draw[black, line width = 1pt] (3,1) .. controls (3,2) and (4,2) .. (4,1);
	\draw[black, line width = 1pt] (1,3) .. controls (1,1.5) and (4,1.5) .. (4,3);
	\draw[black, line width = 1pt] (2,3) .. controls (2,2) and (3,2) .. (3,3);
	\draw[black, line width = 2pt] (.5,1) -- (4.5,1);
	\draw[black, line width = 2pt] (.5,3) -- (4.5,3);
	\filldraw[white] (-.5,.5) -- (.5,.5) -- (.5,3.5) -- (-.5,3.5) -- (-.5,.5);
	\filldraw[white] (5-.5,.5) -- (5.5,.5) -- (5.5,3.5) -- (5-.5,3.5) -- (5-.5,.5);
	\end{tikzpicture},
	\; a_8 = 
	\begin{tikzpicture}[scale = 1/3,baseline={(current bounding box.center)}] 
	\draw[black, line width = 1pt] (1,1) .. controls (1,2) and (2,2) .. (2,1);
	\draw[black, line width = 1pt] (3,1) .. controls (3,2) and (4,2) .. (4,1);
	\draw[black, line width = 1pt] (1,3) .. controls (1,2) and (0,2) .. (0,3);
	\draw[black, line width = 1pt] (2,3) .. controls (2,1.5) and (5,1.5) .. (5,3);
	\draw[black, line width = 1pt] (3,3) .. controls (3,2) and (4,2) .. (4,3);
	\draw[black, line width = 2pt] (.5,1) -- (4.5,1);
	\draw[black, line width = 2pt] (.5,3) -- (4.5,3);
	\filldraw[white] (-.5,.5) -- (.5,.5) -- (.5,3.5) -- (-.5,3.5) -- (-.5,.5);
	\filldraw[white] (5-.5,.5) -- (5.5,.5) -- (5.5,3.5) -- (5-.5,3.5) -- (5-.5,.5);
	\end{tikzpicture},
\end{equation}
\begin{equation}\label{eq:fusion.base3}
	a_9 = 
	\begin{tikzpicture}[scale = 1/3,baseline={(current bounding box.center)}] 
	\clip (-.5,.5) -- (5.5, .5) -- (5.5,3.5) -- (-.5,3.5) -- (-.5,.5);
	\draw[black, line width = 1pt] (1,1) .. controls (1,2) and (2,2) .. (2,1);
	\draw[black, line width = 1pt] (3,1) .. controls (3,2) and (4,2) .. (4,1);
	\draw[black, line width = 1pt] (1,3) .. controls (1,2) and (0,2) .. (0,3);
	\draw[black, line width = 1pt] (2,3) .. controls (2,1.5) and (-1,1.5) .. (-1,3);
	\draw[black, line width = 1pt] (3,3) .. controls (3,1.5) and (6,1.5) .. (6,3);
	\draw[black, line width = 1pt] (4,3) .. controls (4,2) and (5,2) .. (5,3);
	\draw[black, line width = 2pt] (.5,1) -- (4.5,1);
	\draw[black, line width = 2pt] (.5,3) -- (4.5,3);
	\filldraw[white] (-1-.5,.5) -- (.5,.5) -- (.5,3.5) -- (-1-.5,3.5) -- (-1-.5,.5);
	\filldraw[white] (5-.5,.5) -- (7.5,.5) -- (7.5,3.5) -- (5-.5,3.5) -- (5-.5,.5);
	\end{tikzpicture},
	\; a_{10} = 
	\begin{tikzpicture}[scale = 1/3,baseline={(current bounding box.center)},xscale = -1] 
	\clip (-.5,.5) -- (5.5, .5) -- (5.5,3.5) -- (-.5,3.5) -- (-.5,.5);
	\draw[black, line width = 1pt] (1,1) .. controls (1,2) and (2,2) .. (2,1);
	\draw[black, line width = 1pt] (3,1) .. controls (3,2) and (4,2) .. (4,1);
	\draw[black, line width = 1pt] (1,3) .. controls (1,2) and (0,2) .. (0,3);
	\draw[black, line width = 1pt] (2,3) .. controls (2,1.5) and (5,1.5) .. (5,3);
	\draw[black, line width = 1pt] (3,3) .. controls (3,2) and (4,2) .. (4,3);
	\draw[black, line width = 2pt] (.5,1) -- (4.5,1);
	\draw[black, line width = 2pt] (.5,3) -- (4.5,3);
	\filldraw[white] (-.5,.5) -- (.5,.5) -- (.5,3.5) -- (-.5,3.5) -- (-.5,.5);
	\filldraw[white] (5-.5,.5) -- (5.5,.5) -- (5.5,3.5) -- (5-.5,3.5) -- (5-.5,.5);
	\end{tikzpicture}.
\end{equation}	

It is not trivial at all to show that this set is sufficient, for instance, can $e_{4} a_{2}y$ really be expressed as a linear combination of $a_{i}y$? Indeed it can:
\begin{equation}\label{eq:exfusion1}
	e_{4} a_{2} y \; = 
	\begin{tikzpicture}[scale = 1/3,baseline={(current bounding box.center)},yscale=-1] 
	\draw[black, line width = 1pt] (1,3) .. controls (1,2) and (2,2) .. (2,3);
	\draw[black, line width = 1pt] (3,3) .. controls (3,1.5) and (0,1.5) .. (0,3);
	\draw[black, line width = 1pt] (4,3) .. controls (4,2) and (5,2) .. (5,3);
	\foreach \r in {0,2,4}{
		\draw[ black, line width = 1pt] (\r, 0) .. controls (\r, 1) and (\r + 1, 1) .. (\r + 1, 0);
	} ;
	\filldraw[white] (-.5,0) -- (.5,0) -- (.5,3) -- (-.5,3) -- (-.5,0);
	\filldraw[white] (4.5,0) -- (5.5,0) -- (5.5,3) -- (4.5,3) -- (4.5,0);
	\draw[black, line width = 2pt] (.5,0) -- (4.5,0);
	\draw[black, line width = 2pt] (.5,3) -- (4.5,3);
	\draw[black, line width = 1pt] (.5,4) -- (.5,6);
	\draw[black, line width = 1pt] (1.5,4) .. controls (1.5,5) and (2.5,5) .. (2.5,4);
	\node[anchor = north] at (3.5,4) {$\otimes$};
	\draw[black, line width = 1pt] (4.5,4) -- (4.5,6);
	\draw[black, line width = 2pt] (0,4) -- (3,4);
	\draw[black, line width = 2pt] (4,4) -- (5,4);
\end{tikzpicture} \; = z^{-1} \;
	\begin{tikzpicture}[scale = 1/3,baseline={(current bounding box.center)},yscale = -1] 
	\draw[black, line width = 1pt] (1,3) .. controls (1,2) and (2,2) .. (2,3);
	\draw[black, line width = 1pt] (3,3) .. controls (3,1.5) and (0,1.5) .. (0,3);
	\draw[black, line width = 1pt] (4,3) .. controls (4,2) and (5,2) .. (5,3);
	\foreach \r in {0,2,4}{
		\draw[ black, line width = 1pt] (\r, 0) .. controls (\r, 1) and (\r + 1, 1) .. (\r + 1, 0);
	} ;
	\filldraw[white] (-.5,0) -- (.5,0) -- (.5,3) -- (-.5,3) -- (-.5,0);
	\filldraw[white] (4.5,0) -- (5.5,0) -- (5.5,3) -- (4.5,3) -- (4.5,0);
	\draw[black, line width = 2pt] (.5,0) -- (4.5,0);
	\draw[black, line width = 2pt] (.5,3) -- (4.5,3);
	\draw[black, line width = 1pt] (.5,4) .. controls (.5,5) and (-.5,5) .. (-.5,5);
	\draw[black, line width = 1pt] (1.5,4) .. controls (1.5,5) and (2.5,5) .. (2.5,4);
	\draw[black, line width = 1pt] (.5,6) .. controls (.5,5) and (1.5,5) .. (1.5,6);
	\draw[black, line width = 1pt] (2.5,6) .. controls (2.5,5) and (3.5,5) .. (3.5,6);
	\filldraw[white] (-1,4) -- (0,4) -- (0,6) -- (-1,6) -- (-1,4);
	\filldraw[white] (3,4) -- (4,4) -- (4,6) -- (3,6) -- (3,4); 
	\draw[black, line width = 1pt] (.5,6) -- (.5,8);
	\draw[black, line width = 1pt] (1.5,6) .. controls (1.5,7) and (2.5,7) .. (2.5,6);
	\node[anchor = north] at (3.5,4) {$\otimes$};
	\draw[black, line width = 1pt] (4.5,4) -- (4.5,6);
	\draw[black, line width = 2pt] (0,4) -- (3,4);
	\draw[black, line width = 2pt] (0,6) -- (3,6);
	\draw[black, line width = 2pt] (4,4) -- (5,4);
\end{tikzpicture} \; = z^{-1} \;
	\begin{tikzpicture}[scale = 1/3,baseline={(current bounding box.center)},yscale = -1] 
	\draw[black, line width = 1pt] (1,1) .. controls (1,0) and (2,0) .. (2,1);
	\draw[black, line width = 1pt] (3,1) .. controls (3,-.5) and (0,-.5) .. (0,1);
	\draw[black, line width = 1pt] (4,1) .. controls (4,0) and (5,0) .. (5,1);
	\foreach \r in {0,2,4}{
		\draw[ black, line width = 1pt] (\r, -2) .. controls (\r, -1) and (\r + 1, -1) .. (\r + 1, -2);
	} ;
	\draw[black, line width = 1pt] (1,3) .. controls (1,2) and (2,2) .. (2,3);
	\draw[black, line width = 1pt] (3,1) .. controls (3,2) and (2,2) .. (2,1);
	\draw[black, line width = 1pt] (1,1) .. controls (1,2) and (0,2) .. (0,1);
	\draw[black, line width = 1pt] (3,3) .. controls (3,2) and (5,2) .. (5,3);
	\draw[white, line width = 3pt] (4,1) -- (4,3);
	\draw[black, line width = 1pt] (4,1) -- (4,3);
	\filldraw[white] (-.5,-2) -- (.5,-2) -- (.5,3) -- (-.5,3) -- (-.5,-2);
	\filldraw[white] (4.5,-2) -- (5.5,-2) -- (5.5,3) -- (4.5,3) -- (4.5,-2);
	\draw[black, line width = 2pt] (.5,-2) -- (4.5,-2);
	\draw[black, line width = 2pt] (.5,1) -- (4.5,1);
	\draw[black, line width = 2pt] (.5,3) -- (4.5,3);
	\draw[black, line width = 1pt] (.5,4) -- (.5,6);
	\draw[black, line width = 1pt] (1.5,4) .. controls (1.5,5) and (2.5,5) .. (2.5,4);
	\node[anchor = north] at (3.5,4) {$\otimes$};
	\draw[black, line width = 1pt] (4.5,4) -- (4.5,6);
	\draw[black, line width = 2pt] (0,4) -- (3,4);
	\draw[black, line width = 2pt] (4,4) -- (5,4);
\end{tikzpicture} \; = -(-\q)^{3/2} z^{-1} a_{6}y,
\end{equation}
where the last equality was obtained by using the closed braid identity
\begin{equation}
\begin{tikzpicture}[scale = 1/3,baseline={(current bounding box.center)},yscale = -1]
	\draw[black, line width = 1pt] (0,0) -- (2,2);
	\draw[white, line width = 3pt] (0,2) -- (2,0);
	\draw[black, line width = 1pt] (0,2) -- (2,0);	
	\draw[black, line width = 1pt] (2,0) .. controls (3,0) and (3,2) .. (2,2);
\end{tikzpicture} \; = \; -(-\q)^{3/2} \;
\begin{tikzpicture}[scale = 1/3,baseline={(current bounding box.center)}]
	\draw[black, line width = 1pt] (2,0) .. controls (3,0) and (3,2) .. (2,2);
\end{tikzpicture} \;.
\end{equation}
Using similar tricks, every elements can be brought to a linear combination of the $a_{i}y$. 

Note that it is clear that any element $a \in \atl{4}$ acting on a linear combinations of $a_{5}x$, $a_{6}x$, $\hdots$, $a_{10}x$ will result in another linear combination of those same basis elements; in other words, $ \lbrace a_{i} x| i=5,6,\hdots, 10 \rbrace$ generate a submodule of $\mathsf{W}_{1/2,z}(3) \times^{o}_{f} \mathsf{S}_{1/2}(1) $, which we immediately recognize as $\mathsf{W}_{0, z_{-}}(4)$ 
for some $z_{-} \in \mathbb{C}^{*}$. By definition, $z_{-} + z_{-}^{-1}$ is the weight of the non-contractible loops in the standard module, which must be equal to the eigenvalue of $\bar{Y}$; however $\phi^{o}_{3,1}(\bar{Y}^{(3)}) = \bar{Y}^{(4)} $, so this eigenvalue must be $z (-\q)^{-1/2} + z^{-1} ( -\q )^{1/2}$ (the eigenvalue of $\bar{Y}$ on $\mathsf{W}_{1/2,z}(3)$). We thus conclude\footnote{By definition $\mathsf{W}_{0,z} = \mathsf{W}_{0,z^{-1}} $ so there is no ambiguity here.} that $z_{-} = z(-\q)^{-1/2}$.

Similarly, the quotient of $\mathsf{W}_{1/2,z}(3) \times^{o}_{f} \mathsf{S}_{1/2}(1) $ by this submodule yields the standard module $\mathsf{W}_{1,z_{+}}(4) $. By definition, in $\mathsf{W}_{1,z_{+}}(4) $
\begin{equation}
	\begin{tikzpicture}[scale = 1/3,baseline={(current bounding box.center)},yscale = -1] 
	\draw[black, line width = 1pt] (1,1) .. controls (1,2) and (0,2) .. (0,1);
	\draw[black, line width = 1pt] (3,1) .. controls (3,2) and (2,2) .. (2,1);
	\draw[black, line width = 1pt] (4,1) .. controls (4,2) and (1,2) .. (1,3);
	\draw[black, line width = 1pt] (5,1) .. controls (5,2) and (2,2) .. (2,3);
	\filldraw[white] (-.5,1) -- (.5,1) -- (.5,3) -- (-.5,3) -- (-.5,1);
	\filldraw[white] (5.5,1) -- (4.5,1) -- (4.5,3) -- (5.5,3) -- (5.5,1);
	\draw[black, line width = 2pt] (.5,1) -- (4.5,1);
	\draw[black, line width = 2pt] (.5,3) -- (2.5,3);
\end{tikzpicture} \equiv z_{+}\; 
	\begin{tikzpicture}[scale = 1/3,baseline={(current bounding box.center)},yscale = -1] 
	\draw[black, line width = 1pt] (1,1) -- (1,3);
	\draw[black, line width = 1pt] (3,1) .. controls (3,2) and (2,2) .. (2,1);
	\draw[black, line width = 1pt] (4,1) .. controls (4,2) and (2,2) .. (2,3);
	\draw[black, line width = 2pt] (.5,1) -- (4.5,1);
	\draw[black, line width = 2pt] (.5,3) -- (2.5,3);
\end{tikzpicture} \;.
\end{equation}
By contrast, in $\mathsf{W}_{1/2,z}(3) \times^{o}_{f} \mathsf{S}_{1/2}(1) $
\begin{equation}
	\begin{tikzpicture}[scale = 1/3,baseline={(current bounding box.center)},yscale = -1] 
	\draw[black, line width = 1pt] (1,1) .. controls (1,2) and (0,2) .. (0,1);
	\draw[black, line width = 1pt] (3,1) .. controls (3,2) and (2,2) .. (2,1);
	\draw[black, line width = 1pt] (4,1) .. controls (4,2) and (3,2) .. (3,3);
	\draw[black, line width = 1pt] (5,1) .. controls (5,2) and (4,2) .. (4,3);
	\draw[black, line width = 1pt] (1,3) .. controls (1,2) and (2,2) .. (2,3);
	\filldraw[white] (-.5,1) -- (.5,1) -- (.5,3) -- (-.5,3) -- (-.5,1);
	\filldraw[white] (5.5,1) -- (4.5,1) -- (4.5,3) -- (5.5,3) -- (5.5,1);
	\draw[black, line width = 2pt] (.5,1) -- (4.5,1);
	\draw[black, line width = 2pt] (.5,3) -- (4.5,3);
	\draw[black, line width = 1pt] (.5,4) -- (.5,6);
	\draw[black, line width = 1pt] (1.5,4) .. controls (1.5,5) and (2.5,5) .. (2.5,4);
	\node[anchor = north] at (3.5,4) {$\otimes$};
	\draw[black, line width = 1pt] (4.5,4) -- (4.5,6);
	\draw[black, line width = 2pt] (0,4) -- (3,4);
	\draw[black, line width = 2pt] (4,4) -- (5,4);	
\end{tikzpicture} = \q 
	\begin{tikzpicture}[scale = 1/3,baseline={(current bounding box.center)},yscale = -1] 
	\draw[black, line width = 1pt] (1,1) .. controls (1,2) and (0,2) .. (0,1);
	\draw[black, line width = 1pt] (3,1) .. controls (3,2) and (2,2) .. (2,1);
	\draw[black, line width = 1pt] (4,1) .. controls (4,2) and (5,2) .. (5,1);
	\draw[black, line width = 1pt] (3,3) .. controls (3,2) and (4,2) .. (4,3);
	\draw[black, line width = 1pt] (1,3) .. controls (1,2) and (2,2) .. (2,3);
	\filldraw[white] (-.5,1) -- (.5,1) -- (.5,3) -- (-.5,3) -- (-.5,1);
	\filldraw[white] (5.5,1) -- (4.5,1) -- (4.5,3) -- (5.5,3) -- (5.5,1);
	\draw[black, line width = 2pt] (.5,1) -- (4.5,1);
	\draw[black, line width = 2pt] (.5,3) -- (4.5,3);
	\draw[black, line width = 1pt] (.5,4) -- (.5,6);
	\draw[black, line width = 1pt] (1.5,4) .. controls (1.5,5) and (2.5,5) .. (2.5,4);
	\node[anchor = north] at (3.5,4) {$\otimes$};
	\draw[black, line width = 1pt] (4.5,4) -- (4.5,6);
	\draw[black, line width = 2pt] (0,4) -- (3,4);
	\draw[black, line width = 2pt] (4,4) -- (5,4);	
\end{tikzpicture} + z (-\q)^{1/2}
	\begin{tikzpicture}[scale = 1/3,baseline={(current bounding box.center)},yscale = -1] 
	\draw[black, line width = 1pt] (1,1) .. controls (1,2) and (3,2) .. (3,3);
	\draw[black, line width = 1pt] (3,1) .. controls (3,2) and (2,2) .. (2,1);
	\draw[black, line width = 1pt] (4,1) -- (4,3);
	\draw[black, line width = 1pt] (1,3) .. controls (1,2) and (2,2) .. (2,3);
	\filldraw[white] (-.5,1) -- (.5,1) -- (.5,3) -- (-.5,3) -- (-.5,1);
	\filldraw[white] (5.5,1) -- (4.5,1) -- (4.5,3) -- (5.5,3) -- (5.5,1);
	\draw[black, line width = 2pt] (.5,1) -- (4.5,1);
	\draw[black, line width = 2pt] (.5,3) -- (4.5,3);
	\draw[black, line width = 1pt] (.5,4) -- (.5,6);
	\draw[black, line width = 1pt] (1.5,4) .. controls (1.5,5) and (2.5,5) .. (2.5,4);
	\node[anchor = north] at (3.5,4) {$\otimes$};
	\draw[black, line width = 1pt] (4.5,4) -- (4.5,6);
	\draw[black, line width = 2pt] (0,4) -- (3,4);
	\draw[black, line width = 2pt] (4,4) -- (5,4);	
\end{tikzpicture},
\end{equation}
where we used the same trick as in equation \eqref{eq:exfusion1}. We thus conclude that $z_{+} = (-\q)^{1/2}z$.

 Finally, the eigenvalue of $Y^{(4)}$ on $\mathsf{W}_{0,z_{-}}(4)$ is $z_{-}  + z_{-}^{-1}$, while on $\mathsf{W}_{1,z_{+}}$ it is $(-\q) z_{+} +  (-\q)^{-1} z_{+}^{-1}$, so this fusion product cannot be indecomposable unless
 \begin{equation}
 	(-\q) z_{+} +  (-\q)^{-1} z_{+}^{-1} = z_{-} + z_{-}^{-1} \iff z^{2} = (-\q)^{2} \text{ or } (-\q)^{2} = 1.
 \end{equation}
It follows that, if $\q$ and $z$ are \emph{generic}
\begin{equation}\label{eq:fusionprod.ex1}
	\mathsf{W}_{1/2,z}(3) \times_{f}^{o}\mathsf{S}_{1/2} \simeq \mathsf{W}_{0, z (-\q)^{-1/2}}(4) \oplus \mathsf{W}_{1, z (-\q)^{1/2}}(4).
\end{equation}
What if the parameters are not generic? If $z^{2} = (-\q)^{2} $, a direct calculation shows that the defect operator $Y $ has a Jordan block linking the two standard modules, and this fusion product is indecomposable. However, these new indecomposable modules are, for the moment, largely unclassified, and we plan to come back to this question in the close future. 

\newcommand{\ATL}[1]{\mathsf{T}^a_{#1}}
\newcommand{\TL}[1]{TL_{#1}}
\newcommand{\StJTL}[2]{\mathcal{W}_{#1,#2}}
\newcommand{\StTL}[1]{\mathcal{W}_{#1}}
\newcommand{\fus}{\times_{f}}

More generally, we find that: 
\begin{proposition}\label{prop:fus-p}
For all $k \in \mathbb{Z} $, $T \in \mathbb{Z}_{\geq 0} $, $\delta, \q \in \mathbb{C}^{*} $ such that $\delta^{2}$ is not an integer power of $-\q $,
\begin{equation}
\mathsf{W}^{u}_{k/2 ,\delta} \times^{o}_f \mathsf{S}_{T/2} \simeq  \bigoplus_{ \underset{\text{step }=2}{i = k - T} }^{k+T} \mathsf{W}^{u}_{i/2 ,\delta} \simeq \bigoplus_{\underset{\text{step }=2}{i = k - T}}^{k+T} \mathsf{W}_{i/2,\delta (-\q)^{(i-k)}},\label{Jon1}
\end{equation}
\begin{equation}\label{eq:fusionproduct.standard2}
\mathsf{W}^{o}_{k/2 ,\delta} \times^{u}_f \mathsf{S}_{T/2} \simeq  \bigoplus_{\underset{\text{step }=2}{i = k - T}}^{k+T} \mathsf{W}^{o}_{i/2 ,\delta} \simeq \bigoplus_{\underset{\text{step }=2}{i = k - T}}^{k+T}  \mathsf{W}_{i/2,\delta (-\q)^{(k-i)}}.
\end{equation}
\end{proposition}
These formulas were obtained using the technique and the results outlined in Appendix~\ref{sec:calc.fusion}. However, as shown in Appendix~\ref{app:C}, these results can also be derived by following the approach in \cite{GJS}, that is, by first establishing the branching rules from $\atl{n_1+n_2}$ to $\atl{n_1} \otimes \tl{n_2}$ and then inferring the corresponding fusion product from Frobenius reciprocity. Finally, note that the dimension of $\mathsf{W}_{k/2,z}[n]$ is $\binom{n}{|k|}$, so the dimensions of the various fusion products are
\begin{align}
	\mathsf{dim}\big( \mathsf{W}_{k/2 ,z}[n_1] \times^{o/u}_f \mathsf{S}_{T/2}[n_2]\big) & = \sum_{\underset{\text{step }=2}{i = k - T}}^{k+T} \binom{n_1 + n_2}{\frac{|n_1 + n_2 - i|}{2}}.
\end{align}

\subsection{The fusion quotient}
The fusion product defined in the previous section implicitly assumed that the affine module $M$ was only a left $\atl{m}$-module. If $M$ is an $\atl{m}$-bimodule, then the fusion product $M \times^{u/o}_{f} V$ will also be a 
$(\atl{m+k}, \atl{m})$ 
bimodule; given $W$ a left $\atl{m+k}$ module, and $V$ a left $\tl{k}$ module, we define the fusion quotient by
\begin{equation}
	W \div_{f}^{u/o} V \equiv \mathsf{Hom}_{\atl{m+k}} \left( \atl{m}
	\times_{f}^{u/o} V, W \right).
\end{equation}
Because $ \atl{m}\times_{f}^{u/o} V $ is a \emph{right} $\atl{m}$ module, this $\mathsf{Hom} $ space is naturally a \emph{left} $\atl{m}$-module.  Here's an example to show that the construction is actually quite natural despite its abstract definition. Let $V = \tl{k}$ seen as the regular left $\tl{k}$-module; one finds
	\begin{equation*}
		\atl{m} \times^{u/o}_{f}  V= \lbrace a x_{0}| a\in \atl{k+m} \rbrace, \qquad
  x_{0} = \mathbb{I}_{\atl{k+m}}\otimes_{\atl{m}\otimes_{\mathbb{C}} \tl{k}} (\mathbb{I}_{\atl{m}} \otimes_{\mathbb{C}} \mathbb{I}_{\tl{k}}),
	\end{equation*}
and then by definition we have an isomorphism of vector spaces
	\begin{equation*}
		W \div^{u/o}_{f} \tl{k} \equiv\lbrace{ g_{w}:a x_{0} \mapsto  a w\, | \, w \in W \rbrace} \cong W,
	\end{equation*}
where the isomorphism is simply $ g_{w} \mapsto w $. 
We recall then that by definition of the tensor product $  x_0 b \equiv \phi^{u/o}_{m,k}(b) x_0$, and thus the left $\atl{m}$ action is then simply
\begin{equation}
	[b \cdot g_{w}](a x_0) \equiv  g_{w}\bigl(a x_0 b\bigr) =  g_{w}\bigl(a \phi^{u/o}_{m,k}(b) x_0 \bigr) = a \phi^{u/o}_{m,k}(b) w \equiv g_{\phi^{u/o}_{m,k}(b) w }(a x_0).
\end{equation}
It follows that for any $b$ in $\atl{m}$, $ b \cdot g_{w} = g_{\phi^{u/o}_{m,k}(b) w} $, and thus that $W \div^{u/o}_{f} \tl{k}$ is simply the restriction of $W$ to the subalgebra $\phi^{u/o}_{m,k}(\atl{m}) \simeq \atl{m} $.

More generally, if there exists an idempotent $a_{0} \in \tl{k}$ such that $V = \tl{k}a_{0}$ then for the fusion quotient $W \div^{u/o}_{f} \tl{k}a_0$  the previous arguments can be essentially repeated replacing $x_0$ by the cyclic vector $\psi^{o/u}_{k,m}(a_0)$ and we obtain the following result (note that the action of $\atl{m}$ on $W$ commutes with $\psi^{o/u}_{k,m}(a_{0})$):
\begin{proposition}\label{prop:f-quotient}
For an $\atl{m+k}$-module $W$ and a $\tl{k}$-module $V$ such that $V = \tl{k}a_{0}$ for some idempotent $a_0$,
the fusion quotient $W \div^{u/o}_{f} V$  is isomorphic to the $\atl{m}$-submodule $\psi^{o/u}_{k,m}(a_{0})W$ of the restriction of $W$ to $\atl{m}$.
\end{proposition}

As a more concrete example we compute $\mathsf{W}_{1/2,z}(3) \div^{o}_{f} \mathsf{S}_{1/2}(1) $; since $\mathsf{S}_{1/2}(1) \simeq \tl{1}$, this is simply the restriction of $\mathsf{W}_{1/2,z}(3)$ to $\atl{2}$. We start by choosing a basis of the standard module:
\begin{equation}
	x_{1} = \; 
	\begin{tikzpicture}[scale = 1/3,baseline={(current bounding box.center)},yscale = -1] 
	\draw[black, line width = 1pt] (1,1) .. controls (1,2) and (2,2) .. (2,1);
	\draw[black, line width = 1pt] (3,1) .. controls (3,2) and (1,2) .. (1,3);
	\draw[black, line width = 2pt] (.5,1) -- (3.5,1);
	\draw[black, line width = 2pt] (.5,3) -- (1.5,3);	
\end{tikzpicture} \quad
	x_{2} = \; 
	\begin{tikzpicture}[scale = 1/3,baseline={(current bounding box.center)},yscale = -1] 
	\draw[black, line width = 1pt] (1,1) .. controls (1,2) and (0,2) .. (0,1);
	\draw[black, line width = 1pt] (2,1) .. controls (2,2) and (4,2) .. (4,1);
	\draw[white, line width = 3pt] (3,1) .. controls (3,2) and (1,2) .. (1,3);
	\draw[black, line width = 1pt] (3,1) .. controls (3,2) and (1,2) .. (1,3);
	\filldraw[white] (-.5,1) -- (.5,1) -- (.5,3) -- (-.5,3) -- (-.5,1);
	\filldraw[white] (4.5,1) -- (3.5,1) -- (3.5,3) -- (4.5,3) -- (4.5,1);
	\draw[black, line width = 2pt] (.5,1) -- (3.5,1);
	\draw[black, line width = 2pt] (.5,3) -- (1.5,3);	
\end{tikzpicture}
	\;
	x_{3} = \;
	\begin{tikzpicture}[scale = 1/3,baseline={(current bounding box.center)},yscale = -1] 
	\draw[black, line width = 1pt] (3,1) .. controls (3,2) and (2,2) .. (2,1);
	\draw[black, line width = 1pt] (1,1) -- (1,3);
	\draw[black, line width = 2pt] (.5,1) -- (3.5,1);
	\draw[black, line width = 2pt] (.5,3) -- (1.5,3);	
\end{tikzpicture} \; .
\end{equation}
The element $x_{2}$ was chosen so that 
\begin{equation}
	\phi^{o}_{2,1}(e^{(2)}_{1}) x_{2} = ((-\q)^{-1/2}z +(-\q)^{1/2}z^{-1} )x_{1} , \qquad \phi^{o}_{2,1}(e^{(2)}_{2}) x_{2} = (\q + \q^{-1}) x_{2}, \qquad \phi^{o}_{2,1}(u^{(2)}) x_{2} = x_{1}.
\end{equation}
We thus recognize that $\lbrace x_{1}, x_{2} \rbrace  $ span a submodule isomorphic to $\mathsf{W}_{0, z (-\q)^{-1/2}}(2)$. Furthermore,
\begin{equation}
	\phi^{o}_{2,1}(e^{(2)}_{1}) x_{3} = x_{1}, \qquad \phi^{o}_{2,1}(e^{(2)}_{2}) x_{3} = -(-\q)^{-3/2}z^{-1} x_{2},\qquad \phi^{o}_{2,1}(u^{(2)}) x_{3} = (-\q)^{1/2}z x_{3} + \q x_{2},
\end{equation}
so the quotient $\left(\mathsf{W}_{1/2,z}(3) \div^{o}_{f} \mathsf{S}_{1/2}(1)\right)/(\mathsf{W}_{0, z (-\q)^{-1/2}}(2))$ is isomorphic to $\mathsf{W}_{1,(-\q)^{1/2}z}(2)$. Finally, comparing the eigenvalues of $Y$ on the two standard modules yields the conclusion: if $\q$ and $z$ are \emph{generic}
\begin{equation}
	\mathsf{W}_{1/2,z}(3) \div_{f}^{o}\mathsf{S}_{1/2}(1) \simeq \mathsf{W}_{0, z (-\q)^{-1/2}}(2) \oplus \mathsf{W}_{1, z (-\q)^{1/2}}(2).
\end{equation}

More generally, we find that:
\begin{proposition}\label{prop:fus-q}
For all $k \in \mathbb{Z} $, $T,n,m \in \mathbb{Z}_{\geq 0} $, $\delta, \q \in \mathbb{C}^{*} $ such that $\delta^{2}$ is not an integer power of~$-\q $, and 
$ |k| \leq n+m$, $T \leq m $, we have
\begin{equation}\label{eq:fusionquotient.standard1}
\mathsf{W}^{u}_{k/2 ,\delta} (n + m) \div^{o}_f \mathsf{S}_{T/2}(m) \simeq  \bigoplus_{ \underset{\text{step }=2}{i = k - T}}^{k+T} \mathsf{W}^{u}_{i/2 ,\delta}(n) \simeq \bigoplus_{\underset{\text{step }=2}{i = k - T}}^{k+T} \mathsf{W}_{i/2,\delta(-\q)^{(i-k)}}(n),
\end{equation}
\begin{equation}\label{eq:fusionquotient.standard2}
\mathsf{W}^{o}_{k/2 ,\delta}(n+m) \div^{u}_f \mathsf{S}_{T/2} (m)\simeq  \bigoplus_{ \underset{\text{step }=2}{i = k - T}}^{k+T} \mathsf{W}^{o}_{i/2 ,\delta}(n) \simeq \bigoplus_{\underset{\text{step }=2}{i = k - T}}^{k+T}  \mathsf{W}_{i/2,\delta(-\q)^{(k-i)}}(n).
\end{equation}
\end{proposition}
In these expressions it should be understood that all modules of the form $\mathsf{W}_{k/2 ,z}(n)$ with $n < k$ should be identified with the zero module. These formulas were obtained using the technique and the results outlined in Appendix~\ref{sec:calc.fusion}.

\subsection{Dualities between the two fusions}
Aside from their possible interpretation as algebraic realisations of topological defects, the two types of fusion are of independent interest for the representation theory of $\atl{n}$. As such, we mention here certain properties which they have, and which can be used to compute them. The first such property is that the fusion product and the fusion quotients are duals as functors, i.e. for any $\atl{n+m}$ module $W$, $\atl{n}$ module $V$ and $\tl{m}$ module $U$, there 
is a natural isomorphism\footnote{Note that this follows directly from the fact that the tensor product and the $\mathsf{Hom}$ functors form an adjoint pair.}
\begin{equation}\label{eq:fusion.duality}
	\mathsf{Hom}_{\atl{n+m}}\left( V \times^{u/o}_{f} U, W \right) \simeq \mathsf{Hom}_{\atl{n}}\left( V , W \div^{u/o}_{f} U \right) .
\end{equation}
It follows in particular that if one knows every fusion product, one can get back all the fusion quotients by using this duality, and vice versa.

The second property we mention is the associativity: for all  $\atl{n}$ module $W$, $\tl{k}$ module $V$ and $\tl{m}$ module $U$,
\begin{equation}\label{eq:fusionp.assoc}
	(W \times_{f}^{u/o} V) \times_{f}^{u/o} U \simeq  	W \times_{f}^{u/o} ( V \times_{f}^{r} U ) \simeq  (W \times_{f}^{u/o} U) \times_{f}^{u/o} V,
\end{equation}	
where $\times^{r}_{f}$ is the fusion product in the regular Temperley-Lieb algebra, which was studied in detail in \cite{GV,BelleteteFusion}. Similarly, for all  $\atl{n+k+m}$ module $W$, $\tl{k}$ module $V$ and $\tl{m}$ module $U$,
\begin{equation}\label{eq:fusionq.assoc}
	(W \div_{f}^{u/o} V) \div_{f}^{u/o} U \simeq  	W \div_{f}^{u/o} ( V \times_{f}^{r} U ) \simeq  (W \div_{f}^{u/o} U) \div_{f}^{u/o} V.
\end{equation}

As an example, if we assume that $\q$ is generic then for all $k \geq 0$
\begin{equation}
\mathsf{S}_{k}(n) \times^{r}_{f} \mathsf{S}_{0}(2) \simeq   \mathsf{S}_{k}(n+2), \qquad \mathsf{S}_{k}(n) \times^{r}_{f} \mathsf{S}_{1/2}(1) \simeq   \mathsf{S}_{k-1/2}(n+1) \oplus \mathsf{S}_{k + 1/2}(n+1).
\end{equation}
It follows that for a given $\atl{n}$ module $W$, knowing its fusion product (or quotient) with $\mathsf{S}_{0}(2m)$ and $\mathsf{S}_{1/2}(1)$ is enough to compute the fusion with all other standard modules by recurrence. Equations \eqref{Jon1}-\eqref{eq:fusionproduct.standard2} and \eqref{eq:fusionquotient.standard1}-\eqref{eq:fusionquotient.standard2} were obtained in this manner (see appendix \ref{sec:calc.fusion}).

\subsection{Fusion and the Hamiltonian}\label{sec:4.4}

We now go back to the problem of studying the defects in the direct channel, or the spectrum problem of the Hamiltonians $H^{u}_{n,m}$ introduced in the beginning of this section~\ref{sec:4}. Recall that we reduced this problem  to studying $H^{u}_{n,m}[\rho]$ from~\eqref{eq:Hnm-idemp} for an idempotent $\rho$.
  
Let $W$ be an $\atl{n+m}$-module, and $\rho$ be a non-zero idempotent of $\tl{m}$, then the Hamiltonians with impurities (where we also introduced the over lines version)
\begin{align}
		H^{u}_{n,m}[\rho] & =  \phi^{u}_{n,m}\big(H_{n} \big)\psi^{o}_{m,n}(\rho) = \big( \sum_{j = 1}^{n-1} e^{(n+m)}_{j}   +  \mu_{n,m}^{-1} e^{(n+m)}_{n}  \mu_{n,m} \big) \psi^{o}_{m,n}(\rho),\label{eq:hamiltunder}\\
		H^{o}_{n,m}[\rho] & = \phi^{o}_{n,m}\big(H_{n} \big)\psi^{u}_{m,n}(\rho) = \big(\sum_{j = 1}^{n-1} e^{(n+m)}_{j}   +  \nu_{n,m} e_{n}^{(n+m)}  \nu_{n,m}^{-1}\big) \psi^{u}_{m,n}(\rho),\label{eq:hamiltover}
	\end{align}
where $\mu_{n,m} = g_{n}g_{n+1}\hdots g_{n+m}$ and $\nu_{n,m} = g_{n+m}\hdots g_{n}$,
acting on $W$ are the standard periodic TL Hamiltonian (i.e.\ without defects) $H_{n} = \sum_{j=1}^{n} e^{(n)}_{j}$ acting on  the $\atl{n}$-submodule $\psi^{o/u}_{m,n}(\rho)W$ in the restriction of $W$ to $\atl{n}$ under $\phi^{u/o}_{n,m}$. And this submodule is just the fusion quotient $W \div^{u/o}_{f} (\tl{m}\rho)$,
due to Proposition~\ref{prop:f-quotient}.

We note that choosing the idempotent $\rho$ to correspond to a representation of spin $m/2$, or the standard module on $m$ strands with $m$ through lines, e.g.\  the Jones-Wenzl idempotents $\rho=P_{m/2}$ defined in~\eqref{eq:JW-P}, the expressions~\eqref{eq:hamiltunder} and~\eqref{eq:hamiltover}  are precisely the expressions for the Hamiltonians with impurities that correspond in the direct channel to the defect operators  $Y_{m/2}$ and $\bar{Y}_{m/2}$, respectively.

As a corollary of the preceding discussion we formulate our main result on the spectral problem of the defect Hamiltonians:

\begin{Thm}\label{eq:thm-H}
	Let $\rho \in \tl{m}$ be an idempotent such that $\tl{m}\rho \simeq V$, then for any $\atl{n+m}$-module $M$, the Hamiltonian 
	$H^{u/o}_{n,m}[\rho]$ 
	acting on $M$ is similar (as a matrix) to the direct sum of the standard Hamiltonian $H_{n}$ acting on $M \div_{f}^{u/o} V$ and a zero matrix of dimension $\text{dim}((\mathbb{I}_{\atl{n+m}}-\psi^{o/u}_{m,n}(\rho))M)$.
\end{Thm} 
\begin{proof}
	By definition,
	\begin{equation}
		H^{u/o}_{n,m}[\rho] \equiv  \phi^{u/o}_{n,m}(H_{n})\psi^{o/u}_{m,n}(\rho),
	\end{equation}
	and since $\rho$ is an idempotent,
	\begin{equation}
	 M \simeq \big(\psi^{o/u}_{m,n}(\rho) M \big) \oplus  (\mathbb{I}_{\atl{n+m}}-\psi^{o/u}_{m,n}(\rho))M.
	\end{equation} 
	The first summand is isomorphic to $M \div_{f}^{u/o} V$, by Proposition~\ref{prop:f-quotient}, while one quickly sees that $H^{u/o}_{n,m}[\rho]$ is identically zero on the second. Furthermore, by definition $H_{n}$ acts on $\psi^{o/u}_{m,n}(\rho) M$ by left multiplication by $ \phi^{u/o}_{n,m}(H_{n})$.
\end{proof}

\textit{As an application of this theorem, we can calculate the spectrum of the Hamiltonian with impurities corresponding to the defect operator $Y_j$ in the direct channel  as follows: the set of eigenvalues on $M$  is the set of eigenvalues of the standard periodic TL Hamiltonian $H_n$ on the $\atl{n}$-module $M \div_{f}^{u} \mathsf{S}_j$.
The case of the Hamiltonian for $\bar{Y}_j$ is analogous.} Taking into account the general results on the fusion from Proposition~\ref{prop:fus-q} and that the spectrum of $H_n$ is known on all the standard $\atl{n}$-modules $\mathsf{W}_{k,z}$, we thus solved the spectrum problem for the Hamiltonians with the impurities.

 Similarly, the transfer matrix acting on this fused module $M \div_{f}^{u/o} \mathsf{S}_j$ is precisely the one obtained by adding a cluster of lines going under (or over) the other lines in the lattice. This strongly suggests that the fusion quotient is indeed the right algebraic construction for these defects. However it should be mentioned that for generic values of the parameters the fusion product and quotients are equivalent for large values of $n$, in the sense that
\begin{equation}
	\mathsf{W}_{k}[n + 2 m] \div^{u/o}_{f} \mathsf{S}_{t}[m] \simeq \mathsf{W}_{k}[n] \times^{u/o}_{f} \mathsf{S}_{t}[m],
\end{equation}
provided that $\mathsf{W}_{k}[n] \neq 0$. It follows that while the Hamiltonian acting on the fusion product does not have such a simple interpretation it will produce the same spectrum (unless $n$ is too small).
 
\subsection{Example of quotient: the twisted XXZ spin chain}\label{eq:sec4-ex}
The twisted XXZ spin chain on $n$ sites can be realized by the Hamiltonian $H_{n}(Q)$ expressed in terms of the usual Pauli matrices acting on $(\mathbb{C}^{2})^{\otimes n}$:
\begin{equation}
H_{n}(Q) = \sum_{j=1}^{n} \left(\sigma^{-}_{j}\sigma^{+}_{j+1} + \sigma^{-}_{j+1}\sigma^{+}_{j} + \frac{\q +\q^{-1}}{4}\left(\sigma^{z}_{j}\sigma^{z}_{j+1} - 1 \right) \right) = - \sum_{j=1}^{n} e_{j},
 \end{equation}
where $\sigma^{\pm} = 1/2(\sigma^{x}_{j} \pm i \sigma^{y}_{j})$ are the usual ladder operators, $Q$ is a non-zero complex number, and the boundary conditions are
\begin{equation}
	\sigma^{z}_{n+1} \equiv \sigma^{z}_{1}, \qquad \sigma^{\pm}_{n+1} \equiv Q^{\mp 2} \sigma^{\pm}_{1}.
\end{equation}
The model is unitary if $Q$ is on the unit circle in $\mathbb{C}$. The Temperley-Lieb generators are
\begin{equation}
	-e_{j} \equiv  \sigma^{-}_{j}\sigma^{+}_{j+1} + \sigma^{-}_{j+1}\sigma^{+}_{j} + \frac{\q +\q^{-1}}{4}\left(\sigma^{z}_{j}\sigma^{z}_{j+1} - 1 \right) + \frac{\q -\q^{-1}}{4} (\sigma^{z}_{j} - \sigma^{z}_{j+1}),
\end{equation}
with the twist
\begin{equation}
	u = (-1)^{n/2} Q^{-\sigma^{z}_{1}} s_{1}\hdots s_{n-1}, \qquad s_{j} = \sigma^{-}_{j}\sigma^{+}_{j+1} + \sigma^{+}_{j}\sigma^{-}_{j+1} + \frac{1}{2}(\sigma^{z}_{j}\sigma^{z}_{j+1} + 1).
\end{equation}
A quick calculation shows that the hoop operators are\footnote{In everything that follows, one should understand that for all matrix $A$, $\q^{A} \equiv (-\q)^{A}(-1)^{-A}$. We simplify these expressions to lighten the notation but one should be careful when verifying these results numerically.}
\begin{equation}
	Y[n] = (-1)^{n}\left(\q^{S_{z}}Q^{-1} + \q^{-S_{z}}Q \right), \qquad\bar{Y}[n] = \q^{S_{z}}Q  + \q^{-S_{z}}Q^{-1},
\end{equation}
with $S_{z} = \frac{1}{2} \sum_{j=1}^{n}\sigma^{z}_{j}$ the total spin.
Our goal is now to impose a defect of spin $1/2$ on this chain, which according to our formalism (see Theorem~\ref{eq:thm-H}) consist in computing the fusion quotient of its Hilbert space by $\mathsf{S}_{1}(1) = \tl{1}$. This specific defect corresponds to a simple restriction from $\atl{n}$ to $\atl{n-1}$, so the new Hamiltonian with a defect is either
\begin{equation}
	H^{u}_{n-1,1}(Q) 
	= - \sum_{j=1}^{n-1} \phi^{u}_{n-1,1}(e^{(n-1)}_{j}) = -\sum_{j=1}^{n-2}e^{(n)}_{j} - g^{(n)}_{n}e^{(n)}_{n-1}(g^{(n)}_{n})^{-1},
\end{equation}
or
\begin{equation}
	H^{o}_{n-1,1}(Q) = - \sum_{j=1}^{n-1} \phi^{o}_{n-1,1}(e^{(n-1)}_{j}) = -\sum_{j=1}^{n-2}e^{(n)}_{j} - (g^{(n)}_{n})^{-1}e^{(n)}_{n-1}g^{(n)}_{n},
\end{equation}
for a defect that goes under or over the other lines, respectively. Using the explicit construction of $e_{n-1}$ and $g_{n}$ one finds
	\begin{equation*}
		H^{u}_{n-1,1}(Q) = \sum_{j}^{n-1}(a^{-}_{j}a^{+}_{j+1} + a^{-}_{j+1}a^{+}_{j} + \frac{\q +\q^{-1}}{4}\left(a^{z}_{j}a^{z}_{j+1} - 1 \right)) +  \left( (1-\q^{2 a^{z}_{1}})a^{-}_{n-1} + Q^{2}(1-\q^{-2 a^{z}_{n-1}})a^{-}_{1}  \right)\sigma^{+}_{n},
	\end{equation*}
where we defined new operators $a^{k}_{j} = \sigma^{k}_{j}$, $k= z, \pm $, $j = 1, 2, \hdots n-1$, with boundary conditions
\begin{equation}
	a^{z}_{n} \equiv a^{z}_{1}, \qquad a^{\pm}_{n} \equiv (Q^{2} \q^{-\sigma^{z}_{n}} )^{\mp 1} a^{\pm}_{1}.  
\end{equation}
It follows that
\begin{equation}\label{eq:XXZHamilup}
	H^{u}_{n-1,1}(Q) \sim \overset{\begin{array}{cc}
	(\hdots )\otimes | \uparrow \rangle \qquad & \qquad (\hdots )\otimes | \downarrow \rangle
	\end{array}}{\left(
\begin{array}{cc}
H_{n-1}(-Q \q^{-1/2}) & \Delta \\
 0 & H_{n-1}(- Q \q^{1/2})
\end{array}
\right)}, \qquad \Delta = (1-\q^{2 a^{z}_{1}})a^{-}_{n-1} + Q^{2}(1-\q^{-2 a^{z}_{n-1}})a^{-}_{1} .
\end{equation}
A straightforward calculation then shows that the defect operators are now:
\begin{equation}
	Y[n-1] =(-1)^{n} (\q^{S_{z}}Q^{-1} + \q^{-S_{z}}Q) = (-1)^{n-1} \left( \q^{S_{z} - \frac{1}{2}\sigma^{z}_{n}} \left(-Q \q^{-\frac{1}{2}\sigma^{z}_{n}} \right)^{-1} + \q^{-S_{z} + \frac{1}{2}\sigma^{z}_{n}} \left(-Q \q^{-\frac{1}{2}\sigma^{z}_{n}} \right) \right),
\end{equation} 
\begin{equation}
	\bar{Y}[n-1] 
	\sim \overset{\begin{array}{cc}
	(\hdots )\otimes | \uparrow \rangle \qquad & \qquad (\hdots )\otimes | \downarrow \rangle
	\end{array}}{\left(
\begin{array}{cc}
Q_{-}\q^{\tilde{S}_{z}} + Q^{-1}_{-}\q^{-\tilde{S}_{z}} & Q (\q - \q^{-1})^{2} \tilde{S}_{-} \\
0 & Q_{+}\q^{\tilde{S}_{z}} + Q^{-1}_{+}\q^{-\tilde{S}_{z}}
\end{array}
\right)},
\end{equation}
where $Q_{\pm} \equiv -Q \q^{\pm 1/2} $, and $\tilde{S}_{-}$, $\q^{\pm \tilde{S}_{z}} $ are the standard $U_{\q}(\mathfrak{sl}_{2}) $ generators on $n-1$ spins
\begin{equation}
\tilde{S}_{-} = \sum_{i=1}^{n-1} (\q)^{\sum_{j=1}^{i-1}\sigma^{z}_{j}}\sigma^{-}_{i}(\q)^{-\sum_{j=i+1}^{n-1}\sigma^{z}_{j}}, \qquad \q^{\pm \tilde{S}_{z}} = \q^{\sum_{j=1}^{n-1}\sigma^{z}_{j}/2}.
\end{equation}
Note that $\bar{Y}[n-1]$ can be diagonalized if and only if $(Q - \q^{-S_z})(Q + \q^{-S_z})$ is an invertible matrix, which can be verified by comparing its eigenvalues in the $\sigma^{z}_{n} = \pm 1 $ sectors. It follows in particular that the Hamiltonian \eqref{eq:XXZHamilup} cannot have a Jordan block linking the $\sigma^{z}_{n} = \pm 1 $ sectors if $Q$ is generic.

Similarly, one finds
\begin{equation}
	H^{o}_{n-1,1}(Q) \sim \overset{\begin{array}{cc}
	(\hdots )\otimes | \uparrow \rangle \qquad & \qquad (\hdots )\otimes | \downarrow \rangle
	\end{array}}{\left(
\begin{array}{cc}
H_{n-1}(Q \q^{1/2}) & 0 \\
\Delta & H_{n-1}(Q \q^{-1/2})
\end{array}
\right)}, \qquad \Delta = (1-\q^{2 a^{z}_{1}})a^{+}_{n-1} + Q^{-2}(1-\q^{-2 a^{z}_{n-1}})a^{+}_{1} .
\end{equation}
Note that in each of these expressions, the off-diagonal term $\Delta$ can only link sectors of $H_{n-1}(Q \q^{\pm 1/2}) $ corresponding to different total spin ( $\sum_{j=1}^{n-1}a _{j}$) because of the ladder operators appearing in it.

\section{Conclusion: connection to CFT}\label{sec:5}
In order to provide a lattice analogue of CFT topological defects  $X$ satisfying~\eqref{centVir}, we have defined and studied in a model-independent way  operators on the lattice that commute with the local interactions given by the TL elements---the central elements $Y$ and $\bar{Y}$  in $\atl{n}$---and have demonstrated their interesting properties. From the crossed-channel point of view, these defect operators generate an algebra spanned by $Y_j$, $\bar{Y}_j$, and their products, that has structure constants or fusion rules~\eqref{eq:Y-fusion} and~\eqref{eq:bY-fusion} resembling the chiral and anti-chiral fusion rules of Virasoro Kac modules of type $(1,s)$ where $s=2j+1$. We recall that the Kac modules are obtained as quotients of Verma modules of the conformal weight $h_{1,s}$ by the submodule generated by the singular vector at the level $h_{1,s} + s$.

The analogy with CFT goes further: 
Recall that  at least in rational CFT a topological defect can be seen as a map from the set of chiral primary fields to the ring of endomorphisms of the Hilbert space of the full non-chiral CFT.  In much the same way  our maps $Y^m$ and $\bar{Y}^m$ from Fig.~\ref{fig:defectmap} defining the defect operators  send ideals in the open or regular TL algebra (which are known to correspond to chiral primary fields of conformal weight  $h_{1,s}$) to central elements in affine TL algebra which are realized as endomorphisms of the bulk lattice model, e.g.\ of periodic spin-chains. 


\smallskip
We saw that the higher-spin defects $Y_j$ and $\bar{Y}_j$ \eqref{eq:Yj-def}-\eqref{eq:Yj-U} carry some sort of internal structure ``living" on the horizontal non-contractible loops. 
From the direct-channel point of view,  or after a modular transformation, this internal structure was realized  in Section~\ref{sec:4} as some sort of impurities in the spatial direction. Therefore, we have  just rewritten the defects $Y_j$ and $\bar{Y}_j$ in the Hamiltonian formulation. Interestingly,  the problem of spectrum  with  impurities was reformulated in algebraic terms as a rather simple fusion product of affine and regular TL representations which is a combination of the constructions in~\cite{GS,GJS} and~\cite{BSA} that we review in Appendix~\ref{sec:previousfusions}.
 
\smallskip
So far we have defined and studied lattice defects that do not depend on a spectral parameter. Let us call these defects of \textit{first type}. However, there is some evidence that there should be a \textit{second type} of (lattice) defects that do depend on a spectral parameter. Though they are not central in~$\atl{n}$, but possibly become topological defects $X$, i.e.\ they satisfy~\eqref{centVir}, in the continuum limit only. We will address studying these defects of the second type in our next paper where an identification with Virasoro Kac modules of the type $(r,1)$ is expected.

\medskip

It is important for several reasons to try to define what we call lattice defects in a precise mathematical way and in higher generality, for a possible application to more general lattice models not necessarily based on TL interactions.
For the first kind of defects, from the results obtained in this work, we are approaching a mathematical definition of (an algebra of) defects for general lattice algebras (e.g. $\atl{n}(\q)$, Birman-Wenzl-Murakami, Brauer algebras, etc):

\smallskip

\textbf{Definition:}
\textit{
In a lattice algebra $A$, the space of defects $D$ of 1st type is the center of $A$ with the structure of a Verlinde algebra.}

\smallskip

Note that not 
every central element in a lattice algebra corresponds to a defect operator: it should also have nice properties that reflect known properties from the CFT side. That is why we demand that the space of defects forms a Verlinde algebra. First of all this implies the presence of a special basis in this algebra with structure constants being non-negative integers. Secondly, the idea is that these integer numbers should correspond to fusion rules of corresponding representations of an (anti-)chiral  algebra, e.g.\ Virasoro. 

We have indeed recovered these two aspects in our case of $A=\atl{n}(\q)$, where we identified $D$ as the  center $\Z$ of $\atl{n}(\q)$, and the latter as a Verlinde algebra generated by $Y_j$ and $\bar{Y}_j$ where the structure constants do not depend on $n$, for non-mixed products, and  correspond to fusion rules of chiral and anti-chiral Virasoro representations of type $(1,s)$.\footnote{Strictly speaking our algebra of defects is the product of two Verlinde algebras, for chiral and anti-chiral Virasoro representations, 
modulo non-linear algebraic relations between $Y$ and $\bar{Y}$. }
This is here shown to be true for the generic~$\q$ case where the fusion rules might look rather trivial, since they are $sl(2)$ type fusion after all. The situation is not so trivial in degenerate cases (where $\q$ is a root of unity) that we will describe in one of our forthcoming papers on the subject, with applications to minimal models as well as LCFTs. There, a connection to Virasoro fusion rules also holds, although it is much less evident due to more involved representation theory. 

However, this is not the end of the story. Any Verlinde algebra has the third aspect: it admits a modular $S$-transformation that ``diagonalizes" the fusion rules. For the moment we have concentrated on the first two aspects only. It is, of course, an important problem to properly define and analyze such $S$-transformations in a precise algebraic way, and hopefully it will reflect the modular transformation on the lattice. We hope to come back to this problem soon.

\bigskip

\section{Declarations}

\noindent {\bf Funding.}
This work was supported by  the Institut Universitaire de France and the European Research Council (advanced grant NuQFT). We are also grateful to the CRM  in Montreal  and to the organizers of  the conference \emph{Algebraic methods in mathematical physics} in Montreal in 2018 where a part of this work was done.
The work of AMG was supported by CNRS, and partially by ANR grant JCJC ANR-18-CE40-0001 and the RSF Grant No.\ 20-61-46005. 
AMG is also grateful to IPHT Saclay for kind hospitality in 2017 and 2018.

The authors have no competing interests to declare that are relevant to the content of this article.

\noindent {\bf Acknowledgments.} \hspace{2pt} We acknowledge interesting discussions with   Yvan St Aubin, and we thank D. Bulgakova for discussions and early collaboration on this project.  We would also like to thank an anonymous referee for their helpful comments, and for pointing out a problem in the original proof of theorem \ref{thm:main}.

\noindent {\bf Data Availability.} Data sharing not applicable to this article as no datasets were generated or analysed during the current study.

\appendix

\section{Proofs and rigors}
We collect in this appendix the proofs of certain technical results used in this work.

\subsection{Topological defects with a higher spin.}\label{sec:rigor.hspindefect}

We show here how to obtain the expressions for topological defects with higher-spins given in section \ref{sec:higherspin}, i.e. 
	\begin{equation}\label{eq:app.higherspin}
		Y_{j} = \mathsf{U}_{2j} (Y/2).
	\end{equation}
First we show that the result is independent of the choice of idempotent we make.

Let $A$ be some finite dimensional $\mathbb{C}$-algebra, $\rho_{1}$, $\rho_{2} $ be two idempotents such that $A \rho_{1} \simeq A \rho_{2} $ as left $A$ modules, and let $F$ be any function defined on $A$ such that for all  $a,b \in A $, $F(ab) = F(ba)$ (in other words $F$ is cyclic).
 We know that
\begin{equation}
	\mathsf{Hom}_{A}(A\rho_{1}, A \rho_{2}) \simeq \rho_{1}A\rho_{2}, \qquad \mathsf{Hom}_{A}(A\rho_{2}, A \rho_{1}) \simeq \rho_{2}A\rho_{1},
\end{equation}
where the isomorphism is obtained by right-multiplication. For instance,
\begin{equation}
	(f: A\rho_{1} \to A\rho_{2}) 
\mapsto
 \rho_{1}f(\rho_{1}) \rho_{2}, \qquad \rho_{1}a\rho_{2}
 \mapsto
 (\underbrace{ b \rho_{1} \to b \rho_{1} a \rho_{2}}_{\in \mathsf{Hom}_{A}(A\rho_{1}, A \rho_{2}) }).
\end{equation}
Because $A \rho_{1} \simeq A \rho_{2} $, it follows that there exists $a,b  \in A $ such that $\rho_{1} a \rho_{2}$ gives the $A$-linear isomorphism 
$A \rho_{1}  \to A \rho_{2}$ and $ \rho_{2} b \rho_{1}$ 
gives  the isomorphism
$A\rho_{2} \to A\rho_{1}$. In particular, if $A \rho_{1}$ is irreducible this means that the compositions of the two isomorphisms applied to $\rho_1$ and $\rho_2$ are respectively
$\rho_{1} a \rho_{2} b \rho_{1} = \alpha \rho_{1}$ and $ \rho_{2} b \rho_{1} a \rho_{2} = \gamma \rho_{2}$ for some non-zero complex numbers $\alpha, \gamma $. However one quickly verifies that
\begin{equation*}
	\gamma^{2} \rho_{2} = (\rho_{2}b \rho_1 a \rho_{2})^{2} =  \rho_{2}b ( \rho_{1} a \rho_{2} b \rho_{1}) a \rho_{2} = \alpha \gamma \rho_{2},
\end{equation*}
so $\alpha = \gamma $ and we can thus choose $a,b $ such that $\alpha= \gamma = 1 $. Now by hypothesis the function $F$ is cyclic and $\rho_{1}$ and $\rho_{2}$ are idempotents so
\begin{equation}
	F(\rho_{1}) = F(\rho_{1} a \rho_{2} b \rho_{1}) = F(\rho_{2} b\rho_{1} a \rho_{2}) = F(\rho_{2}).
\end{equation}

Next, we remark that for any elements $a \in \tl{n}$, $b\in \tl{m}$, $Y^{n+m}(a \otimes^{\tl{}} b) = Y^{n}(a)Y^{m}(b)$ where $\otimes^{\tl{}}$ is the tensor product in the Temperley-Lieb category, obtained by joining diagrams side by side. For instance,
\begin{equation}
	e_{1} \otimes^{\tl{}} e_{1} \equiv 
	\begin{tikzpicture}[scale = 1/3,baseline={(current bounding box.center)}] 
	\draw[black, line width = 2pt] (1,.5) -- (1,2.5);
	\draw[black, line width = 2pt] (3,.5) -- (3,2.5);
	\draw[black, line width = 1pt] (1,1) .. controls (2,1) and (2,2) .. (1,2);
	\draw[black, line width = 1pt] (3,1) .. controls (2,1) and (2,2) .. (3,2);
	\node[anchor = south] at (4.5,1) 	{$\otimes^{\tl{}}$};
	\draw[black, line width = 2pt] (6,.5) -- (6,2.5);
	\draw[black, line width = 2pt] (8,.5) -- (8,2.5);
	\draw[black, line width = 1pt] (6,1) .. controls (7,1) and (7,2) .. (6,2);
	\draw[black, line width = 1pt] (8,1) .. controls (7,1) and (7,2) .. (8,2);
\end{tikzpicture} \; \equiv \;
	\begin{tikzpicture}[scale = 1/3,baseline={(current bounding box.center)}] 
	\draw[black, line width = 2pt] (1,.5) -- (1,4.5);
	\draw[black, line width = 2pt] (3,.5) -- (3,4.5);
	\draw[black, line width = 1pt] (1,1) .. controls (2,1) and (2,2) .. (1,2);
	\draw[black, line width = 1pt] (3,1) .. controls (2,1) and (2,2) .. (3,2);
	\draw[black, line width = 1pt] (1,3) .. controls (2,3) and (2,4) .. (1,4);
	\draw[black, line width = 1pt] (3,3) .. controls (2,3) and (2,4) .. (3,4);
\end{tikzpicture} \;
\end{equation}
It follows in particular that for any idempotent $a\in \tl{n} $, 
\begin{equation}
	Y^{n+1}(a \otimes^{\tl{}} \mathbb{I}_{\tl{1}}) = Y_{1/2} Y^{n}(a). 
\end{equation}
Now if the idempotent $a$ is such that $\tl{n}a \simeq \mathsf{S}_{k/2}(n)$,
 one can show that there exists a decomposition of the idempotent $(a \otimes^{\tl{}} \mathbb{I}_{\tl{1}}) = a_{-} + a_{+} $ in $\tl{n+1}$, where $a_{\pm} $ are orthogonal idempotents such that 
 $\tl{n+1} a_{\pm} \simeq
  \mathsf{S}_{(k \pm 1)/2}(n+1)$.
This follows from the decomposition of the induced $\tl{n+1}$-module (see e.g.~\cite{GV,BelleteteFusion}). We therefore have for generic values of $\q$ the following relations:
\begin{equation}
	Y_{(k+1)/2} = Y_{1/2}Y_{k/2} - Y_{(k-1)/2}, \qquad Y_{0} \equiv 1_{\atl{n}},
\end{equation}
which is the Chebyshev recurrence relation, giving \eqref{eq:app.higherspin}.

\subsection{The Jucys-Murphy elements}\label{sec:rigor.JM}

Here,  we  prove the identities \eqref{eq:JMandTopD1} and \eqref{eq:JMandTopD2} involving $C_{k}(n)$ and $\bar{C}_{k}(n)$ defined in~\eqref{def:Ck}. As the proofs for the identities involving $M$s are identical to those involving the $J$s,  we  only prove the two identities involving $\bar{C}_{k}(n)$. The proof of this result relies on two key observations; the first is the identity
\begin{equation}
	[i]_{k}(-\q)^{ \pm k} - [i-1]_{k} = (-\q)^{\pm k i}, \qquad i = 0, 1, \hdots.
\end{equation}

Furthermore, defining for $i = 1, \hdots, n-1$, 
\begin{align}
	\bar{X}_{i+1} &\equiv (-\q)^{2} J_{i}  + (-\q)^{-2} J_{i}^{-1},\\
	X_{i+1}   &\equiv (-\q)^{2} M_{i}  + (-\q)^{-2} M_{i}^{-1},
\end{align}
the second observation is that for all $i= 1, \hdots ,n -1$ we have the relations in $\atl{n}(\q)$:
\begin{align}
	 \bar{X}_{i+1} &  = (-\q) J_{i+1}  + (-\q)^{-1} J_{i+1}^{-1} ,\label{eq:prop.recurrence} \\ 
	X_{i+1}   & = (-\q) M_{i+1}  + (-\q)^{-1} M_{i+1}^{-1}  , \\
Y   &=	(-\q)^{2} J_{n} + (-\q)^{-2} J_{n}^{-1} , \\
 \bar{Y} &= (-\q)^{2} M_{n} + (-\q)^{-2} M_{n}^{-1} .
\end{align}
These can all be proven in the same way, by showing that both sides of these equality correspond to the same diagram. For instance
\begin{equation}
	(-\q)^{2} J_{n} + (-\q)^{-2} J_{n}^{-1}  = (-\q)^{1/2} \; 
	\begin{tikzpicture}[scale = 1/3,baseline={(current bounding box.center)}, yscale = -1] 
	\draw[black, line width = 1pt] (1,0) -- (1,2);
	\draw[black, line width = 1pt] (3,0) -- (3,2);
	\draw[white, line width = 3pt] (4,0) .. controls (4,1) .. (0,1);
	\draw[black, line width = 1pt] (4,0) .. controls (4,1) .. (0,1);
	\draw[black, line width = 1pt] (4,2) .. controls (4,1) .. (5,1);
	\node[anchor = south] at (2,2) {$\hdots$};
	\node[anchor = north] at (2,0) {$\hdots$};
	\filldraw[white] (0,0) -- (1/2,0) -- (1/2,2) -- (0,2) -- (0,0);
	\filldraw[white] (5,0) -- (9/2,0) -- (9/2,2) -- (5,2) -- (5,0);
	\draw[black, line width = 2pt] (.5,0) -- (4.5,0);
	\draw[black, line width = 2pt] (.5,2) -- (4.5,2);
	\draw[decorate, decoration = {brace, mirror, amplitude = 4 pt}, yshift = 3pt] (.5,2.5) -- (3.5,2.5) node [midway,yshift = -7pt] {\footnotesize{n-1}};
	\end{tikzpicture} \; + (-\q)^{-1/2}\; 
	\begin{tikzpicture}[scale = 1/3,baseline={(current bounding box.center)},yscale = -1] 
	\draw[black, line width = 1pt] (1,0) -- (1,2);
	\draw[black, line width = 1pt] (3,0) -- (3,2);
	\draw[white, line width = 3pt] (4,2) .. controls (4,1) .. (0,1);
	\draw[black, line width = 1pt] (4,2) .. controls (4,1) .. (0,1);
	\draw[black, line width = 1pt] (4,0) .. controls (4,1) .. (5,1);
	\node[anchor = south] at (2,2) {$\hdots$};
	\node[anchor = north] at (2,0) {$\hdots$};
	\filldraw[white] (0,0) -- (1/2,0) -- (1/2,2) -- (0,2) -- (0,0);
	\filldraw[white] (5,0) -- (9/2,0) -- (9/2,2) -- (5,2) -- (5,0);
	\draw[black, line width = 2pt] (.5,0) -- (4.5,0);
	\draw[black, line width = 2pt] (.5,2) -- (4.5,2);
	\draw[decorate, decoration = {brace, mirror, amplitude = 4 pt}, yshift = 3pt] (.5,2.5) -- (3.5,2.5) node [midway,yshift = -7pt] {\footnotesize{n-1}};
	\end{tikzpicture} \; = \;
	\begin{tikzpicture}[scale = 1/3,baseline={(current bounding box.center)},yscale=-1] 
	\draw[black, line width = 1pt] (1,0) -- (1,2);
	\draw[black, line width = 1pt] (3,0) -- (3,2);
	\draw[black, line width = 1pt] (4,0) -- (4,2);
	\draw[white, line width = 3pt] (5,1) -- (0,1);
	\draw[black, line width = 1pt] (5,1) -- (0,1);
	\node[anchor = south] at (2,2) {$\hdots$};
	\node[anchor = north] at (2,0) {$\hdots$};
	\filldraw[white] (0,0) -- (1/2,0) -- (1/2,2) -- (0,2) -- (0,0);
	\filldraw[white] (5,0) -- (9/2,0) -- (9/2,2) -- (5,2) -- (5,0);
	\draw[black, line width = 2pt] (.5,0) -- (4.5,0);
	\draw[black, line width = 2pt] (.5,2) -- (4.5,2);
	\draw[decorate, decoration = {brace, mirror, amplitude = 4 pt}, yshift = 3pt] (.5,2.5) -- (4.5,2.5) node [midway,yshift = -8pt] {\footnotesize{ = Y}};
	\end{tikzpicture},
\end{equation}
where we used the definition of the braids.
Putting the two observations together gives the following relations, for all $i = 1, \hdots, n-1$, 
\begin{align*}
((-\q)^{i+1} J_{i})^{k} + ((-\q)^{i+1} J_{i})^{-k} & = [i]_{k}\left(((-\q)^{2}J_{i})^{k} + ((-\q)^{2}J_{i})^{-k} \right)- [i-1]_{k}\left(((-\q)J_{i})^{k} + ((-\q)J_{i})^{-k} \right),\\
	& = 2[i]_{k}\mathsf{T}_{k}\left(\ffrac{(-\q)^{2}J_{i} + (-\q)^{-2}J_{i}^{-1}}{2} \right) - 2[i-1]_{k}\mathsf{T}_{k}\left(\ffrac{(-\q)J_{i} + (-\q)^{-1}J_{i}^{-1}}{2} \right),\\
	& = 2[i]_{k}\mathsf{T}_{k}\left(\frac{\bar{X}_{i+1}}{2} \right) - 2[i-1]_{k}\mathsf{T}_{k}\left(\frac{ \bar{X}_{i}}{2} \right),\\
 ((-\q)^{n+1} J_{n})^{k} + ((-\q)^{n+1} J_{n})^{-k} & = 2[n]_{k}\mathsf{T}_{k}\left(\frac{Y}{2} \right) - 2[n-1]_{k}\mathsf{T}_{k}\left(\frac{\bar{X}_{n}}{2} \right),
\end{align*}
where we used the fact that  for all non-zero $x$
$$
2 \mathsf{T}_{k}((x+x^{-1})/2) = x^{k} + x^{-k}.
$$
Then, it follows that 
\begin{align*}
\bar{C}_{k}(n) + \bar{C}_{-k}(n) & = \sum_{i = 1}^{n}\left(((-\q)^{i+1} J_{i})^{k} + ((-\q)^{i+1} J_{i})^{-k} \right) \\
	& = 2 [1]_{k} \mathsf{T}_{k}(\bar{X}_{2}/2) + \sum_{i = 2}^{n}\left(((-\q)^{i+1} J_{i})^{k} + ((-\q)^{i+1} J_{i})^{-k} \right)\\
	& = 2[2]_{k}\mathsf{T}_{k}(\bar{X}_{3}/2) + \sum_{i = 3}^{n}\left(((-\q)^{i+1} J_{i})^{k} + ((-\q)^{i+1} J_{i})^{-k} \right)\\
	& = 2[n]_{k} \mathsf{T}_{k}(Y/2).
\end{align*}

Using very similar arguments, we have 
\begin{align*}
((-\q)^{i+1 - n} J_{i})^{k} + ((-\q)^{i+1-n} J_{i})^{-k} & = - [n-i]_{k}\left(((-\q)^{2}J_{i})^{k} + ((-\q)^{2}J_{i})^{-k} \right) \\
& \qquad + [n + 1 - i]_{k}\left(((-\q)J_{i})^{k} + ((-\q)J_{i})^{-k} \right) \\
	& = 2[n + 1 - i]_{k} \mathsf{T}_{k}(\bar{X}_{i}/2) - 2 [n-i]_{k} \mathsf{T}_{k}(\bar{X}_{i+1}/2)\\
((-\q)^{1} J_{n})^{k} + ((-\q)^{1} J_{n})^{-k} & = 2 \mathsf{T}_{k}(\bar{X}_{n}/2),
\end{align*}
which give
\begin{align*}
	(-\q)^{-n k }\bar{C}_{k}(n) + (-\q)^{n k }\bar{C}_{-k}(n)
	 & = \sum_{i = 1}^{n}\left(((-\q)^{i+1 - n} J_{i})^{k} + ((-\q)^{i+1- n} J_{i})^{-k} \right) \\ 
		& = 2[n+1-n]_{k} \mathsf{T}_{k}(\bar{X}_{n}/2) + \sum_{i = 1}^{n-1}\left(((-\q)^{i+1 - n} J_{i})^{k} + ((-\q)^{i+1- n} J_{i})^{-k} \right)\\
		& = 2 [n+1 - 1]_{k} \mathsf{T}_{k}(\bar{X}_{1}/2) \equiv 2 [n]_{k}  \mathsf{T}_{k}(\bar{Y}/2),
\end{align*}
where we used the fact that $X_{1} \equiv \bar{Y}$ by definition.

\subsection{Fusion with standard modules}\label{sec:calc.fusion}
We explain here how to compute the fusion product/quotient of standard modules.
 The final results from Propositions~\ref{prop:fus-p} and~\ref{prop:fus-q} rely on two key facts:
\begin{enumerate}
	\item Fusions of the same type ($u/o$) are associative
	 (see \eqref{eq:fusionp.assoc} and \eqref{eq:fusionq.assoc}),
 i.e.\ for any $\atl{}$-module~$\mathsf{M}  $, and $\tl{}$-modules $\mathsf{V}_1$, $\mathsf{V}_2$, we have isomorphisms
	\begin{equation}
		\big( \mathsf{M} \times^{u/o}_f \mathsf{V}_1 \big)\times^{u/o}_f \mathsf{V}_2 \simeq  \mathsf{M} \times^{u/o}_f \big( \mathsf{V}_1  \times_f \mathsf{V}_2 \big),
	\end{equation}
	\begin{equation}
		\big( \mathsf{M} \div^{u/o}_f \mathsf{V}_1 \big) \div^{u/o}_f \mathsf{V}_2 \simeq  \mathsf{M} \div^{u/o}_f \big( \mathsf{V}_1  \times_f \mathsf{V}_2 \big).
	\end{equation}
	\item For any $ t \in \mathbb{N} $	and for generic $\q$:
	\begin{equation}
		\mathsf{S}_{t/2}(n) \times_f \mathsf{S}_{0}(2) \simeq \mathsf{S}_{t/2}(n+2),
	\end{equation}
	\begin{equation}
		\mathsf{S}_{t/2}(n) \times_f \mathsf{S}_{1/2}(1) \simeq \mathsf{S}_{(t-1)/2}(n+1) \oplus \mathsf{S}_{(t+1)/2}(n+1) ,
	\end{equation}
	where it is understood that $\mathsf{S}_{(-1)/2}(n+1) \equiv 0$.
\end{enumerate}
Putting these two facts together, it follows that if we can compute the fusions of $\atl{}$-modules with $\mathsf{S}_{0}(2)$ and $\mathsf{S}_{1}(1)$, fusions with $\mathsf{S}_{t/2}(m)$, $ t \geq 0, m \geq 1 $ can be computed easily by recurrence. In this appendix, we thus present how these four necessary fusions are calculated.
\subsubsection{$\mathsf{W}_{k,z}(n+2) \div^{u/o}_f \mathsf{S}_{0}(2) $ }
Assuming that $\q^{2} \neq -1$, the primitive idempotent corresponding to the projective module $\mathsf{S}_{0}(2)$ is $ \rho_{0} \equiv (\q+\q^{-1})^{-1} e_{1}$. According to our definition of the fusion quotient, we must consider the subspace $W_{0} \equiv  \rho_{0}\mathsf{W}_{k,z}(n+2)$ with the action of $\atl{n}$ obtained from the morphism of algebras $\phi^{u/o}_{n,2} $; however, in this case there is a map  $\psi: \mathsf{W}_{k,z}(n) \to W_{0} $ which consists in adding two positions linked with an arc on the right of every diagram in $\mathsf{W}_{k,z}(n)$, for instance
\begin{equation}
 \begin{tikzpicture}[scale = 1/3,baseline={(current bounding box.center)},yscale= -1] 
	\draw[black, line width = 1pt] (1,1) .. controls (1,2) and (0,2) .. (0,1);
	\draw[black, line width = 1pt] (2,1) .. controls (2,2) and (1,2) .. (1,3);
	\draw[black, line width = 1pt] (3,1) .. controls (3,2) and (4,2) .. (4,1);
	\draw[black, line width = 1pt] (5,1) .. controls (5,2) and (2,2) .. (2,3);
	\draw[black, line width = 1pt] (6,1) .. controls (6,2) and (7,2) .. (7,1);
	\filldraw[white] (-.5,1) -- (.5,1) -- (.5,3) -- (-.5,3) -- (-.5,1);
	\filldraw[white] (7-.5,1) -- (7.5,1) -- (7.5,3) -- (7-.5,3) -- (7-.5,1);
	\draw[black, line width = 2pt] (.5,1) -- (6.5,1);
	\draw[black, line width = 2pt] (.5,3) -- (2.5,3);
\end{tikzpicture} \; \to \;
\begin{tikzpicture}[scale = 1/3,baseline={(current bounding box.center)},yscale = -1] 
	\draw[black, line width = 1pt] (1,1) .. controls (1,2) and (0,2) .. (0,1);
	\draw[black, line width = 1pt] (2,1) .. controls (2,2) and (1,2) .. (1,3);
	\draw[black, line width = 1pt] (3,1) .. controls (3,2) and (4,2) .. (4,1);
	\draw[black, line width = 1pt] (5,1) .. controls (5,2) and (2,2) .. (2,3);
	\draw[black, line width = 1pt] (6,1) .. controls (6,2.5) and (9,2.5) .. (9,1);
	\draw[black, line width = 1pt] (7,1) .. controls (7,2) and (8,2) .. (8,1);
	\filldraw[white] (-.5,1) -- (.5,1) -- (.5,3) -- (-.5,3) -- (-.5,1);
	\filldraw[white] (9-.5,1) -- (9.5,1) -- (9.5,3) -- (9-.5,3) -- (9-.5,1);
	\draw[black, line width = 2pt] (.5,1) -- (8.5,1);
	\draw[black, line width = 2pt] (.5,3) -- (2.5,3);
\end{tikzpicture}\;.
\end{equation} 
One sees directly that this map defines a morphism of modules, and that the resulting sub-module of $W_{0}$ is the same for both types of fusion. Furthermore, any diagram in $\mathsf{W}_{k,z}(n+2)$ is sent to one of the form $\psi(x)$ by the action of the idempotent $\rho_{0}$, so this morphism is surjective. Because the map $\psi$ is obviously injective as well, it must be an isomorphism and we thus get
\begin{equation}
	\mathsf{W}_{k,z}(n+2) \div^{u/o}_{f} \mathsf{S}_{0}(2) \simeq \mathsf{W}_{k,z}(n),
\end{equation}
where it is understood that $k \leq n/2 $ because otherwise the $e_{n+1}$ would acts as zero on $\mathsf{W}_{k,z}(n+2)$.

\subsubsection{$\mathsf{W}_{k,z}(n) \times_{f}^{u/o} \mathsf{S}_{0}(2) $ }
Because both $\mathsf{W}_{k,z}(n)$ and $\mathsf{S}_{0}(2)$ are cyclic, so is their fusion product; in particular one can write $\mathsf{W}_{k,z}(n) \times_{f}^{u/o} \mathsf{S}_{0}(2) = \atl{n+2}x$ with 
\begin{equation}
	x \equiv \;
	\begin{tikzpicture}[scale = 1/3,baseline={(current bounding box.center)},yscale = -1]
	\draw[black, line width = 2pt] (.5,0) -- (11.5,0);
	\draw[black, line width = 2pt] (.5,3) -- (11.5,3);
	\draw[black, line width = 1pt] (1,0) -- (1,3);
	\draw[black, line width = 1pt] (3,0) -- (3,3);
	\draw[black, line width = 1pt] (4,0) .. controls (4,2) and (9,1) .. (9,3);
	\draw[black, line width = 1pt] (5,0) .. controls (5,1) and (6,1) .. (6,0);
	\draw[black, line width = 1pt] (8,0) .. controls (8,1) and (9,1) .. (9,0);
	\draw[black, line width = 1pt] (10,0) .. controls (10,1) and (11,1) .. (11,0);
	\draw[black, line width = 1pt] (4,3) .. controls (4,2) and (5,2) .. (5,3);
	\draw[black, line width = 1pt] (7,3) .. controls (7,2) and (8,2) .. (8,3);
	\draw[black, line width = 1pt] (11,3) .. controls (11,2) and (10,2) .. (10,3);
	\node[anchor = north] at (2,1) { $\hdots$};
	\node[anchor = north] at (7,0) { $\hdots$};
	\node[anchor = south] at (6,3) { $\hdots$};
	\draw[decorate, decoration = {brace, amplitude = 4 pt}, yshift = 3pt] (.5,-0.5) -- (4.5,-0.5) node [midway,yshift = 7pt] {\footnotesize{2k}};
	\draw[decorate, decoration = {brace, amplitude = 4 pt}, yshift = 3pt] (4.5,-0.5) -- (9.5,-0.5) node [midway,yshift = 7pt] {\footnotesize{n-2k}};
	\node[anchor = north ] at (5.5,3) { $ \otimes_{\atl{n}} $};
	\draw[black, line width = 2pt] (0.5,5) -- (9.25,5); 
	\draw[black, line width = 2pt] (0.5,8) -- (4.5,8); 
	\foreach \r in {1,3,4}{
		\draw[black, line width = 1pt] (\r, 5) -- (\r, 8);
	};
	\draw[black, line width = 1pt] (5,5) .. controls (5,6) and (6,6) .. (6,5);
	\draw[black, line width = 1pt] (8,5) .. controls (8,6) and (9,6) .. (9,5);
	\node[anchor = north] at (2,6) {$\hdots $};
	\node[anchor = north] at (7,5) {$\hdots $};
	\draw[black, line width = 2pt] (9.75,5) -- (11.25,5);
	\draw[black, line width = 1pt] (10,5) .. controls (10,6) and (11,6) .. (11,5);
	\draw[decorate, decoration = {brace, mirror, amplitude = 4 pt}, yshift = 3pt] (.5,8.5) -- (4.5,8.5) node [midway,yshift = -7pt] {\footnotesize{2k}};
\end{tikzpicture} \;.
\end{equation}
We also recall our diagram notation for the fusion product: the diagram at the top is the element of $\atl{n+2}$, the one on the bottom left corner is the element of $\mathsf{W}_{k,z}(n) $, and the one on the bottom right corner is the element of $\mathsf{S}_{0}(2)$. Note that closing together any of the through lines in the top diagram of $x$ will yield the zero element, because it will be able to pass through the tensor product. It follows that this fusion product is isomorphic to a standard module of the form $\mathsf{W}_{k,z'}(n+2)$ for some $z' $. Note also that because of the duality between the two types of fusion
\begin{equation}
	\text{Hom}_{\atl{n+2}}\left(\mathsf{W}_{k,z}(n) \times_{f}^{u/o} \mathsf{S}_{0}(2), \mathsf{W}_{k,z}(n+2)\right) \simeq \text{Hom}_{\atl{n}}\big(\mathsf{W}_{k,z}(n) , \underbrace{\mathsf{W}_{k,z}(n+2)\div_{f}^{u/o} \mathsf{S}_{0}(2)}_{\simeq \mathsf{W}_{k,z}(n)} \big) \simeq \mathbb{C},
\end{equation}
so we conclude that $z' = z$, and thus
\begin{equation}
	\mathsf{W}_{k,z}(n) \times_{f}^{u/o} \mathsf{S}_{0}(2) \simeq \mathsf{W}_{k,z}(n+2).
\end{equation}

\subsubsection{$\mathsf{W}_{k,z}(n+1) \div^{u/o}_{f} \mathsf{S}_{1/2}(1)$, ($k \neq 0 $)}
Both types of fusion are very similar so we focus on the $u$-type. The primitive idempotent corresponding to $\mathsf{S}_{1/2}(1)$ is simply the identity so the fusion quotient is the full restriction of $\mathsf{W}_{k,z}(n+1)$. There is then a map $\psi: \mathsf{W}_{k-1/2,z'}(n)  \to \mathsf{W}_{k,z}(n+1)$ which consists in adding a single position on the right of every diagram in $\mathsf{W}_{k-1/2,z'}(n) $ and adding a through line to it which passes under every other line. For instance, 
\begin{equation}
	\begin{tikzpicture}[scale = 1/3,baseline={(current bounding box.center)},yscale = -1] 
	\draw[black, line width = 1pt] (1,1) .. controls (1,2) and (0,2) .. (0,1);
	\draw[black, line width = 1pt] (2,1) .. controls (2,2) and (1,2) .. (1,3);
	\draw[black, line width = 1pt] (3,1) .. controls (3,2) and (4,2) .. (4,1);
	\draw[black, line width = 1pt] (5,1) .. controls (5,2) and (2,2) .. (2,3);
	\draw[black, line width = 1pt] (6,1) .. controls (6,2) and (7,2) .. (7,1);
	\filldraw[white] (-.5,1) -- (.5,1) -- (.5,3) -- (-.5,3) -- (-.5,1);
	\filldraw[white] (7-.5,1) -- (7.5,1) -- (7.5,3) -- (7-.5,3) -- (7-.5,1);
	\draw[black, line width = 2pt] (.5,1) -- (6.5,1);
	\draw[black, line width = 2pt] (.5,3) -- (2.5,3);
\end{tikzpicture} \; \to \;
	\begin{tikzpicture}[scale = 1/3,baseline={(current bounding box.center)},yscale = -1] 
	\draw[black, line width = 1pt] (1,1) .. controls (1,2) and (0,2) .. (0,1);
	\draw[black, line width = 1pt] (2,1) .. controls (2,2) and (1,2) .. (1,3);
	\draw[black, line width = 1pt] (3,1) .. controls (3,2) and (4,2) .. (4,1);
	\draw[black, line width = 1pt] (5,1) .. controls (5,2) and (2,2) .. (2,3);
	\draw[black, line width = 1pt] (7,1) .. controls (7,2) and (3,2) .. (3,3);
	\draw[white, line width = 3pt] (6,1) .. controls (6,2) and (8,2) .. (8,1);
	\draw[black, line width = 1pt] (6,1) .. controls (6,2) and (8,2) .. (8,1);
	\filldraw[white] (-.5,1) -- (.5,1) -- (.5,3) -- (-.5,3) -- (-.5,1);
	\filldraw[white] (8-.5,1) -- (8.5,1) -- (8.5,3) -- (8-.5,3) -- (8-.5,1);
	\draw[black, line width = 2pt] (.5,1) -- (7.5,1);
	\draw[black, line width = 2pt] (.5,3) -- (3.5,3);
\end{tikzpicture}\;.
\end{equation}
One can see that this map (extended linearly) indeed defines an injective morphism of $\atl{n}$ module if $z' = (-\q)^{1/2}z $; this condition on $z$ can be seen by observing that
\begin{equation}
	\begin{tikzpicture}[scale = 1/3,baseline={(current bounding box.center)},yscale = -1] 
	\draw[black, line width = 1pt] (1,1) .. controls (1,2) and (0,2) .. (0,1);
	\draw[black, line width = 1pt] (2,1) .. controls (2,2) and (1,2) .. (1,3);
	\draw[black, line width = 1pt] (3,1) .. controls (3,2) and (4,2) .. (4,1);
	\draw[black, line width = 1pt] (5,1) .. controls (5,2) and (2,2) .. (2,3);
	\draw[black, line width = 1pt] (6,1) .. controls (6,2) and (3,2) .. (3,3); 
	\filldraw[white] (-.5,1) -- (.5,1) -- (.5,3) -- (-.5,3) -- (-.5,1);
	\filldraw[white] (6-.5,1) -- (6.5,1) -- (6.5,3) -- (6-.5,3) -- (6-.5,1);
	\draw[black, line width = 2pt] (.5,1) -- (5.5,1);
	\draw[black, line width = 2pt] (.5,3) -- (3.5,3);
\end{tikzpicture} = 
	z' \begin{tikzpicture}[scale = 1/3,baseline={(current bounding box.center)},yscale = -1] 
	\draw[black, line width = 1pt] (1,1) -- (1,3);
	\draw[black, line width = 1pt] (2,1) -- (2,3);
	\draw[black, line width = 1pt] (3,1) .. controls (3,2) and (4,2) .. (4,1);
	\draw[black, line width = 1pt] (5,1) .. controls (5,2) and (3,2) .. (3,3);
	\filldraw[white] (-.5,1) -- (.5,1) -- (.5,3) -- (-.5,3) -- (-.5,1);
	\filldraw[white] (7-.5,1) -- (7.5,1) -- (7.5,3) -- (7-.5,3) -- (7-.5,1);
	\draw[black, line width = 2pt] (.5,1) -- (5.5,1);
	\draw[black, line width = 2pt] (.5,3) -- (3.5,3);
\end{tikzpicture} \to 
	\begin{tikzpicture}[scale = 1/3,baseline={(current bounding box.center)},yscale = -1] 
	\draw[black, line width = 1pt] (1,1) .. controls (1,2) and (0,2) .. (0,1);
	\draw[black, line width = 1pt] (2,1) .. controls (2,2) and (1,2) .. (1,3);
	\draw[black, line width = 1pt] (3,1) .. controls (3,2) and (4,2) .. (4,1);
	\draw[black, line width = 1pt] (5,1) .. controls (5,2) and (2,2) .. (2,3);
	\draw[black, line width = 1pt] (6,1) .. controls (6,2) and (4,2) .. (4,3);
	\draw[white, line width = 3pt] (7,1) .. controls (7,2) and (3,2) .. (3,3); 
	\draw[black, line width = 1pt] (7,1) .. controls (7,2) and (3,2) .. (3,3); 
	\filldraw[white] (-.5,1) -- (.5,1) -- (.5,3) -- (-.5,3) -- (-.5,1);
	\filldraw[white] (7-.5,1) -- (7.5,1) -- (7.5,3) -- (7-.5,3) -- (7-.5,1);
	\draw[black, line width = 2pt] (.5,1) -- (6.5,1);
	\draw[black, line width = 2pt] (.5,3) -- (4.5,3);
\end{tikzpicture} = 
	(-\q)^{1/2}z 
	\begin{tikzpicture}[scale = 1/3,baseline={(current bounding box.center)},yscale = -1] 
	\draw[black, line width = 1pt] (1,1) -- (1,3);
	\draw[black, line width = 1pt] (2,1) -- (2,3);
	\draw[black, line width = 1pt] (3,1) .. controls (3,2) and (4,2) .. (4,1);
	\draw[black, line width = 1pt] (5,1) .. controls (5,2) and (3,2) .. (3,3);
	\draw[black, line width = 1pt] (6,1) .. controls (6,2) and (4,2) .. (4,3); 
	\filldraw[white] (-.5,1) -- (.5,1) -- (.5,3) -- (-.5,3) -- (-.5,1);
	\filldraw[white] (7-.5,1) -- (7.5,1) -- (7.5,3) -- (7-.5,3) -- (7-.5,1);
	\draw[black, line width = 2pt] (.5,1) -- (6.5,1);
	\draw[black, line width = 2pt] (.5,3) -- (4.5,3);
\end{tikzpicture},
\end{equation}

\noindent Next, consider the map $\phi: \mathsf{W}_{k,z}(n+1)  \to \mathsf{W}_{k+1/2,z(-\q)^{-1/2}}(n+2)$, defined by $\phi = (\q+\q^{-1})^{-1}e_{n+1}\psi$. In other words we add an extra through line, going under all others, at the right of each diagram in $\mathsf{W}_{k,z}(n+1)$ then multiply the result by the idempotent $(\q+\q^{-1})^{-1}e_{n+1} $. For instance, we get
 \begin{equation}
 	\begin{tikzpicture}[scale = 1/3,baseline={(current bounding box.center)},yscale = -1] 
	\draw[black, line width = 1pt] (1,1) --  (1,3);
	\draw[black, line width = 1pt] (2,1) -- (2,3);
	\draw[black, line width = 1pt] (3,1) .. controls (3,2) and (4,2) .. (4,1);
	\draw[black, line width = 1pt] (5,1) .. controls (5,2) and (6,2) .. (6,1); 
	\filldraw[white] (-.5,1) -- (.5,1) -- (.5,3) -- (-.5,3) -- (-.5,1);
	\filldraw[white] (7-.5,1) -- (7.5,1) -- (7.5,3) -- (7-.5,3) -- (7-.5,1);
	\draw[black, line width = 2pt] (.5,1) -- (6.5,1);
	\draw[black, line width = 2pt] (.5,3) -- (2.5,3);
\end{tikzpicture} \to \frac{1}{\q + \q^{-1}}
\begin{tikzpicture}[scale = 1/3,baseline={(current bounding box.center)},yscale = -1] 
	\draw[black, line width = 1pt] (1,1) --  (1,3);
	\draw[black, line width = 1pt] (2,1) -- (2,3);
	\draw[black, line width = 1pt] (3,1) .. controls (3,2) and (4,2) .. (4,1);
	\draw[black, line width = 1pt] (5,1) .. controls (5,2) and (6,2) .. (6,1);
	\draw[black, line width = 1pt] (7,1) .. controls (7,2) and (3,2) .. (3,3); 
	\filldraw[white] (-.5,1) -- (.5,1) -- (.5,3) -- (-.5,3) -- (-.5,1);
	\filldraw[white] (8-.5,1) -- (8.5,1) -- (8.5,3) -- (8-.5,3) -- (8-.5,1);
	\draw[black, line width = 2pt] (.5,1) -- (7.5,1);
	\draw[black, line width = 2pt] (.5,3) -- (3.5,3);
	\foreach \r in {1,...,5}{
		\draw[black, line width = 1pt] (\r,-1) -- (\r,1);
	}
	\draw[black, line width = 1pt] (6,-1) .. controls (6,0) and (7,0) .. (7,-1);
	\draw[black, line width = 1pt] (6,1) .. controls (6,0) and (7,0) .. (7,1); 
	\draw[black, line width = 2pt] (.5,1) -- (7.5,1);
	\draw[black, line width = 2pt] (.5,-1) -- (7.5,-1);
\end{tikzpicture} =  \frac{1}{\q + \q^{-1}}
\begin{tikzpicture}[scale = 1/3,baseline={(current bounding box.center)},yscale = -1] 
	\draw[black, line width = 1pt] (1,1) --  (1,3);
	\draw[black, line width = 1pt] (2,1) -- (2,3);
	\draw[black, line width = 1pt] (3,1) .. controls (3,2) and (4,2) .. (4,1);
	\draw[black, line width = 1pt] (5,1) .. controls (5,2) and (3,2) .. (3,3);
	\draw[black, line width = 1pt] (7,1) .. controls (7,2) and (6,2) .. (6,1); 
	\filldraw[white] (-.5,1) -- (.5,1) -- (.5,3) -- (-.5,3) -- (-.5,1);
	\filldraw[white] (8-.5,1) -- (8.5,1) -- (8.5,3) -- (8-.5,3) -- (8-.5,1);
	\draw[black, line width = 2pt] (.5,1) -- (7.5,1);
	\draw[black, line width = 2pt] (.5,3) -- (3.5,3);
\end{tikzpicture}.
 \end{equation}
 Based on the previous results of this section, we recognize the the image of this map is $\mathsf{W}_{k+1/2,z(-\q)^{-1/2}}(n+2) \div_{f}^{u} \mathsf{S}_{0}(2) \simeq \mathsf{W}_{k+1/2,z(-\q)^{-1/2}}(n)$. It can then be shown that this map (extended linearly) defines a surjective morphism of $\atl{n}$ modules. Furthermore, one can see that the image of the first map $\psi $ is contained in the kernel of the second map $\phi$:
 \begin{equation}
 \begin{tikzpicture}[scale = 1/3,baseline={(current bounding box.center)},yscale = -1] 
	\draw[black, line width = 1pt] (1,1) .. controls (1,2) and (0,2) .. (0,1);
	\draw[black, line width = 1pt] (2,1) .. controls (2,2) and (1,2) .. (1,3);
	\draw[black, line width = 1pt] (3,1) .. controls (3,2) and (4,2) .. (4,1);
	\draw[black, line width = 1pt] (5,1) .. controls (5,2) and (2,2) .. (2,3);
	\draw[black, line width = 1pt] (7,1) .. controls (7,2) and (3,2) .. (3,3);
	\draw[white, line width = 3pt] (6,1) .. controls (6,2) and (8,2) .. (8,1);
	\draw[black, line width = 1pt] (6,1) .. controls (6,2) and (8,2) .. (8,1);
	\filldraw[white] (-.5,1) -- (.5,1) -- (.5,3) -- (-.5,3) -- (-.5,1);
	\filldraw[white] (8-.5,1) -- (8.5,1) -- (8.5,3) -- (8-.5,3) -- (8-.5,1);
	\draw[black, line width = 2pt] (.5,1) -- (7.5,1);
	\draw[black, line width = 2pt] (.5,3) -- (3.5,3);
\end{tikzpicture} \to
	\frac{1}{\q + \q^{-1}}
	\begin{tikzpicture}[scale = 1/3,baseline={(current bounding box.center)},yscale = -1] 
	\draw[black, line width = 1pt] (1,1) .. controls (1,2) and (0,2) .. (0,1);
	\draw[black, line width = 1pt] (2,1) .. controls (2,2) and (1,2) .. (1,3);
	\draw[black, line width = 1pt] (3,1) .. controls (3,2) and (4,2) .. (4,1);
	\draw[black, line width = 1pt] (5,1) .. controls (5,2) and (2,2) .. (2,3);
	\draw[black, line width = 1pt] (7,1) .. controls (7,2) and (3,2) .. (3,3);
	\draw[black, line width = 1pt] (8,1) .. controls (8,2) and (4,2) .. (4,3);
	\draw[white, line width = 3pt] (6,1) .. controls (6,3) and (9,3) .. (9,1);
	\draw[black, line width = 1pt] (6,1) .. controls (6,3) and (9,3) .. (9,1);
	\filldraw[white] (-.5,1) -- (.5,1) -- (.5,3) -- (-.5,3) -- (-.5,1);
	\filldraw[white] (9-.5,1) -- (9.5,1) -- (9.5,3) -- (9-.5,3) -- (9-.5,1);
	\draw[black, line width = 2pt] (.5,1) -- (8.5,1);
	\draw[black, line width = 2pt] (.5,3) -- (4.5,3);
	\foreach \r in {1,...,6}{
		\draw[black, line width = 1pt] (\r,-1) -- (\r,1);
	};
	\draw[black, line width = 1pt] (7,1) .. controls (7,0) and (8,0) .. (8,1);
	\draw[black, line width = 1pt] (7,-1) .. controls (7,0) and (8,0) .. (8,-1);
	\draw[black, line width = 2pt] (.5,-1) -- (8.5,-1);
\end{tikzpicture} = \frac{1}{\q + \q^{-1}}
\begin{tikzpicture}[scale = 1/3,baseline={(current bounding box.center)},yscale = -1] 
	\draw[black, line width = 1pt] (1,1) .. controls (1,2) and (0,2) .. (0,1);
	\draw[black, line width = 1pt] (2,1) .. controls (2,2) and (1,2) .. (1,3);
	\draw[black, line width = 1pt] (3,1) .. controls (3,2) and (4,2) .. (4,1);
	\draw[black, line width = 1pt] (5,1) .. controls (5,2) and (2,2) .. (2,3);
	\draw[black, line width = 1pt] (7,1) .. controls (7,2) and (8,2) .. (8,1);
	\draw[black, line width = 1pt] (3,3) .. controls (3,2) and (4,2) .. (4,3);
	\draw[white, line width = 3pt] (6,1) .. controls (6,3) and (9,3) .. (9,1);
	\draw[black, line width = 1pt] (6,1) .. controls (6,3) and (9,3) .. (9,1);
	\filldraw[white] (-.5,1) -- (.5,1) -- (.5,3) -- (-.5,3) -- (-.5,1);
	\filldraw[white] (9-.5,1) -- (9.5,1) -- (9.5,3) -- (9-.5,3) -- (9-.5,1);
	\draw[black, line width = 2pt] (.5,1) -- (8.5,1);
	\draw[black, line width = 2pt] (.5,3) -- (4.5,3);
\end{tikzpicture} = 0,
 \end{equation}
 where we used the fact that diagrams with less than $k/2$ through lines are equivalent to the zero element in $\mathsf{W}_{k,z}(n)$.
 
 Finally, note that 
 	\begin{equation}
 		 \mathsf{Dim}(\mathsf{W}_{k,z}(n+1) \div^{u}_{f} \mathsf{S}_{1}(1)) = \underbrace{\mathsf{Dim}(\mathsf{W}_{k,z}(n+1))}_{\binom{n+1}{(n+1)/2 - k}} = \underbrace{\mathsf{Dim}(\mathsf{W}_{k-1/2,z'}(n))}_{\binom{n}{(n+1)/2 - k}} + \underbrace{\mathsf{Dim}(\mathsf{W}_{k+1/2,z''}(n))}_{\binom{n}{(n-1)/2 - k}},
 	\end{equation}
 	so the image of $\psi$ is exactly the kernel of $\phi $. Since the eigenvalues of $\bar{Y}$ are different on  $ \mathsf{W}_{k\pm 1/2,z(-\q)^{\mp 1/2}}(n)$, if $z$ is generic, it follows that
 	\begin{equation}
 		\mathsf{W}_{k,z}(n+1) \div^{u}_{f} \mathsf{S}_{1/2}(1) \simeq \mathsf{W}_{k- 1/2,z(-\q)^{ 1/2}}(n) \oplus \mathsf{W}_{k+ 1/2,z(-\q)^{- 1/2}}(n).
 	\end{equation}
 	What about when $z$ is not generic?
	 The eigenvalues of $\bar{Y}$ will be the same on the modules appearing in the previous direct sum if and only if $(-\q)^{2} = 1 $ or $z^{2} = (-\q)^{k/2} $; one can verify directly that $\bar{Y}$ has a Jordan block when acting on the fusion quotient only in the later case. This can be seen by taking any $x \in \mathsf{W}_{k,z}(n+1) \div^{u}_{f} \mathsf{S}_{1/2}(1)$ that is not in the kernel of $\phi$ and verifying that $(\bar{Y}- \lambda)x \neq 0$, where $\lambda$ is the eigenvalue of $\bar{Y}$ on $\mathsf{W}_{k+1/2,z(-\q)^{-1/2}}(n)$. The modules appearing in the fusion quotient is then the indecomposable extension of two standard modules; while these are not yet classified, we shall show that they are indeed unique in our forthcoming work.
 	
\subsubsection{$\mathsf{W}_{k,z}(n+1) \times^{u/o}_{f} \mathsf{S}_{1/2}(1)$, ($k \neq 0 $)}

Both types of fusion are very similar so we again focus on the $u$-type. Since standard modules are always cyclic, so is their fusion and $\mathsf{W}_{k,z}(n) \times^{u}_{f} \mathsf{S}_{1/2}(1) = \atl{n+1}x $ with
\begin{equation}
	x \equiv \;
	\begin{tikzpicture}[scale = 1/3,baseline={(current bounding box.center)},yscale = -1]
	\draw[black, line width = 2pt] (.5,0) -- (10.5,0);
	\draw[black, line width = 2pt] (.5,3) -- (10.5,3);
	\draw[black, line width = 1pt] (1,0) -- (1,3);
	\draw[black, line width = 1pt] (3,0) -- (3,3);
	\draw[black, line width = 1pt] (4,0) .. controls (4,2) and (9,1) .. (9,3);
	\draw[black, line width = 1pt] (5,0) .. controls (5,1) and (6,1) .. (6,0);
	\draw[black, line width = 1pt] (8,0) .. controls (8,1) and (9,1) .. (9,0);
	\draw[black, line width = 1pt] (10,0) -- (10,3);
	\draw[black, line width = 1pt] (4,3) .. controls (4,2) and (5,2) .. (5,3);
	\draw[black, line width = 1pt] (7,3) .. controls (7,2) and (8,2) .. (8,3);
	\node[anchor = north] at (2,1) { $\hdots$};
	\node[anchor = north] at (7,0) { $\hdots$};
	\node[anchor = south] at (6,3) { $\hdots$};
	\draw[decorate, decoration = {brace, amplitude = 4 pt}, yshift = 3pt] (.5,-.5) -- (4.5,-.5) node [midway,yshift = 7pt] {\footnotesize{2k}};
	\draw[decorate, decoration = {brace, amplitude = 4 pt}, yshift = 3pt] (4.5,-.5) -- (9.5,-.5) node [midway,yshift = 7pt] {\footnotesize{n-2k}};
	\node[anchor = north] at (5.5,3) { $ \otimes_{\atl{n}} $};
	\draw[black, line width = 2pt] (0.5,5) -- (9.25,5); 
	\draw[black, line width = 2pt] (0.5,8) -- (4.5,8); 
	\foreach \r in {1,3,4}{
		\draw[black, line width = 1pt] (\r, 5) -- (\r, 8);
	};
	\draw[black, line width = 1pt] (5,5) .. controls (5,6) and (6,6) .. (6,5);
	\draw[black, line width = 1pt] (8,5) .. controls (8,6) and (9,6) .. (9,5);
	\node[anchor = north] at (2,6) {$\hdots $};
	\node[anchor = north] at (7,5) {$\hdots $};
	\draw[black, line width = 2pt] (9.75,5) -- (10.25,5);
	\draw[black, line width = 2pt] (9.75,8) -- (10.25,8);
	\draw[black, line width = 1pt] (10,5) -- (10,8);
	\draw[decorate, decoration = {brace, mirror, amplitude = 4 pt}, yshift = -3pt] (.5,8.5) -- (4.5,8.5) node [midway,yshift = -7pt] {\footnotesize{2k}};
\end{tikzpicture} \; .
\end{equation}

\noindent The module can be decomposed by defining diagrams
\begin{equation}
 v_{+} = e_{2k+2}e_{2k+4} \hdots e_{n}, \qquad v_{-} = e_{2k}e_{2k+2}\hdots e_{n}. 
\end{equation}
These were chosen such that
\begin{equation*}
	v_{+}x = \;
	\begin{tikzpicture}[scale = 1/3,baseline={(current bounding box.center)},yscale = -1]
	\draw[black, line width = 2pt] (.5,0) -- (10.5,0);
	\draw[black, line width = 2pt] (.5,3) -- (10.5,3);
	\draw[black, line width = 1pt] (1,0) -- (1,3);
	\draw[black, line width = 1pt] (3,0) -- (3,3);
	\draw[black, line width = 1pt] (4,0) .. controls (4,2) and (9,1) .. (9,3);
	\draw[black, line width = 1pt] (5,0) .. controls (5,1.5) and (10,.5) .. (10,3);
	\draw[black, line width = 1pt] (6,0) .. controls (6,1) and (7,1) .. (7,0);
	\draw[black, line width = 1pt] (10,0) .. controls (10,1) and (9,1) .. (9,0);
	\draw[black, line width = 1pt] (4,3) .. controls (4,2) and (5,2) .. (5,3);
	\draw[black, line width = 1pt] (7,3) .. controls (7,2) and (8,2) .. (8,3);
	\node[anchor = north] at (2,1) { $\hdots$};
	\node[anchor = north] at (8,0) { $\hdots$};
	\node[anchor = south] at (6,3) { $\hdots$};
	\draw[decorate, decoration = {brace, amplitude = 4 pt}, yshift = 3pt] (.5,-.5) -- (5.5,-.5) node [midway,yshift = 7pt] {\footnotesize{2k+1}};
	\draw[decorate, decoration = {brace, amplitude = 4 pt}, yshift = 3pt] (5.5,-.5) -- (10.5,-.5) node [midway,yshift = 7pt] {\footnotesize{n-2k}};
	\node[anchor = north ] at (5.5,3) { $ \otimes_{\atl{n}} $};
	\draw[black, line width = 2pt] (0.5,5) -- (9.25,5); 
	\draw[black, line width = 2pt] (0.5,8) -- (4.5,8); 
	\foreach \r in {1,3,4}{
		\draw[black, line width = 1pt] (\r, 5) -- (\r, 8);
	};
	\draw[black, line width = 1pt] (5,5) .. controls (5,6) and (6,6) .. (6,5);
	\draw[black, line width = 1pt] (8,5) .. controls (8,6) and (9,6) .. (9,5);
	\node[anchor = north] at (2,6) {$\hdots $};
	\node[anchor = north] at (7,5) {$\hdots $};
	\draw[black, line width = 2pt] (9.75,5) -- (10.25,5);
	\draw[black, line width = 2pt] (9.75,8) -- (10.25,8);
	\draw[black, line width = 1pt] (10,5) -- (10,8);
	\draw[decorate, decoration = {brace, mirror, amplitude = 4 pt}, yshift = -3pt] (.5,8.5) -- (4.5,8.5) node [midway,yshift = -7pt] {\footnotesize{2k}};
	\end{tikzpicture} \; , \qquad
	v_{-}x = \;
	\begin{tikzpicture}[scale = 1/3,baseline={(current bounding box.center)},yscale = -1]
	\draw[black, line width = 2pt] (.5,0) -- (10.5,0);
	\draw[black, line width = 2pt] (.5,3) -- (10.5,3);
	\draw[black, line width = 1pt] (1,0) -- (1,3);
	\draw[black, line width = 1pt] (3,0) -- (3,3);
	\draw[black, line width = 1pt] (4,0) .. controls (4,1) and (5,1) .. (5,0);
	\draw[black, line width = 1pt] (6,0) .. controls (6,1) and (7,1) .. (7,0);
	\draw[black, line width = 1pt] (10,0) .. controls (10,1) and (9,1) .. (9,0);
	\draw[black, line width = 1pt] (4,3) .. controls (4,2) and (5,2) .. (5,3);
	\draw[black, line width = 1pt] (7,3) .. controls (7,2) and (8,2) .. (8,3);
	\draw[black, line width = 1pt] (10,3) .. controls (10,2) and (9,2) .. (9,3);
	\node[anchor = north] at (2,1) { $\hdots$};
	\node[anchor = north] at (8,0) { $\hdots$};
	\node[anchor = south] at (6,3) { $\hdots$};
	\draw[decorate, decoration = {brace, amplitude = 4 pt}, yshift = 3pt] (.5,-.5) -- (3.5,-.5) node [midway,yshift = 7pt] {\footnotesize{2k-1}};
	\draw[decorate, decoration = {brace, amplitude = 4 pt}, yshift = 3pt] (3.5,-.5) -- (10.5,-.5) node [midway,yshift = 7pt] {\footnotesize{n-2k+2}};
	\node[anchor = north ] at (5.5,3) { $ \otimes_{\atl{n}} $};
	\draw[black, line width = 2pt] (0.5,5) -- (9.25,5); 
	\draw[black, line width = 2pt] (0.5,8) -- (4.5,8); 
	\foreach \r in {1,3,4}{
		\draw[black, line width = 1pt] (\r, 5) -- (\r, 8);
	};
	\draw[black, line width = 1pt] (5,5) .. controls (5,6) and (6,6) .. (6,5);
	\draw[black, line width = 1pt] (8,5) .. controls (8,6) and (9,6) .. (9,5);
	\node[anchor = north] at (2,6) {$\hdots $};
	\node[anchor = north] at (7,5) {$\hdots $};
	\draw[black, line width = 2pt] (9.75,5) -- (10.25,5);
	\draw[black, line width = 2pt] (9.75,8) -- (10.25,8);
	\draw[black, line width = 1pt] (10,5) -- (10,8);
	\draw[decorate, decoration = {brace, mirror, amplitude = 4 pt}, yshift = 3pt] (.5,8.5) -- (4.5,8.5) node [midway,yshift = -7pt] {\footnotesize{2k}};
\end{tikzpicture} \; .
\end{equation*}
One verifies easily that $\atl{n+1}v_{-}x$ is a sub-module: any diagram acting on $v_{-}x$ will either contract two (or more) through lines together or shuffle around the arcs on the bottom boundary. However, the former will be trivial because, for instance,
\begin{equation}
e_{1}v_{-}x = \;
\begin{tikzpicture}[scale = 1/3,baseline={(current bounding box.center)},yscale = -1]
	\draw[black, line width = 2pt] (-1.5,0) -- (10.5,0);
	\draw[black, line width = 2pt] (-1.5,3) -- (10.5,3);
	\draw[black, line width = 1pt] (-1,0) .. controls (-1,1) and (0,1) .. (0,0);
	\draw[black, line width = 1pt] (-1,3) .. controls (-1,2) and (0,2) .. (0,3);
	\draw[black, line width = 1pt] (1,0) -- (1,3);
	\draw[black, line width = 1pt] (3,0) -- (3,3);
	\draw[black, line width = 1pt] (4,0) .. controls (4,1) and (5,1) .. (5,0);
	\draw[black, line width = 1pt] (6,0) .. controls (6,1) and (7,1) .. (7,0);
	\draw[black, line width = 1pt] (10,0) .. controls (10,1) and (9,1) .. (9,0);
	\draw[black, line width = 1pt] (4,3) .. controls (4,2) and (5,2) .. (5,3);
	\draw[black, line width = 1pt] (7,3) .. controls (7,2) and (8,2) .. (8,3);
	\draw[black, line width = 1pt] (10,3) .. controls (10,2) and (9,2) .. (9,3);
	\node[anchor = north] at (2,1) { $\hdots$};
	\node[anchor = north] at (8,0) { $\hdots$};
	\node[anchor = south] at (6,3) { $\hdots$};
	\draw[decorate, decoration = {brace, amplitude = 4 pt}, yshift = 3pt] (-1.5,-.5) -- (3.5,-.5) node [midway,yshift = 7pt] {\footnotesize{2k-1}};
	\draw[decorate, decoration = {brace, amplitude = 4 pt}, yshift = 3pt] (3.5,-.5) -- (10.5,-.5) node [midway,yshift = 7pt] {\footnotesize{n-2k+2}};
	\node[anchor = north ] at (5.5,3) { $ \otimes_{\atl{n}} $};
	\draw[black, line width = 2pt] (-1.5,5) -- (9.25,5); 
	\draw[black, line width = 2pt] (-1.5,8) -- (4.5,8); 
	\foreach \r in {-1,0,1,3,4}{
		\draw[black, line width = 1pt] (\r, 5) -- (\r, 8);
	};
	\draw[black, line width = 1pt] (5,5) .. controls (5,6) and (6,6) .. (6,5);
	\draw[black, line width = 1pt] (8,5) .. controls (8,6) and (9,6) .. (9,5);
	\node[anchor = north] at (2,6) {$\hdots $};
	\node[anchor = north] at (7,5) {$\hdots $};
	\draw[black, line width = 2pt] (9.75,5) -- (10.25,5);
	\draw[black, line width = 2pt] (9.75,8) -- (10.25,8);
	\draw[black, line width = 1pt] (10,5) -- (10,8);
	\draw[decorate, decoration = {brace, mirror, amplitude = 4 pt}, yshift = -3pt] (-1.5,8.5) -- (4.5,8.5) node [midway,yshift = -7pt] {\footnotesize{2k}};
\end{tikzpicture} \; = \;
\begin{tikzpicture}[scale = 1/3,baseline={(current bounding box.center)},yscale = -1]
	\draw[black, line width = 2pt] (-1.5,0) -- (10.5,0);
	\draw[black, line width = 2pt] (-1.5,3) -- (10.5,3);
	\draw[black, line width = 1pt] (-1,0) --  (-1,3);
	\draw[black, line width = 1pt] (0,0) -- (0,3);
	\draw[black, line width = 1pt] (1,0) -- (1,3);
	\draw[black, line width = 1pt] (3,0) -- (3,3);
	\draw[black, line width = 1pt] (4,0) .. controls (4,1) and (5,1) .. (5,0);
	\draw[black, line width = 1pt] (6,0) .. controls (6,1) and (7,1) .. (7,0);
	\draw[black, line width = 1pt] (10,0) .. controls (10,1) and (9,1) .. (9,0);
	\draw[black, line width = 1pt] (4,3) .. controls (4,2) and (5,2) .. (5,3);
	\draw[black, line width = 1pt] (7,3) .. controls (7,2) and (8,2) .. (8,3);
	\draw[black, line width = 1pt] (10,3) .. controls (10,2) and (9,2) .. (9,3);
	\node[anchor = north] at (2,1) { $\hdots$};
	\node[anchor = north] at (8,0) { $\hdots$};
	\node[anchor = south] at (6,3) { $\hdots$};
	\draw[decorate, decoration = {brace, amplitude = 4 pt}, yshift = 3pt] (-1.5,-.5) -- (3.5,-.5) node [midway,yshift = 7pt] {\footnotesize{2k-1}};
	\draw[decorate, decoration = {brace, amplitude = 4 pt}, yshift = 3pt] (3.5,-.5) -- (10.5,-.5) node [midway,yshift = 7pt] {\footnotesize{n-2k+2}};
	\node[anchor = north ] at (5.5,3) { $ \otimes_{\atl{n}} $};
	\draw[black, line width = 2pt] (-1.5,5) -- (9.25,5); 
	\draw[black, line width = 2pt] (-1.5,8) -- (4.5,8); 
	\foreach \r in {1,3,4}{
		\draw[black, line width = 1pt] (\r, 5) -- (\r, 8);
	};
	\draw[black, line width = 1pt] (-1,5) .. controls (-1,6) and (0,6) .. (0,5);
	\draw[black, line width = 1pt] (-1,8) .. controls (-1,7) and (0,7) .. (0,8);
	\draw[black, line width = 1pt] (5,5) .. controls (5,6) and (6,6) .. (6,5);
	\draw[black, line width = 1pt] (8,5) .. controls (8,6) and (9,6) .. (9,5);
	\node[anchor = north] at (2,6) {$\hdots $};
	\node[anchor = north] at (7,5) {$\hdots $};
	\draw[black, line width = 2pt] (9.75,5) -- (10.25,5);
	\draw[black, line width = 2pt] (9.75,8) -- (10.25,8);
	\draw[black, line width = 1pt] (10,5) -- (10,8);
	\draw[decorate, decoration = {brace, mirror, amplitude = 4 pt}, yshift = -3pt] (-1.5,8.5) -- (4.5,8.5) node [midway,yshift = -7pt] {\footnotesize{2k}};
\end{tikzpicture} \; = 0,
\end{equation}
where we used the definition of the tensor product, and the definition of the standard modules. We thus conclude that $\atl{n+1}v_{-}x$ is a sub module isomorphic to $\mathsf{W}_{k-1/2 , z'} (n+1)$ for some $z'$, where the isomorphism is obtained by simply \emph{cutting off} the top of the diagrams, i.e.
\begin{equation}
	\begin{tikzpicture}[scale = 1/3,baseline={(current bounding box.center)},yscale = -1]
	\draw[black, line width = 2pt] (.5,0) -- (10.5,0);
	\draw[black, line width = 2pt] (.5,3) -- (10.5,3);
	\draw[black, line width = 1pt] (1,0) -- (1,3);
	\draw[black, line width = 1pt] (3,0) -- (3,3);
	\draw[black, line width = 1pt] (4,0) .. controls (4,1) and (5,1) .. (5,0);
	\draw[black, line width = 1pt] (6,0) .. controls (6,1) and (7,1) .. (7,0);
	\draw[black, line width = 1pt] (10,0) .. controls (10,1) and (9,1) .. (9,0);
	\draw[black, line width = 1pt] (4,3) .. controls (4,2) and (5,2) .. (5,3);
	\draw[black, line width = 1pt] (7,3) .. controls (7,2) and (8,2) .. (8,3);
	\draw[black, line width = 1pt] (10,3) .. controls (10,2) and (9,2) .. (9,3);
	\node[anchor = north] at (2,1) { $\hdots$};
	\node[anchor = north] at (8,0) { $\hdots$};
	\node[anchor = south] at (6,3) { $\hdots$};
	\draw[decorate, decoration = {brace, amplitude = 4 pt}, yshift = 3pt] (.5,-.5) -- (3.5,-.5) node [midway,yshift = 7pt] {\footnotesize{2k-1}};
	\draw[decorate, decoration = {brace, amplitude = 4 pt}, yshift = 3pt] (3.5,-.5) -- (10.5,-.5) node [midway,yshift = 7pt] {\footnotesize{n-2k+2}};
	\node[anchor = north ] at (5.5,3) { $ \otimes_{\atl{n}} $};
	\draw[black, line width = 2pt] (0.5,5) -- (9.25,5); 
	\draw[black, line width = 2pt] (0.5,8) -- (4.5,8); 
	\foreach \r in {1,3,4}{
		\draw[black, line width = 1pt] (\r, 5) -- (\r, 8);
	};
	\draw[black, line width = 1pt] (5,5) .. controls (5,6) and (6,6) .. (6,5);
	\draw[black, line width = 1pt] (8,5) .. controls (8,6) and (9,6) .. (9,5);
	\node[anchor = north] at (2,6) {$\hdots $};
	\node[anchor = north] at (7,5) {$\hdots $};
	\draw[black, line width = 2pt] (9.75,5) -- (10.25,5);
	\draw[black, line width = 2pt] (9.75,8) -- (10.25,8);
	\draw[black, line width = 1pt] (10,5) -- (10,8);
	\draw[decorate, decoration = {brace, mirror, amplitude = 4 pt}, yshift = -3pt] (.5,8.5) -- (4.5,8.5) node [midway,yshift = -7pt] {\footnotesize{2k}};
\end{tikzpicture} \; \to \;
\begin{tikzpicture}[scale = 1/3,baseline={(current bounding box.center)},yscale = -1]
	\draw[black, line width = 2pt] (.5,0) -- (10.5,0);
	\draw[black, line width = 2pt] (.5,3) -- (3.5,3);
	\draw[black, line width = 1pt] (1,0) -- (1,3);
	\draw[black, line width = 1pt] (3,0) -- (3,3);
	\draw[black, line width = 1pt] (4,0) .. controls (4,1) and (5,1) .. (5,0);
	\draw[black, line width = 1pt] (6,0) .. controls (6,1) and (7,1) .. (7,0);
	\draw[black, line width = 1pt] (10,0) .. controls (10,1) and (9,1) .. (9,0);
	\node[anchor = north] at (2,1) { $\hdots$};
	\node[anchor = north] at (8,0) { $\hdots$};
	\draw[decorate, decoration = {brace, amplitude = 4 pt}, yshift = 3pt] (.5,-.5) -- (3.5,-.5) node [midway,yshift = 7pt] {\footnotesize{2k-1}};
	\draw[decorate, decoration = {brace, amplitude = 4 pt}, yshift = 3pt] (3.5,-.5) -- (10.5,-.5) node [midway,yshift = 7pt] {\footnotesize{n-2k+2}};
\end{tikzpicture} \; .
\end{equation}
To find $z'$ we use the duality between the two fusion \eqref{eq:fusion.duality}:
\begin{align*}
	\text{Hom}_{\atl{n+1}}\left(\mathsf{W}_{k,z}(n) \times_{f}^{u} \mathsf{S}_{\half}(1), \mathsf{W}_{k-\half,z'}(n+1)\right) & \simeq \text{Hom}_{\atl{n}}\big(\mathsf{W}_{k,z}(n) , \underbrace{\mathsf{W}_{k-\half,z'}(n+1)\div_{f}^{u} \mathsf{S}_{\half}(1)}_{\simeq \mathsf{W}_{k,z' (-\q)^{-1/2}}(n) \oplus \mathsf{W}_{k-1,z' (-\q)^{1/2}}(n)}\!\!\!\big)\notag \\
	& \simeq \delta_{z, z' (-\q)^{-\half}}\mathbb{C},
\end{align*}
so it follows that $z' = z (-\q)^{1/2}$.

Note now that acting on $v_{+}x$ with some diagram $a$ can do two things: it can shuffle the arcs on the bottom boundary, and it can close pairs of through lines. One can verify that closing the two rightmost through lines together will produce an element of the form $a v_{-}x$, and the same thing happens when closing the leftmost line with the rightmost one. To see this last assertion, observe that in $\mathsf{W}_{k,z}$ 
\begin{equation}
	\begin{tikzpicture}[scale = 1/3,baseline={(current bounding box.center)},yscale = -1]
	\draw[black, line width = 2pt] (.5,0) -- (9.5,0);
	\draw[black, line width = 2pt] (.5,3) -- (4.5,3);
	\draw[black, line width = 1pt] (1,0) -- (1,3);
	\draw[black, line width = 1pt] (3,0) -- (3,3);
	\draw[black, line width = 1pt] (4,0) -- (4,3);
	\draw[black, line width = 1pt] (5,0) .. controls (5,1) and (6,1) .. (6,0);
	\draw[black, line width = 1pt] (8,0) .. controls (8,1) and (9,1) .. (9,0);
	\node[anchor = north] at (2,1) {$\hdots $};
	\node[anchor = north] at (7,0) {$\hdots $};
\end{tikzpicture} \; = z^{-1} \;
\begin{tikzpicture}[scale = 1/3,baseline={(current bounding box.center)},yscale = -1]
	\draw[black, line width = 1pt] (1,6) -- (1,3);
	\draw[black, line width = 1pt] (3,6) -- (3,3);
	\draw[black, line width = 1pt] (4,6) -- (4,3);
	\draw[black, line width = 1pt] (5,3) .. controls (5,4) and (6,4) .. (6,3);
	\draw[black, line width = 1pt] (8,3) .. controls (8,4) and (9,4) .. (9,3);
	\node[anchor = north] at (2,4) {$\hdots $};
	\node[anchor = north] at (7,3) {$\hdots $};
	\draw[black, line width = 1pt] (1,0) .. controls (1,1) and (0,1) .. (0,0);
	\draw[black, line width = 1pt] (2,0) .. controls (2,1) and (1,2) .. (1,3);
	\draw[black, line width = 1pt] (4,0) .. controls (4,1) and (3,2) .. (3,3);
	\draw[black, line width = 1pt] (5,0) .. controls (5,1) and (6,1) .. (6,0);
	\draw[black, line width = 1pt] (8,0) .. controls (8,1) and (9,1) .. (9,0);
	\draw[black, line width = 1pt] (4,3) .. controls (4,2) and (5,2) .. (5,3);
	\draw[black, line width = 1pt] (7,3) .. controls (7,2) and (8,2) .. (8,3);
	\draw[black, line width = 1pt] (9,3) .. controls (9,2) and (10,2) .. (10,3);
	\filldraw[white] (-.5,0) -- (.5,0) -- (.5,3) -- (-.5,3) -- (-.5,0);
	\filldraw[white] (10.5,0) -- (9.5,0) -- (9.5,3) -- (10.5,3) -- (10.5,0);
	\draw[black, line width = 2pt] (.5,0) -- (9.5,0);
	\draw[black, line width = 2pt] (.5,3) -- (9.5,3);
	\draw[black, line width = 2pt] (.5,6) -- (4.5,6);
	\node[anchor = north] at (2.5, 1) {$\hdots $};
	\node[anchor = north] at (7,0) {$\hdots $};
	\node[anchor = south] at (6,3) {$\hdots $};
\end{tikzpicture} \;,
\end{equation}
it thus follows that
\begin{align}
	e_{2k+1} \hdots e_{n-1}e_{n+1} v_{+}x & = \;
	\begin{tikzpicture}[scale = 1/3,baseline={(current bounding box.center)},yscale = -1]
	\draw[black, line width = 1pt] (0,0) .. controls (0,1) and (-1,1) .. (-1,0);
	\draw[black, line width = 1pt] (0,3) .. controls (0,2) and (-1,2) .. (-1,3);
	\draw[black, line width = 1pt] (1,0) -- (1,3);
	\draw[black, line width = 1pt] (3,0) -- (3,3);
	\draw[black, line width = 1pt] (4,0) .. controls (4,2) and (9,1) .. (9,3);
	\draw[black, line width = 1pt] (5,0) .. controls (5,1) and (6,1) .. (6,0);
	\draw[black, line width = 1pt] (8,0) .. controls (8,1) and (9,1) .. (9,0);
	\draw[black, line width = 1pt] (10,0) .. controls (10,1) and (11,1) .. (11,0);
	\draw[black, line width = 1pt] (10,3) .. controls (10,2) and (11,2) .. (11,3);
	\draw[black, line width = 1pt] (4,3) .. controls (4,2) and (5,2) .. (5,3);
	\draw[black, line width = 1pt] (7,3) .. controls (7,2) and (8,2) .. (8,3);
	\node[anchor = north] at (2,1) { $\hdots$};
	\node[anchor = north] at (7,0) { $\hdots$};
	\node[anchor = south] at (6,3) { $\hdots$};
	\draw[decorate, decoration = {brace, amplitude = 4 pt}, yshift = 3pt] (-.5,-.5) -- (5.5,-.5) node [midway,yshift = 7pt] {\footnotesize{2k+1}};
	\draw[decorate, decoration = {brace, amplitude = 4 pt}, yshift = 3pt] (5.5,-.5) -- (10.5,-.5) node [midway,yshift = 7pt] {\footnotesize{n-2k}};
	\node[anchor = north ] at (5.5,3) { $ \otimes_{\atl{n}} $};
	\draw[black, line width = 2pt] (-0.5,5) -- (9.25,5); 
	\draw[black, line width = 2pt] (-0.5,8) -- (4.5,8); 
	\foreach \r in {0,1,3,4}{
		\draw[black, line width = 1pt] (\r, 5) -- (\r, 8);
	};
	\draw[black, line width = 1pt] (5,5) .. controls (5,6) and (6,6) .. (6,5);
	\draw[black, line width = 1pt] (8,5) .. controls (8,6) and (9,6) .. (9,5);
	\node[anchor = north] at (2,6) {$\hdots $};
	\node[anchor = north] at (7,5) {$\hdots $};
	\draw[black, line width = 2pt] (9.75,5) -- (10.25,5);
	\draw[black, line width = 2pt] (9.75,8) -- (10.25,8);
	\draw[black, line width = 1pt] (10,5) -- (10,8);
	\draw[decorate, decoration = {brace, mirror, amplitude = 4 pt}, yshift = -3pt] (-.5,8.5) -- (4.5,8.5) node [midway,yshift = -7pt] {\footnotesize{2k}};
	\filldraw[white] (-.5,0) -- (-1.5,0) -- (-1.5,3) -- (-.5,3) -- (-.5,0);
	\filldraw[white] (10.5,0) -- (11.5,0) -- (11.5,3) -- (10.5,3) -- (10.5,0);
	\draw[black, line width = 2pt] (-.5,0) -- (10.5,0);
	\draw[black, line width = 2pt] (-.5,3) -- (10.5,3);
	\end{tikzpicture} \; = z^{-1}\;
		\begin{tikzpicture}[scale = 1/3,baseline={(current bounding box.center)},yscale = -1]
	\draw[black, line width = 1pt] (0,0) .. controls (0,1) and (-1,1) .. (-1,0);
	\draw[black, line width = 1pt] (0,3) .. controls (0,2) and (-1,2) .. (-1,3);
	\draw[black, line width = 1pt] (1,0) -- (1,3);
	\draw[black, line width = 1pt] (3,0) -- (3,3);
	\draw[black, line width = 1pt] (4,0) .. controls (4,2) and (9,1) .. (9,3);
	\draw[black, line width = 1pt] (5,0) .. controls (5,1) and (6,1) .. (6,0);
	\draw[black, line width = 1pt] (8,0) .. controls (8,1) and (9,1) .. (9,0);
	\draw[black, line width = 1pt] (10,0) .. controls (10,1) and (11,1) .. (11,0);
	\draw[black, line width = 1pt] (10,3) .. controls (10,2) and (11,2) .. (11,3);
	\draw[black, line width = 1pt] (4,3) .. controls (4,2) and (5,2) .. (5,3);
	\draw[black, line width = 1pt] (7,3) .. controls (7,2) and (8,2) .. (8,3);
	\node[anchor = north] at (2,1) { $\hdots$};
	\node[anchor = north] at (7,0) { $\hdots$};
	\node[anchor = south] at (6,3) { $\hdots$};
	\draw[decorate, decoration = {brace, amplitude = 4 pt}, yshift = 3pt] (-.5,-.5) -- (5.5,-.5) node [midway,yshift = 7pt] {\footnotesize{2k+1}};
	\draw[decorate, decoration = {brace, amplitude = 4 pt}, yshift = 3pt] (5.5,-.5) -- (10.5,-.5) node [midway,yshift = 7pt] {\footnotesize{n-2k}};
	\node[anchor = north ] at (5.5,6) { $ \otimes_{\atl{n}} $};
	\draw[black, line width = 2pt] (-0.5,8) -- (9.25,8); 
	\draw[black, line width = 2pt] (-0.5,11) -- (4.5,11); 
	\foreach \r in {0,1,3,4}{
		\draw[black, line width = 1pt] (\r, 8) -- (\r, 11);
	};
	\draw[black, line width = 1pt] (5,8) .. controls (5,9) and (6,9) .. (6,8);
	\draw[black, line width = 1pt] (8,8) .. controls (8,9) and (9,9) .. (9,8);
	\node[anchor = north] at (2,9) {$\hdots $};
	\node[anchor = north] at (7,8) {$\hdots $};
	\draw[black, line width = 2pt] (9.75,8) -- (10.25,8);
	\draw[black, line width = 2pt] (9.75,11) -- (10.25,11);
	\draw[black, line width = 1pt] (10,8) -- (10,11);
	\draw[decorate, decoration = {brace, mirror, amplitude = 4 pt}, yshift = -3pt] (-.5,11.5) -- (4.5,11.5) node [midway,yshift = -7pt] {\footnotesize{2k}};
	\filldraw[white] (-.5,0) -- (-1.5,0) -- (-1.5,3) -- (-.5,3) -- (-.5,0);
	\filldraw[white] (10.5,0) -- (11.5,0) -- (11.5,3) -- (10.5,3) -- (10.5,0);
	\draw[black, line width = 2pt] (-.5,0) -- (10.5,0);
	\draw[black, line width = 2pt] (-.5,3) -- (10.5,3);
	%
	%
	\draw[black, line width = 1pt] (0,3) .. controls (0,4) and (-1,4) .. (-1,3);
	\draw[black, line width = 1pt] (1,3) .. controls (1,4) and (0,5) .. (0,6);
	\draw[black, line width = 1pt] (3,3) .. controls (3,4) and (2,5) .. (2,6);
	\draw[black, line width = 1pt] (4,3) .. controls (4,4) and (3,5) .. (3,6);
	\draw[black, line width = 1pt] (5,3) .. controls (5,4) and (6,4) .. (6,3);
	\draw[black, line width = 1pt] (8,3) .. controls (8,4) and (9,4) .. (9,3);
	\draw[black, line width = 1pt] (4,6) .. controls (4,5) and (5,5) .. (5,6);
	\draw[black, line width = 1pt] (7,6) .. controls (7,5) and (8,5) .. (8,6);
	\draw[black, line width = 1pt] (9,6) .. controls (9,5) and (11,5) .. (11,6);
	\draw[white, line width = 3pt] (10,3) -- (10,6);
	\draw[black, line width = 1pt] (10,3) -- (10,6);
	\filldraw[white] (-.5,3) -- (-2.5,3) -- (-2.5,6) -- (-.5,6) -- (-.5,3);
	\filldraw[white] (10.5,3) -- (11.5,3) -- (11.5,6) -- (10.5,6) -- (10.5,3);
	\draw[black, line width = 2pt] (-.5,3) -- (10.5,3);
	\draw[black, line width = 2pt] (-.5,6) -- (10.5,6);
	\node[anchor = north] at (1.5, 4) {$\hdots $};
	\node[anchor = north] at (7,3) {$\hdots $};
	\node[anchor = south] at (6,6) {$\hdots $};
	\end{tikzpicture} \\
	& = -(-\q)^{3/2}z^{-1} \;
\begin{tikzpicture}[scale = 1/3,baseline={(current bounding box.center)},yscale = -1]
	\draw[black, line width = 1pt] (0,0) .. controls (0,1) and (-1,1) .. (-1,0);
	\draw[black, line width = 1pt] (0,3) .. controls (0,2) and (1,1) .. (1,0);
	\draw[black, line width = 1pt] (3,0) .. controls (3,1) and (2,2) .. (2,3);
	\draw[black, line width = 1pt] (4,0) .. controls (4,1) and (3,2) .. (3,3);
	\draw[black, line width = 1pt] (5,0) .. controls (5,1) and (6,1) .. (6,0);
	\draw[black, line width = 1pt] (8,0) .. controls (8,1) and (9,1) .. (9,0);
	\draw[black, line width = 1pt] (10,0) .. controls (10,1) and (11,1) .. (11,0);
	\draw[black, line width = 1pt] (10,3) .. controls (10,2) and (9,2) .. (9,3);
	\draw[black, line width = 1pt] (4,3) .. controls (4,2) and (5,2) .. (5,3);
	\draw[black, line width = 1pt] (7,3) .. controls (7,2) and (8,2) .. (8,3);
	\node[anchor = north] at (1.5,1) { $\hdots$};
	\node[anchor = north] at (7,0) { $\hdots$};
	\node[anchor = south] at (6,3) { $\hdots$};
	\draw[decorate, decoration = {brace, amplitude = 4 pt}, yshift = 3pt] (-.5,-.5) -- (5.5,-.5) node [midway,yshift = 7pt] {\footnotesize{2k+1}};
	\draw[decorate, decoration = {brace, amplitude = 4 pt}, yshift = 3pt] (5.5,-.5) -- (10.5,-.5) node [midway,yshift = 7pt] {\footnotesize{n-2k}};
	\node[anchor = north ] at (5.5,3) { $ \otimes_{\atl{n}} $};
	\draw[black, line width = 2pt] (-0.5,5) -- (9.25,5); 
	\draw[black, line width = 2pt] (-0.5,8) -- (4.5,8); 
	\foreach \r in {0,1,3,4}{
		\draw[black, line width = 1pt] (\r, 5) -- (\r, 8);
	};
	\draw[black, line width = 1pt] (5,5) .. controls (5,6) and (6,6) .. (6,5);
	\draw[black, line width = 1pt] (8,5) .. controls (8,6) and (9,6) .. (9,5);
	\node[anchor = north] at (2,6) {$\hdots $};
	\node[anchor = north] at (7,5) {$\hdots $};
	\draw[black, line width = 2pt] (9.75,5) -- (10.25,5);
	\draw[black, line width = 2pt] (9.75,8) -- (10.25,8);
	\draw[black, line width = 1pt] (10,5) -- (10,8);
	\draw[decorate, decoration = {brace, mirror, amplitude = 4 pt}, yshift = -3pt] (-.5,8.5) -- (4.5,8.5) node [midway,yshift = -7pt] {\footnotesize{2k}};
	\filldraw[white] (-.5,0) -- (-1.5,0) -- (-1.5,3) -- (-.5,3) -- (-.5,0);
	\filldraw[white] (10.5,0) -- (11.5,0) -- (11.5,3) -- (10.5,3) -- (10.5,0);
	\draw[black, line width = 2pt] (-.5,0) -- (10.5,0);
	\draw[black, line width = 2pt] (-.5,3) -- (10.5,3);
	\end{tikzpicture}\; ,
\end{align}
where we used the closed braid identity
\begin{equation}
\begin{tikzpicture}[scale = 1/3,baseline={(current bounding box.center)}, yscale = -1]
	\draw[black, line width = 1pt] (0,0) -- (2,2);
	\draw[white, line width = 3pt] (0,2) -- (2,0);
	\draw[black, line width = 1pt] (0,2) -- (2,0);	
	\draw[black, line width = 1pt] (2,0) .. controls (3,0) and (3,2) .. (2,2);
\end{tikzpicture} \; = \; -(-\q)^{3/2} \;
\begin{tikzpicture}[scale = 1/3,baseline={(current bounding box.center)}]
	\draw[black, line width = 1pt] (2,0) .. controls (3,0) and (3,2) .. (2,2);
\end{tikzpicture} \;.
\end{equation}
Repeating the arguments leading to the identification of $\atl{n+1}v_{-}x$, we finally obtain 
\begin{equation}
	(\mathsf{W}_{k,z}(n) \times_{f}^{u} \mathsf{S}_{1/2}(1))/(\atl{n+1}v_{-}x) = \atl{n+1}(v_{+}x + \atl{n+1}v_{-}x) \simeq \mathsf{W}_{k+1/2,z (-\q)^{-1/2}}(n+1).
\end{equation}
To complete the decomposition, we need to figure out if this quotient splits, i.e. if the fusion product is the direct sum of two standard modules, or if it's their indecomposable extension. However, if $z $ is generic then $\bar{Y} $ has distinct eigenvalues on $\mathsf{W}_{k \pm 1/2, z (-\q)^{\mp 1/2}} (n+1)$ so the quotient must split, and thus
\begin{equation}
	\mathsf{W}_{k,z}(n) \times_{f}^{u} \mathsf{S}_{1/2}(1) \simeq \mathsf{W}_{k-1/2,z(-\q)^{1/2}}(n+1) \oplus \mathsf{W}_{k+1/2,z(-\q)^{-1/2}}(n+1).
\end{equation}
If $z^{2} = (-\q)^{k} $ then the quotient does not split and the fusion product is then the indecomposable extension of the two standard modules appearing in the previous direct sum.

\subsubsection{$\mathsf{W}_{0,z}(n) \times^{u}_{f} \mathsf{S}_{{1/2}}(1)$}
By the results in the previous sections and associativity of the fusion product \eqref{eq:fusionp.assoc}, we know that
\begin{align*}
	\mathsf{W}_{0,z}(2m) \times^{u}_{f} \mathsf{S}_{1/2}(1) 
	&\simeq (\mathsf{W}_{0,z}(2) \times^{u}_{f} \mathsf{S}_{0}(2(m-1)))\times^{u}_{f} \mathsf{S}_{1/2}(1)\\
	& \simeq (\mathsf{W}_{0,z}(2) \times^{u}_{f} \mathsf{S}_{1/2}(1))\times^{u}_{f} \mathsf{S}_{0}(2(m-1)),
\end{align*}
so we focus on the case $n=2$ and the other cases will follow directly from it. Using the same reasoning as in the $k\neq 0 $ cases, we find that $ \mathsf{W}_{0,z}(2) \times^{u}_{f} \mathsf{S}_{1/2}(1) = \atl{3}x$, with
\begin{equation}
	 x \equiv 
	 \begin{tikzpicture}[scale = 1/3,baseline={(current bounding box.center)},yscale = -1]
	\draw[black, line width = 2pt] (.5,1) -- (3.5,1);
	\draw[black, line width = 2pt] (.5,3) -- (3.5,3);
	\draw[black, line width = 1pt] (1,1) .. controls (1,2) and (2,2) .. (2,1);
	\draw[black, line width = 1pt] (1,3) .. controls (1,2) and (2,2) .. (2,3);
	\draw[black, line width = 1pt] (3,1) -- (3,3);
	\node[anchor = north ] at (2,3) { $ \otimes_{\atl{2}} $};
	\draw[black, line width = 2pt] (0.5,5) -- (2.25,5); 
	\draw[black, line width = 1pt] (1,5) .. controls (1,6) and (2,6) .. (2,5);
	\draw[black, line width = 2pt] (2.75,5) -- (3.25,5);
	\draw[black, line width = 2pt] (2.75,7) -- (3.25,7);
	\draw[black, line width = 1pt] (3,5) -- (3,7);
\end{tikzpicture}.
\end{equation}
Now acting on $x$ with any diagram can only move around the arc at the bottom of $x$, so the fusion product is generated by elements of the form $u^{i}x$ for $i \in \mathbb{Z}$. However,
\begin{equation}
	Y e_{1} = \;
	\begin{tikzpicture}[scale = 1/3,baseline={(current bounding box.center)},yscale = -1]
	\draw[black, line width = 1pt] (1,1) .. controls (1,2) and (2,2) .. (2,1);
	\draw[black, line width = 1pt] (1,3) .. controls (1,2) and (2,2) .. (2,3);
	\draw[black, line width = 1pt] (3,1) -- (3,3);
	\draw[white, line width = 3pt] (.5,2) -- (3.5,2);
	\draw[black, line width = 1pt] (.5,2) -- (3.5,2);
	\draw[black, line width = 2pt] (.5,1) -- (3.5,1);
	\draw[black, line width = 2pt] (.5,3) -- (3.5,3);
\end{tikzpicture} = (-\q)^{1/2} \;
\begin{tikzpicture}[scale = 1/3,baseline={(current bounding box.center)},yscale = -1]
	\draw[black, line width = 1pt] (1,1) .. controls (1,2) and (2,2) .. (2,1);
	\draw[black, line width = 1pt] (1,3) .. controls (1,2) and (2,2) .. (2,3);
	\draw[black, line width = 1pt] (3,1) .. controls (3,2) and (1,2) .. (.5,2);
	\draw[black, line width = 1pt] (3,3) .. controls (3,2) and (4,2) .. (4,3);
	\filldraw[white] (3.5,1) -- (4.5,1) -- (4.5,3) -- (3.5,3) -- (3.5,1);
	\draw[black, line width = 2pt] (.5,1) -- (3.5,1);
	\draw[black, line width = 2pt] (.5,3) -- (3.5,3);
\end{tikzpicture} + (-\q)^{-1/2}
\begin{tikzpicture}[scale = 1/3,baseline={(current bounding box.center)},yscale = -1]
	\draw[black, line width = 1pt] (1,1) .. controls (1,2) and (2,2) .. (2,1);
	\draw[black, line width = 1pt] (1,3) .. controls (1,2) and (2,2) .. (2,3);
	\draw[black, line width = 1pt] (3,1) .. controls (3,2) and (4,2) .. (4,1);
	\draw[black, line width = 1pt] (3,3) .. controls (3,2) and (1,2) .. (.5,2);
	\filldraw[white] (3.5,1) -- (4.5,1) -- (4.5,3) -- (3.5,3) -- (3.5,1);
	\draw[black, line width = 2pt] (.5,1) -- (3.5,1);
	\draw[black, line width = 2pt] (.5,3) -- (3.5,3);
\end{tikzpicture} = ((-\q)^{-1/2}u^{3} + (-\q)^{1/2}u^{-3}) e_{1}.
\end{equation}
Since by construction $Y x = (z+z^{-1})x$, it follows that $(u^{3} - (-\q)^{-1/2}z)(u^{3} - (-\q)^{-1/2}z^{-1})x = 0$, and thus that the fusion product has dimension six. Furthermore, if $z^{2} \neq 1$ then $u^{3}$ must have the two eigenvalues $(-\q)^{-1/2}z^{\pm 1}$ so 
\begin{equation}
	 \mathsf{W}_{0,z}(2) \times^{u}_{f} \mathsf{S}_{1/2}(1) \simeq  \mathsf{W}_{1/2,(-\q)^{-1/2}z}(3) \oplus \mathsf{W}_{1/2,(-\q)^{-1/2}z^{-1}}(3).
\end{equation}
Note that if $z^{2} = 1 $ then the two standard modules $\mathsf{W}_{1/2,(-\q)^{-1/2}z^{\pm 1}}(3)$ are isomorphic, and one can show that the fusion product is then the self-extension of $\mathsf{W}_{1/2,(-\q)^{-1/2}z }(3)$.

\subsubsection{$\mathsf{W}_{0,z}(n+1) \div^{u/o}_{f} \mathsf{S}_{1/2}(1)$}
Here again the two fusion types are similar so we focus only on the $u$-type. Here we must restrict to $z$ generic right from the start; it is possible to present a unified proof for the $z$ generic or not cases, but doing so requires more sophisticated tools which we haven't introduced here. We thus start by noticing that
\begin{align*}
	\text{Hom}_{\atl{n}} \big( \mathsf{W}_{k,z'}(n), \mathsf{W}_{0,z}(n+1) \div^{u}_{f} \mathsf{S}_{1/2}(1)\big) & \simeq 
	\text{Hom}_{\atl{n+1}} \big( \mathsf{W}_{k,z'}(n) \times^{u}_{f} \mathsf{S}_{1/2}(1), \mathsf{W}_{0,z}(n+1)\big)\\
		& \simeq \delta_{k,1/2}(\delta_{z',(-\q)^{1/2} z^1}+\delta_{z',(-\q)^{1/2} z^{-1}}) \mathbb{C},
\end{align*}
where we used the duality between the fusion product and the fusion quotient together with the formulas for the fusion product of standard modules obtained in the previous section. Furthermore, for generic values of $z$ the standard modules $ \mathsf{W}_{1/2,(-\q)^{1/2}z^{\pm 1}}(n) $ are simple and non-isomorphic, so these morphisms must be injective. Finally, we have
\begin{equation}
	\underbrace{\text{Dim}(\mathsf{W}_{0,z}(n+1))}_{\binom{n+1}{(n+1)/2}} = 2 \underbrace{\text{Dim}(\mathsf{W}_{1/2,z'}(n))}_{\binom{n}{(n-1)/2}},
\end{equation}
and we thus conclude that
\begin{equation}
	\mathsf{W}_{0,z}(n+1)\div^{u}_{f}\mathsf{S}_{1/2}(1) \simeq \mathsf{W}_{1/2,z(-\q)^{1/2}}(n) \oplus \mathsf{W}_{1/2,z^{-1}(-\q)^{1/2}}(n).
\end{equation}

\section{Comparison with the other fusion types}\label{sec:previousfusions}
We discuss briefly how the defect operators -- the hoop elements $Y$ and $\bar{Y}$ -- introduced in this work behave with respect  to the various previously defined fusion products, in particular the one introduced by Gainutdinov and Saleur in~\cite{GS} and the one introduced by Bellet\^ete and Saint-Aubin~\cite{BSA}.
\subsection{The GS fusion}
\begin{figure}
\begin{center}
\subfloat[ Left garter to belt\label{fig:leftgarter}]{
\begin{tikzpicture}[scale = 1/4, baseline = {( current bounding box.center)}]
	\foreach \r in {1,2,3}{
		\draw[blue, line width = 1pt] (\r, -2) -- (\r,-1)  .. controls (\r,1) and (\r + 2, 3) .. (\r + 2,5) -- (\r + 2,6);
		\draw[red, line width = 1pt] (11 - \r,-2) -- (11 - \r,-1)  .. controls (11 - \r,1) and (9 - \r, 3) .. (9 - \r,5) -- (9-\r, 6);
	};
	\draw[white, line width = 3pt] (0,-.5) .. controls (1.25,-1.75) and (2.75, -1.75) .. (4,-.5);
	\draw[black, line width = 2pt] (0,-.5) .. controls (1.25,-1.75) and (2.75, -1.75) .. (4,-.5);
	\draw[black, line width = 3pt] (0,-2) -- (0,-1) .. controls (0,1) and (2,3) .. (2,5) -- (2,6);
	\draw[black, line width = 3pt] (11,-2) -- (11,-1) .. controls (11,1) and (9,3) .. (9,5) -- (9,6);
	\draw[black, line width = 3pt] (7,-2) -- (7,-1) .. controls (7,.5) and (5,1.5) .. (5,2);
	\filldraw[white, line width = 3pt] (5,2) circle  (.5);
	\draw[black, line width = 3pt] (4,-2) -- (4,-1) .. controls (4,.5) and (6,1.5) .. (6,2);
\end{tikzpicture} $\qquad \equiv \qquad  $
\begin{tikzpicture}[scale = 1/4, baseline = {( current bounding box.center)}]
	\foreach \r in {1,2,3}{
		\draw[blue, line width = 1pt] (\r, -2) -- (\r,-1)  .. controls (\r,1) and (\r + 2, 3) .. (\r + 2,5) -- (\r + 2,6);
		\draw[red, line width = 1pt] (11 - \r,-2) -- (11 - \r,-1)  .. controls (11 - \r,1) and (9 - \r, 3) .. (9 - \r,5) -- (9-\r, 6);
	};
	\draw[white, line width = 3pt] (2,5) .. controls (4.25,4) and (6.75, 4) .. (9,5);
	\draw[black, line width = 2pt] (2,5) .. controls (4.25,4) and (6.75, 4) .. (9,5);
	\draw[black, line width = 3pt] (0,-2) -- (0,-1) .. controls (0,1) and (2,3) .. (2,5) -- (2,6);
	\draw[black, line width = 3pt] (11,-2) -- (11,-1) .. controls (11,1) and (9,3) .. (9,5) -- (9,6);
	\draw[black, line width = 3pt] (7,-2) -- (7,-1) .. controls (7,.5) and (5,1.5) .. (5,2);
	\filldraw[white, line width = 3pt] (5,2) circle  (.5);
	\draw[black, line width = 3pt] (4,-2) -- (4,-1) .. controls (4,.5) and (6,1.5) .. (6,2);
\end{tikzpicture}} $\qquad $
\subfloat[ Right garter to belt\label{fig:rightgarter}]{
\begin{tikzpicture}[scale = 1/4, baseline = {( current bounding box.center)}]
	\draw[black, line width = 2pt] (7,-.5) .. controls (8.25,-1.75) and (9.75, -1.75) .. (11,-.5);
	\foreach \r in {1,2,3}{
		\draw[white, line width = 3pt] (\r, -2) -- (\r,-1)  .. controls (\r,1) and (\r + 2, 3) .. (\r + 2,5) -- (\r + 2,6);
		\draw[blue, line width = 1pt] (\r, -2) -- (\r,-1)  .. controls (\r,1) and (\r + 2, 3) .. (\r + 2,5) -- (\r + 2,6);
		\draw[white, line width = 3pt] (11 - \r,-2) -- (11 - \r,-1)  .. controls (11 - \r,1) and (9 - \r, 3) .. (9 - \r,5) -- (9-\r, 6);
		\draw[red, line width = 1pt] (11 - \r,-2) -- (11 - \r,-1)  .. controls (11 - \r,1) and (9 - \r, 3) .. (9 - \r,5) -- (9-\r, 6);
	};
	\draw[black, line width = 3pt] (0,-2) -- (0,-1) .. controls (0,1) and (2,3) .. (2,5) -- (2,6);
	\draw[black, line width = 3pt] (11,-2) -- (11,-1) .. controls (11,1) and (9,3) .. (9,5) -- (9,6);
	\draw[black, line width = 3pt] (7,-2) -- (7,-1) .. controls (7,.5) and (5,1.5) .. (5,2);
	\filldraw[white, line width = 3pt] (5,2) circle  (.5);
	\draw[black, line width = 3pt] (4,-2) -- (4,-1) .. controls (4,.5) and (6,1.5) .. (6,2);
\end{tikzpicture} $\qquad \equiv \qquad  $
\begin{tikzpicture}[scale = 1/4, baseline = {( current bounding box.center)}]
	\draw[black, line width = 2pt] (2,5) .. controls (4.25,4) and (6.75, 4) .. (9,5);
	\foreach \r in {1,2,3}{
		\draw[white, line width = 3pt] (\r, -2) -- (\r,-1)  .. controls (\r,1) and (\r + 2, 3) .. (\r + 2,5) -- (\r + 2,6);
		\draw[blue, line width = 1pt] (\r, -2) -- (\r,-1)  .. controls (\r,1) and (\r + 2, 3) .. (\r + 2,5) -- (\r + 2,6);
		\draw[white, line width = 3pt] (11 - \r,-2) -- (11 - \r,-1)  .. controls (11 - \r,1) and (9 - \r, 3) .. (9 - \r,5) -- (9-\r, 6);
		\draw[red, line width = 1pt] (11 - \r,-2) -- (11 - \r,-1)  .. controls (11 - \r,1) and (9 - \r, 3) .. (9 - \r,5) -- (9-\r, 6);
	};
	\draw[black, line width = 3pt] (0,-2) -- (0,-1) .. controls (0,1) and (2,3) .. (2,5) -- (2,6);
	\draw[black, line width = 3pt] (11,-2) -- (11,-1) .. controls (11,1) and (9,3) .. (9,5) -- (9,6);
	\draw[black, line width = 3pt] (7,-2) -- (7,-1) .. controls (7,.5) and (5,1.5) .. (5,2);
	\filldraw[white, line width = 3pt] (5,2) circle  (.5);
	\draw[black, line width = 3pt] (4,-2) -- (4,-1) .. controls (4,.5) and (6,1.5) .. (6,2);
\end{tikzpicture}}\\
\subfloat[Garters to uncomfortable belt\label{fig:mixedgarter}]{
\begin{tikzpicture}[scale = 1/4, baseline = {( current bounding box.center)}]
	\draw[black, line width = 2pt] (0,-.5) .. controls (1.25,-1.75) and (2.75, -1.75) .. (4,-.5);
	\foreach \r in {1,2,3}{
		\draw[white, line width = 3pt] (\r, -2) -- (\r,-1)  .. controls (\r,1) and (\r + 2, 3) .. (\r + 2,5) -- (\r + 2,6);
		\draw[blue, line width = 1pt] (\r, -2) -- (\r,-1)  .. controls (\r,1) and (\r + 2, 3) .. (\r + 2,5) -- (\r + 2,6);
		\draw[white, line width = 3pt] (11 - \r,-2) -- (11 - \r,-1)  .. controls (11 - \r,1) and (9 - \r, 3) .. (9 - \r,5) -- (9-\r, 6);
		\draw[red, line width = 1pt] (11 - \r,-2) -- (11 - \r,-1)  .. controls (11 - \r,1) and (9 - \r, 3) .. (9 - \r,5) -- (9-\r, 6);
	};
	\draw[black, line width = 3pt] (0,-2) -- (0,-1) .. controls (0,1) and (2,3) .. (2,5) -- (2,6);
	\draw[black, line width = 3pt] (11,-2) -- (11,-1) .. controls (11,1) and (9,3) .. (9,5) -- (9,6);
	\draw[black, line width = 3pt] (7,-2) -- (7,-1) .. controls (7,.5) and (5,1.5) .. (5,2);
	\filldraw[white, line width = 3pt] (5,2) circle  (.5);
	\draw[black, line width = 3pt] (4,-2) -- (4,-1) .. controls (4,.5) and (6,1.5) .. (6,2);
\end{tikzpicture} $\qquad \equiv \qquad  $
\begin{tikzpicture}[scale = 1/4, baseline = {( current bounding box.center)}]
	\foreach \r in {1,2,3}{
		\draw[white, line width = 3pt] (11 - \r,-2) -- (11 - \r,-1)  .. controls (11 - \r,1) and (9 - \r, 3) .. (9 - \r,5) -- (9-\r, 6);
		\draw[red, line width = 1pt] (11 - \r,-2) -- (11 - \r,-1)  .. controls (11 - \r,1) and (9 - \r, 3) .. (9 - \r,5) -- (9-\r, 6);
	};
	\draw[white, line width = 3pt] (2,5) .. controls (4.25,4) and (6.75, 4) .. (9,5);
	\draw[black, line width = 2pt] (2,5) .. controls (4.25,4) and (6.75, 4) .. (9,5);
	\foreach \r in {1,2,3}{
		\draw[white, line width = 3pt] (\r, -2) -- (\r,-1)  .. controls (\r,1) and (\r + 2, 3) .. (\r + 2,5) -- (\r + 2,6);
		\draw[blue, line width = 1pt] (\r, -2) -- (\r,-1)  .. controls (\r,1) and (\r + 2, 3) .. (\r + 2,5) -- (\r + 2,6);
	};
	\draw[black, line width = 3pt] (0,-2) -- (0,-1) .. controls (0,1) and (2,3) .. (2,5) -- (2,6);
	\draw[black, line width = 3pt] (11,-2) -- (11,-1) .. controls (11,1) and (9,3) .. (9,5) -- (9,6);
	\draw[black, line width = 3pt] (7,-2) -- (7,-1) .. controls (7,.5) and (5,1.5) .. (5,2);
	\filldraw[white, line width = 3pt] (5,2) circle  (.5);
	\draw[black, line width = 3pt] (4,-2) -- (4,-1) .. controls (4,.5) and (6,1.5) .. (6,2);
\end{tikzpicture} $\qquad \equiv \qquad $
\begin{tikzpicture}[scale = 1/4, baseline = {( current bounding box.center)}]
	\foreach \r in {1,2,3}{
		\draw[white, line width = 3pt] (\r, -2) -- (\r,-1)  .. controls (\r,1) and (\r + 2, 3) .. (\r + 2,5) -- (\r + 2,6);
		\draw[blue, line width = 1pt] (\r, -2) -- (\r,-1)  .. controls (\r,1) and (\r + 2, 3) .. (\r + 2,5) -- (\r + 2,6);
		\draw[white, line width = 3pt] (11 - \r,-2) -- (11 - \r,-1)  .. controls (11 - \r,1) and (9 - \r, 3) .. (9 - \r,5) -- (9-\r, 6);
		\draw[red, line width = 1pt] (11 - \r,-2) -- (11 - \r,-1)  .. controls (11 - \r,1) and (9 - \r, 3) .. (9 - \r,5) -- (9-\r, 6);
	};
	\draw[white, line width = 3pt] (7,-.5) .. controls (8.25,-1.75) and (9.75, -1.75) .. (11,-.5);
	\draw[black, line width = 2pt] (7,-.5) .. controls (8.25,-1.75) and (9.75, -1.75) .. (11,-.5);
	\draw[black, line width = 3pt] (0,-2) -- (0,-1) .. controls (0,1) and (2,3) .. (2,5) -- (2,6);
	\draw[black, line width = 3pt] (11,-2) -- (11,-1) .. controls (11,1) and (9,3) .. (9,5) -- (9,6);
	\draw[black, line width = 3pt] (7,-2) -- (7,-1) .. controls (7,.5) and (5,1.5) .. (5,2);
	\filldraw[white, line width = 3pt] (5,2) circle  (.5);
	\draw[black, line width = 3pt] (4,-2) -- (4,-1) .. controls (4,.5) and (6,1.5) .. (6,2);
\end{tikzpicture}
}
\end{center}
\caption{The behaviour of the hoop operators under the fusion product.}
 \label{fig:garter.belt.relations}
\end{figure}
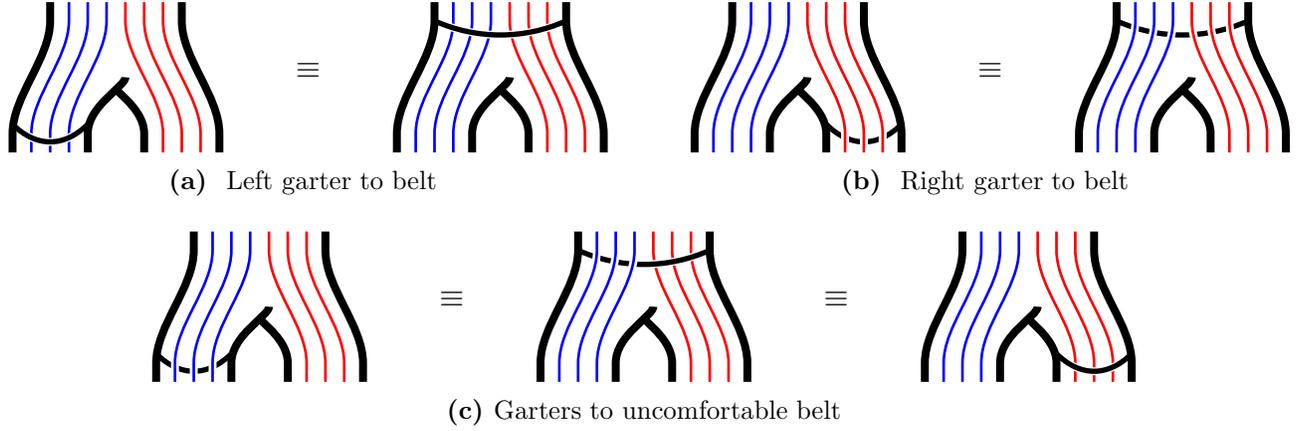
For this fusion, we endow $\atl{n+m} $ ($n,m $ positive integers) with the structure of a left $(\atl{n},\atl{m})$ module through the injection $\Phi \equiv \phi^{u}_{n,m} \otimes \psi^{o}_{m,n} : \atl{n} \otimes \atl{m} \to \atl{n+m}$. In terms of diagrams, this corresponds to gluing the two cylinders on which $\atl{n}$ and $\atl{m}$ lives into a pair of \emph{pants};  Fig.~\ref{fig:garter.belt.relations} illustrates how the hoop operators behave under this gluing.

Given an $\atl{n}$-module  $U$ and an $\atl{m}$-module $V$, their fusion is defined as
\begin{equation}
	U \times_{GS} V \equiv \atl{n+m} \otimes_{\atl{n} \otimes \atl{m}} U \otimes_{\mathbb{C}} V.
\end{equation}
Note that by construction (see Figs.~\ref{fig:leftgarter} and~\ref{fig:rightgarter})
\begin{equation}\label{eq:garterstobelt}
	Y^{(n+m)} = \phi^{u}_{n,m}(Y^{(n)}), \qquad \bar{Y}^{(n+m)} = \psi^{o}_{m,n}(\bar{Y}^{(m)}),
\end{equation}
so that  eigenvalues of the central elements $Y$, $\bar{Y}$ on $U \times_{GS} V$
are fully determined from their value on $U$ and $V$, respectively. It follows in particular that this fusion product is not commutative in general. Furthermore, the construction also gives (see  Fig.~\ref{fig:mixedgarter})
\begin{equation}\label{eq:mixedgarters}
	\phi^{u}_{n,m}(\bar{Y}^{(n)}) = \psi^{o}_{m,n}(Y^{(m)}),
\end{equation}
which imposes severe constraints on the combinations of modules leading to non-trivial fusion products. 

To illustrate just how restrictive equations \eqref{eq:garterstobelt} and \eqref{eq:mixedgarters} are, we compute the fusion of two standard modules  $U \simeq \mathsf{W}^{u}_{r,\delta}$, and $V \simeq \mathsf{W}^{o}_{s,\mu} $. First, equation \eqref{eq:mixedgarters} imposes  $\mu = \delta^{\pm 1}$. Second, if the fusion product is non-zero then it must have at least one non-trivial simple quotient\footnote{This follows because the fusion of cyclic modules is also cyclic.} and, for generic values of $\q$ and $\delta$, all of the simple modules of $\atl{n}$ are isomorphic to standard modules $\mathsf{W}^{o}_{k,\nu}$ for some integer $k$ and $\nu \in \mathbb{C}^{*}$. 
 Combining this observation with equation \eqref{eq:garterstobelt} gives the conditions
\begin{equation}
		\nu = (\delta (-\q)^{2r})^{\pm 1}, \qquad \text{ and } \qquad \nu = (\mu (-\q)^{-2s})^{ \pm 1}(-\q)^{2k},
\end{equation}
where all the $\pm $ are independent. In the generic cases where $\delta^{2}$ is not an integer power of $(-\q)$, and $ \q $ is not a root of unity, this leaves the possibilities:
\begin{equation}
	k = (r + \epsilon s), \qquad \delta = \mu^{\epsilon},
\end{equation}
or, using a more usual notation
\begin{equation}
		 \mathsf{W}_{r,z}(n) \times_{GS} \mathsf{W}_{s,w}(m) \simeq 
		 \begin{cases}
		 	a_{+}   \mathsf{W}_{r+s,z (-\q)^{-s}}(n) & \text{ if } z = w (-\q)^{r + s}\\
		 	a_{-}   \mathsf{W}_{r-s,z (-\q)^{s}}(n) & \text{ if } z = w^{-1} (-\q)^{r - s}\\
		 	0 & \text{otherwise}
		 \end{cases},
	\end{equation} 
where $a_{\pm}$ are (unknown) non-negative integers (because each module could appear multiple times) and it is understood that if $ s>r $, $\mathsf{W}_{r-s,z (-\q)^{s}}(n) \equiv \mathsf{W}_{s-r, z^{-1} (-\q)^{-s}}(n) $. It therefore simply remains to find the value of $a_{\pm}$; more extensive calculations \cite{GS} yields $a_{\pm} = 1$.

\subsection{The BSA fusion }
 
 For this fusion, we endow $\atl{n+m} $ ($n,m $ positive integers) with the structure of a left $(\tl{n},\tl{m})$ module through the injection $\Phi \equiv \phi^{u}_{n,m} \otimes \psi^{o}_{m,n} : \tl{n} \otimes \tl{m} \to \atl{n+m}$, where we identified the regular algebras with their images inside the affine algebra. In terms of diagrams, this corresponds to taking the two strips on which the regular algebras live and stitching them into a cylinder, introducing no particular relations for the hoop operators. Given $U, V$ a $\tl{n}$ and a $\tl{m}$ module, respectively, their fusion is defined as
\begin{equation}
	U \times_{BSA} V \equiv \atl{n+m} \otimes_{\tl{n} \otimes \tl{m}} U \otimes_{\mathbb{C}} V.
\end{equation}
 Note that because no relations were introduced for the hoop operators, the fusion of any non-zero module is always infinite dimensional. For the particular case of standard modules (and $\q$ generic), one finds \cite{BSA}
 \begin{equation}
 	\mathsf{S}_{r} \times_{BSA} \mathsf{S}_{s} \simeq \bigoplus_{k = |r-s|}^{r+s} \mathsf{P}_{k},
 \end{equation}
 where the $\mathsf{P}_{k}$ are the projective indecomposable modules of the affine Temperley-Lieb algebra. The algebraic structure of these modules is quite complicated and described in some details in \cite{BSA}. We simply mention that there is a family of inclusions $ \mathsf{P}_{k} \subset \mathsf{P}_{k+2} \subset \mathsf{P}_{k+4} \subset \hdots $ that is such that $ \mathsf{P}_{k}/\mathsf{P}_{k-2}$ is an indecomposable submodule of the direct product of the standard modules $\mathsf{W}_{k,z}$ for all non-zero $z$. In other (more heuristic) words, $ \mathsf{P}_{k}$ is an indecomposable \emph{collage} of all $\mathsf{W}_{r,z}$ with $r \leq k $, $z \in \mathbb{C}^{*} $.

\section{Alternative proof of the relations~\eqref{Jon1}-\eqref{eq:fusionproduct.standard2}}
\label{app:C}
We first establish the branching rules from $\atl{n_1+n_2}$ to $\atl{n_1} \otimes \tl{n_2}$,
by an adaptation of the working of \cite{GJS} that dealt with the case of $\atl{n_1} \otimes \atl{n_2}$.
The algebras $\atl{n_1}$ and $\tl{n_2}$ can be embedded into $\atl{n}$ on $n = n_1 + n_2$ sites,
by defining the periodic generator $e_0^{(1)}$ and the shift operator $u^{(1)}$ for the first subalgebra
by braid translation in $\atl{n}$ (see \cite{GJS}), while all other generators simply carry over.

Using the techniques of \cite{GJS}, and in particular the operator $\tau_j^{(1)}$ defined there,
we then find the branching rules (actually, one has just to restrict the second tensor factor in branching rules of~\cite{GJS} to the subalgebra $\tl{n_2}\subset\atl{n_2}$)
\begin{equation}
 \mathsf{W}_{j,z}(n)|_{\atl{n_1} \otimes \tl{n_2}} \cong \bigoplus_{j_1,j_2} \mathsf{W}_{j_1,z_1}(n_1) \otimes \left( \bigoplus_{k \ge j_2} \mathsf{S}_{k}(n_2) \right) \,,
\label{general_ATL_decomposition}
\end{equation}
with the following values of the momenta:
\begin{itemize}
 \item For $j = j_1 + j_2$ and any values of $j_1,j_2$: $z_1 = (i \sqrt{\q})^{-2 j_2} z^{+1}$.
 \item For $j = j_1 - j_2$ and either $j=0$ or $j_2 > 0$: $z_1 = (i \sqrt{\q})^{+2 j_2} z^{+1}$.
 \item For $j = j_2 - j_1$ and either $j=0$ or $j_1 > 0$: $z_1 = (i \sqrt{\q})^{+2 j_2} z^{-1}$.
\end{itemize}
Notice that this closely parallels the main result of \cite{GJS}, the only difference being that the
right tensorands $\mathsf{W}_{j_2,z_2}$ there have been repaced by $\bigoplus_{k \ge j_2} \mathsf{S}_{k}$, which is exactly the restriction of $\mathsf{W}_{j_2,z_2}$ to the subalgebra $\tl{n_2}\subset\atl{n_2}$.
In particular, these modules have  the same dimensions.

We can now read off the corresponding fusion rules by Frobenius reciprocity:
\begin{equation*}
	\text{Hom}_{\atl{n + m}}\big( \mathsf{W}_{k,w}(n) \times_{f} \mathsf{S}_{j}(m), \mathsf{W}_{j,z}(n+m) \big) \simeq \text{Hom}_{\atl{n} \otimes \tl{m}}\big( \mathsf{W}_{k,w}(n) \otimes \mathsf{S}_{j}(m), \mathsf{W}_{j,z}(n+m) \big), 
\end{equation*} 
where on the right side of the equation $\mathsf{W}_{j,z}(n+m)$ is seen as a $\atl{n} \otimes \tl{m}$-module, i.e. is identified with the branching rules \eqref{general_ATL_decomposition}. If $\q$ and $z$ are generic, the standard modules $\mathsf{W}_{j,z}(n+m)$ are simple, so the dimension of this homomorphism group is the number of copies of the standard module appearing as direct summands of the fusion product.%
\footnote{Note that this implicitly assumes that the fusion product of standard modules is semisimple, this can be shown by using the fact that the fusion product preserves the spectrum of at least one of the hoop operators. It follows that if a module factors through the blob algebra, then so will its fusion with any other module; for generic values of $\q, w $, the blob algebra through which the standard module $\mathsf{W}_{k,w}(n)$ factors through is semisimple, therefore so is its fusion.}
There is however a slight subtlety in 
computing these dimensions, which is conveniently illustrated in the example $(n_1,n_2,n) = (2,4,6)$.
The branching rules read in this case
\begin{eqnarray}
 \mathsf{W}_{0,z}(6) &=& \mathsf{W}_{0,z}(2) \otimes \big(\mathsf{S}_{2}(4) \oplus \mathsf{S}_{1}(4) \oplus \mathsf{S}_{0}(4)\big) \oplus \nonumber \\
   & & \mathsf{W}_{1,(i \sqrt{\q})^2 z}(2) \otimes \big(\mathsf{S}_{2}(4) \oplus \mathsf{S}_{1}(4)\big) \oplus \mathsf{W}_{1,\frac{(i \sqrt{\q})^2}{z}}(2) \otimes \big(\mathsf{S}_{2}(4) \oplus \mathsf{S}_{1}(4)\big) \,, \nonumber \\ 
 \mathsf{W}_{1,z}(6) &=& \mathsf{W}_{0,\frac{z}{(i \sqrt{\q})^2}}(2) \otimes \big(\mathsf{S}_{2}(4) \oplus \mathsf{S}_{1}(4)\big) \oplus \nonumber \\
   & & \mathsf{W}_{1,z}(2) \otimes \big(\mathsf{S}_{2}(4) \oplus \mathsf{S}_{1}(4) \oplus \mathsf{S}_{0}(4)\big) \oplus \mathsf{W}_{1,\frac{(i \sqrt{\q})^4}{z}}(2) \otimes \mathsf{S}_{2}(4) \,, \nonumber \\ 
 \mathsf{W}_{2,z}(6) &=& \mathsf{W}_{0,\frac{z}{(i \sqrt{\q})^4}}(2) \otimes \mathsf{S}_{2}(4) \oplus \mathsf{W}_{1,\frac{z}{(i \sqrt{\q})^2}}(2) \otimes \big(\mathsf{S}_{2}(4) \oplus \mathsf{S}_{1}(4)\big) \nonumber \\ 
 \mathsf{W}_{3,z}(6) &=& \mathsf{W}_{1,\frac{z}{(i \sqrt{\q})^4}}(2) \otimes \mathsf{S}_{2}(4) \,. 
 \label{branching-rules}
\end{eqnarray}
At first sight it appears that one would have fusion rules like
\begin{eqnarray}
 \mathsf{W}_{0,z} \fus \mathsf{S}_{1} &=& \mathsf{W}_{0,z} \oplus \mathsf{W}_{1,(-\q)z} \,,\nonumber \\
 \mathsf{W}_{0,z} \fus \mathsf{S}_{2} &=& \mathsf{W}_{0,z} \oplus \mathsf{W}_{1,(-\q)z} \oplus \mathsf{W}_{2,(-\q)^2 z} \,.
 \label{fusion-rules}
\end{eqnarray}
This is however not quite correct. Indeed, we should be careful when the left tensorand in the fusion
product is $\mathsf{W}_{0,z}$, since we have to take into account the isomorphism
$\mathsf{W}_{0,z} \simeq \mathsf{W}_{0,z^{-1}}$. Therefore the corresponding terms in the
branching rules (\ref{branching-rules}) can also be written
\begin{eqnarray}
 \mathsf{W}_{1,z}(6) &=& \mathsf{W}_{0,\frac{(i \sqrt{\q})^2}{z}}(2) \otimes
  \big(\mathsf{S}_{2}(4) \oplus \mathsf{S}_{1}(4)\big) \oplus \ldots \,, \nonumber \\
 \mathsf{W}_{2,z}(6) &=& \mathsf{W}_{0,\frac{(i \sqrt{\q})^4}{z}}(2) \otimes
 \mathsf{S}_{2}(4) \oplus \ldots \,.
\end{eqnarray}
This implies that we have a few extra terms, and (\ref{fusion-rules}) should be corrected into
\begin{eqnarray}
 \mathsf{W}_{0,z} \fus \mathsf{S}_{1} &=& \mathsf{W}_{0,z} \oplus \mathsf{W}_{1,(-\q)z} \oplus \mathsf{W}_{1,(-\q) z^{-1}} \,, \nonumber \\
 \mathsf{W}_{0,z} \fus \mathsf{S}_{2} &=& \mathsf{W}_{0,z} \oplus \mathsf{W}_{1,(-\q)z} \oplus \mathsf{W}_{2,(-\q)^2 z} \oplus \mathsf{W}_{1,(-\q) z^{-1}} \oplus \mathsf{W}_{2,(-\q)^2 z^{-1}} \,. \nonumber
\end{eqnarray}

Taking into account this subtlety, the general result comes out as
\begin{equation}
 \mathsf{W}_{j_1,z} \fus \mathsf{S}_{j_2} =
 \bigoplus_{j={\rm max}(j_1-j_2,j_{12}^\star)}^{j_1+j_2} \hspace{-0.7cm} \mathsf{W}_{j,(-\q)^{j-j_1} z} \oplus
 \bigoplus_{j=j_{21}^\star}^{j_2-j_1} \mathsf{W}_{j,(-\q)^{j+j_1} z^{-1}} \,,
 \label{genresATL-TL}
\end{equation}
where we have defined $j_{12}^\star = (j_1 - j_2) \mbox{ mod } 1$, and $j_{21}^\star = (j_2 - j_1) \mbox{ mod } 1$.
After some amount of rewriting, this can be shown to lead to~\eqref{Jon1}-\eqref{eq:fusionproduct.standard2} in the main text, as claimed.

\section{Finding the basis in \eqref{eq:fusion.base1} to \eqref{eq:fusion.base3} }\label{app:fusion.basis}

We explain here how we arrived at the basis of $\mathsf{W}_{1/2,z}(3) \times^{o}_{f} \mathsf{S}_{1/2}(1) = \atl{4}x$ shown in section \ref{sec:the_fusion_product}. The general procedure can be used to find bases of the fusion product of arbitrary standard modules, but the calculations quickly become very difficult to track when the size of the modules increases.

 Recall that since $\mathsf{W}_{1/2,z}(3)$ and $\mathsf{S}_{1/2}(1)$ are both cyclic, so is their fusion, and we can thus write $\mathsf{W}_{1/2,z}(3) \times^{o}_{f} \mathsf{S}_{1/2}(1) = \atl{4}y$ with 
\begin{equation}
	y =\; 
	\begin{tikzpicture}[scale = 1/3,baseline={(current bounding box.center)},yscale = -1] 
	\draw[black, line width = 1pt] (3,3) --  (3,1);
	\draw[black, line width = 1pt] (1,3) -- (1,1);
	\draw[black, line width = 1pt] (2,3) -- (2,1);
	\draw[black, line width = 1pt] (4,1) -- (4,3);
	\draw[black, line width = 2pt] (.5,1) -- (4.5,1);
	\draw[black, line width = 2pt] (.5,3) -- (4.5,3);
	\draw[black, line width = 1pt] (.5,4) -- (.5,6);
	\draw[black, line width = 1pt] (1.5,4) .. controls (1.5,5) and (2.5,5) .. (2.5,4);
	\node[anchor = north] at (3.5,4) {$\otimes$};
	\draw[black, line width = 1pt] (4.5,4) -- (4.5,6);
	\draw[black, line width = 2pt] (0,4) -- (3,4);
	\draw[black, line width = 2pt] (4,4) -- (5,4);
	\end{tikzpicture} \;.
\end{equation}
To start, observe that 
\begin{equation}
	\begin{tikzpicture}[scale = 1/3,baseline={(current bounding box.center)},yscale = -1] 
	\draw[black, line width = 2pt] (.5,3) -- (3.5,3);
	\draw[black, line width = 1pt] (1,3) -- (1,5);
	\draw[black, line width = 1pt] (2,3) .. controls (2,4) and (3,4) .. (3,3);
	\filldraw[white] (-.5,.5) -- (.5,.5) -- (.5,5.5) -- (-.5,5.5) -- (-.5,.5);
	\filldraw[white] (4-.5,.5) -- (4.5,.5) -- (4.5,5.5) -- (4-.5,5.5) -- (4-.5,.5);
	\end{tikzpicture} = 
	\begin{tikzpicture}[scale = 1/3,baseline={(current bounding box.center)},yscale = -1] 
	\draw[black, line width = 1pt] (3,3) .. controls (3,2) and (1,2) .. (1,1);
	\draw[black, line width = 1pt] (2,3) .. controls (2,2) and (1,2) .. (1,3);
	\draw[black, line width = 1pt] (3,1) .. controls (3,2) and (2,2) .. (2,1);
	\draw[black, line width = 2pt] (.5,1) -- (3.5,1);
	\draw[black, line width = 2pt] (.5,3) -- (3.5,3);
	\draw[black, line width = 1pt] (1,3) -- (1,5);
	\draw[black, line width = 1pt] (2,3) .. controls (2,4) and (3,4) .. (3,3);
	\filldraw[white] (-.5,.5) -- (.5,.5) -- (.5,5.5) -- (-.5,5.5) -- (-.5,.5);
	\filldraw[white] (4-.5,.5) -- (4.5,.5) -- (4.5,5.5) -- (4-.5,5.5) -- (4-.5,.5);
	\end{tikzpicture}.
\end{equation}
It thus follows that $e_2e_1 y = y$, and thus, when looking for a basis of $\mathsf{W}_{1/2,z}(3) \times^{o}_{f} \mathsf{S}_{1/2}(1)$, we can restrict our search to elements of the form $a y$ where $a \in \atl{4}$ is a sum of diagrams with at most $2$ through lines (lines connecting the top and bottom of the diagram), and such that $a e_1 = \beta a$ (since $a y = a e_2 e_1 y$).

Furthermore, one also has
 \begin{equation}
	\begin{tikzpicture}[scale = 1/3,baseline={(current bounding box.center)},yscale = -1] 
	\draw[black, line width = 2pt] (.5,3) -- (3.5,3);
	\draw[black, line width = 1pt] (1,3) -- (1,5);
	\draw[black, line width = 1pt] (2,3) .. controls (2,4) and (3,4) .. (3,3);
	\filldraw[white] (-.5,.5) -- (.5,.5) -- (.5,5.5) -- (-.5,5.5) -- (-.5,.5);
	\filldraw[white] (4-.5,.5) -- (4.5,.5) -- (4.5,5.5) -- (4-.5,5.5) -- (4-.5,.5);
	\end{tikzpicture} = \frac{1}{z}
	\begin{tikzpicture}[scale = 1/3,baseline={(current bounding box.center)},yscale = -1] 
	\draw[black, line width = 1pt] (3,3) .. controls (3,2) and (4,2) .. (4,3);
	\draw[black, line width = 1pt] (0,1) .. controls (0,2) and (1,2) .. (1,1);
	\draw[black, line width = 1pt] (2,3) .. controls (2,2) and (1,2) .. (1,3);
	\draw[black, line width = 1pt] (3,1) .. controls (3,2) and (2,2) .. (2,1);
	\draw[black, line width = 2pt] (.5,1) -- (3.5,1);
	\draw[black, line width = 2pt] (.5,3) -- (3.5,3);
	\draw[black, line width = 1pt] (1,3) -- (1,5);
	\draw[black, line width = 1pt] (2,3) .. controls (2,4) and (3,4) .. (3,3);
	\filldraw[white] (-.5,.5) -- (.5,.5) -- (.5,5.5) -- (-.5,5.5) -- (-.5,.5);
	\filldraw[white] (4-.5,.5) -- (4.5,.5) -- (4.5,5.5) -- (4-.5,5.5) -- (4-.5,.5);
	\end{tikzpicture} = z
	\begin{tikzpicture}[scale = 1/3,baseline={(current bounding box.center)},yscale = -1] 
	\draw[black, line width = 1pt] (3,3) .. controls (3,1.5) and (0,1.5) .. (0,3);
	\draw[black, line width = 1pt] (2,0) .. controls (2,1) and (3,1) .. (3,0);
	\draw[black, line width = 1pt] (2,3) .. controls (2,2) and (1,2) .. (1,3);
	\draw[black, line width = 1pt] (1,0) .. controls (1,1.5) and (4,1.5) .. (4,0);
	\draw[black, line width = 2pt] (.5,0) -- (3.5,0);
	\draw[black, line width = 2pt] (.5,3) -- (3.5,3);
	\draw[black, line width = 1pt] (1,3) -- (1,5);
	\draw[black, line width = 1pt] (2,3) .. controls (2,4) and (3,4) .. (3,3);
	\filldraw[white] (-.5,-.5) -- (.5,-.5) -- (.5,5.5) -- (-.5,5.5) -- (-.5,.5);
	\filldraw[white] (4-.5,-.5) -- (4.5,-.5) -- (4.5,5.5) -- (4-.5,5.5) -- (4-.5,.5);
	\end{tikzpicture},
\end{equation}
and since
\begin{align}
	\phi^{o}_{3,1}\Large(
	\begin{tikzpicture}[scale = 1/3,baseline={(current bounding box.center)},yscale = -1] 
	\draw[black, line width = 1pt] (3,3) .. controls (3,2) and (4,2) .. (4,3);
	\draw[black, line width = 1pt] (0,1) .. controls (0,2) and (1,2) .. (1,1);
	\draw[black, line width = 1pt] (2,3) .. controls (2,2) and (1,2) .. (1,3);
	\draw[black, line width = 1pt] (3,1) .. controls (3,2) and (2,2) .. (2,1);
	\draw[black, line width = 2pt] (.5,1) -- (3.5,1);
	\draw[black, line width = 2pt] (.5,3) -- (3.5,3);
	\filldraw[white] (-.5,.5) -- (.5,.5) -- (.5,3.5) -- (-.5,4.5) -- (-.5,.5);
	\filldraw[white] (4-.5,.5) -- (4.5,.5) -- (4.5,3.5) -- (4-.5,3.5) -- (4-.5,.5);
	\end{tikzpicture} \Large) & =
	\begin{tikzpicture}[scale = 1/3,baseline={(current bounding box.center)},yscale = -1] 
	\draw[black, line width = 1pt] (3,3) .. controls (3,2) and (5,2) .. (5,3);
	\draw[white, line width = 4pt] (4,1) -- (4,3);
	\draw[black, line width = 1pt] (4,1) -- (4,3);
	\draw[black, line width = 1pt] (0,1) .. controls (0,2) and (1,2) .. (1,1);
	\draw[black, line width = 1pt] (2,3) .. controls (2,2) and (1,2) .. (1,3);
	\draw[black, line width = 1pt] (3,1) .. controls (3,2) and (2,2) .. (2,1);
	\draw[black, line width = 2pt] (.5,1) -- (4.5,1);
	\draw[black, line width = 2pt] (.5,3) -- (4.5,3);
	\filldraw[white] (-.5,.5) -- (.5,.5) -- (.5,3.5) -- (-.5,4.5) -- (-.5,.5);
	\filldraw[white] (6-.5,.5) -- (4.5,.5) -- (4.5,3.5) -- (6-.5,3.5) -- (6-.5,.5);
	\end{tikzpicture} = (-\q)^{-1/2}
	\begin{tikzpicture}[scale = 1/3,baseline={(current bounding box.center)},yscale = -1] 
	\draw[black, line width = 1pt] (0,1) .. controls (0,2) and (1,2) .. (1,1);
	\draw[black, line width = 1pt] (2,1) .. controls (2,2) and (3,2) .. (3,1);
	\draw[black, line width = 1pt] (4,1) .. controls (4,2) and (3,2) .. (3,3);
	\draw[black, line width = 1pt] (1,3) .. controls (1,2) and (2,2) .. (2,3);
	\draw[black, line width = 1pt] (4,3) .. controls (4,2) and (5,2) .. (5,3);
	\draw[black, line width = 2pt] (.5,1) -- (4.5,1);
	\draw[black, line width = 2pt] (.5,3) -- (4.5,3);
	\filldraw[white] (-.5,.5) -- (.5,.5) -- (.5,3.5) -- (-.5,4.5) -- (-.5,.5);
	\filldraw[white] (6-.5,.5) -- (4.5,.5) -- (4.5,3.5) -- (6-.5,3.5) -- (6-.5,.5);
	\end{tikzpicture} + (-\q)^{1/2}
	\begin{tikzpicture}[scale = 1/3,baseline={(current bounding box.center)},yscale = -1] 
	\draw[black, line width = 1pt] (0,1) .. controls (0,2) and (1,2) .. (1,1);
	\draw[black, line width = 1pt] (2,1) .. controls (2,2) and (3,2) .. (3,1);
	\draw[black, line width = 1pt] (4,1) .. controls (4,2) and (5,2) .. (5,1);
	\draw[black, line width = 1pt] (1,3) .. controls (1,2) and (2,2) .. (2,3);
	\draw[black, line width = 1pt] (3,3) .. controls (3,2) and (4,2) .. (4,3);
	\draw[black, line width = 2pt] (.5,1) -- (4.5,1);
	\draw[black, line width = 2pt] (.5,3) -- (4.5,3);
	\filldraw[white] (-.5,.5) -- (.5,.5) -- (.5,3.5) -- (-.5,4.5) -- (-.5,.5);
	\filldraw[white] (6-.5,.5) -- (4.5,.5) -- (4.5,3.5) -- (6-.5,3.5) -- (6-.5,.5);
	\end{tikzpicture},
\end{align}
it follows that
\begin{equation}\label{eq:appendix.fusion.twist}
	z y = \left((-\q)^{-1/2}u e_1 + (-\q)^{1/2} u e_1e_3 \right) y.
\end{equation}

Let's consider then $M_0$, the submodule of $\mathsf{W}_{1/2,z}(3) \times^{o}_{f} \mathsf{S}_{1/2}(1)$ generated by elements of the form $a y$, where $a$ is a diagram with no through lines, no closed loop wrapping around the cylinder\footnote{Note that for any diagram $a$ with no through lines, $Y a =  a Y = \bar{Y}a = a \bar{Y}$, therefore the weight of non-contractible loops is always fixed by the parameter $z$ in $\mathsf{W}_{1/2,z}(3)$.}, and such that $ae_1 = \beta a$. We find $12$ such diagrams:
\begin{equation}\label{eq:appendix.diag.fusion.1}
	\overbrace{
	\begin{tikzpicture}[scale = 1/3,baseline={(current bounding box.center)}] 
	\draw[black, line width = 1pt] (1,1) .. controls (1,2) and (2,2) .. (2,1);
	\draw[black, line width = 1pt] (3,1) .. controls (3,2) and (4,2) .. (4,1);
	\draw[black, line width = 1pt] (1,3) .. controls (1,2) and (2,2) .. (2,3);
	\draw[black, line width = 1pt] (3,3) .. controls (3,2) and (4,2) .. (4,3);
	\draw[black, line width = 2pt] (.5,1) -- (4.5,1);
	\draw[black, line width = 2pt] (.5,3) -- (4.5,3);
	\filldraw[white] (-.5,.5) -- (.5,.5) -- (.5,3.5) -- (-.5,3.5) -- (-.5,.5);
	\filldraw[white] (5-.5,.5) -- (5.5,.5) -- (5.5,3.5) -- (5-.5,3.5) -- (5-.5,.5);
	\end{tikzpicture}}^{a_5}, 
	\; \overbrace{
	\begin{tikzpicture}[scale = 1/3,baseline={(current bounding box.center)}] 
	\draw[black, line width = 1pt] (1,1) .. controls (1,2) and (2,2) .. (2,1);
	\draw[black, line width = 1pt] (3,1) .. controls (3,2) and (4,2) .. (4,1);
	\draw[black, line width = 1pt] (1,3) .. controls (1,2) and (0,2) .. (0,3);
	\draw[black, line width = 1pt] (5,3) .. controls (5,2) and (4,2) .. (4,3);
	\draw[black, line width = 1pt] (3,3) .. controls (3,2) and (2,2) .. (2,3);
	\draw[black, line width = 2pt] (.5,1) -- (4.5,1);
	\draw[black, line width = 2pt] (.5,3) -- (4.5,3);
	\filldraw[white] (-.5,.5) -- (.5,.5) -- (.5,3.5) -- (-.5,3.5) -- (-.5,.5);
	\filldraw[white] (5-.5,.5) -- (5.5,.5) -- (5.5,3.5) -- (5-.5,3.5) -- (5-.5,.5);
	\end{tikzpicture}}^{a_6},
	\; \overbrace{
	\begin{tikzpicture}[scale = 1/3,baseline={(current bounding box.center)}] 
	\draw[black, line width = 1pt] (1,1) .. controls (1,2) and (2,2) .. (2,1);
	\draw[black, line width = 1pt] (3,1) .. controls (3,2) and (4,2) .. (4,1);
	\draw[black, line width = 1pt] (1,3) .. controls (1,1.5) and (4,1.5) .. (4,3);
	\draw[black, line width = 1pt] (2,3) .. controls (2,2) and (3,2) .. (3,3);
	\draw[black, line width = 2pt] (.5,1) -- (4.5,1);
	\draw[black, line width = 2pt] (.5,3) -- (4.5,3);
	\filldraw[white] (-.5,.5) -- (.5,.5) -- (.5,3.5) -- (-.5,3.5) -- (-.5,.5);
	\filldraw[white] (5-.5,.5) -- (5.5,.5) -- (5.5,3.5) -- (5-.5,3.5) -- (5-.5,.5);
	\end{tikzpicture}}^{a_7},
	\; \overbrace{
	\begin{tikzpicture}[scale = 1/3,baseline={(current bounding box.center)}] 
	\draw[black, line width = 1pt] (1,1) .. controls (1,2) and (2,2) .. (2,1);
	\draw[black, line width = 1pt] (3,1) .. controls (3,2) and (4,2) .. (4,1);
	\draw[black, line width = 1pt] (1,3) .. controls (1,2) and (0,2) .. (0,3);
	\draw[black, line width = 1pt] (2,3) .. controls (2,1.5) and (5,1.5) .. (5,3);
	\draw[black, line width = 1pt] (3,3) .. controls (3,2) and (4,2) .. (4,3);
	\draw[black, line width = 2pt] (.5,1) -- (4.5,1);
	\draw[black, line width = 2pt] (.5,3) -- (4.5,3);
	\filldraw[white] (-.5,.5) -- (.5,.5) -- (.5,3.5) -- (-.5,3.5) -- (-.5,.5);
	\filldraw[white] (5-.5,.5) -- (5.5,.5) -- (5.5,3.5) -- (5-.5,3.5) -- (5-.5,.5);
	\end{tikzpicture}}^{a_8},
	\; \overbrace{
	\begin{tikzpicture}[scale = 1/3,baseline={(current bounding box.center)}] 
	\clip (-.5,.5) -- (5.5, .5) -- (5.5,3.5) -- (-.5,3.5) -- (-.5,.5);
	\draw[black, line width = 1pt] (1,1) .. controls (1,2) and (2,2) .. (2,1);
	\draw[black, line width = 1pt] (3,1) .. controls (3,2) and (4,2) .. (4,1);
	\draw[black, line width = 1pt] (1,3) .. controls (1,2) and (0,2) .. (0,3);
	\draw[black, line width = 1pt] (2,3) .. controls (2,1.5) and (-1,1.5) .. (-1,3);
	\draw[black, line width = 1pt] (3,3) .. controls (3,1.5) and (6,1.5) .. (6,3);
	\draw[black, line width = 1pt] (4,3) .. controls (4,2) and (5,2) .. (5,3);
	\draw[black, line width = 2pt] (.5,1) -- (4.5,1);
	\draw[black, line width = 2pt] (.5,3) -- (4.5,3);
	\filldraw[white] (-1-.5,.5) -- (.5,.5) -- (.5,3.5) -- (-1-.5,3.5) -- (-1-.5,.5);
	\filldraw[white] (5-.5,.5) -- (7.5,.5) -- (7.5,3.5) -- (5-.5,3.5) -- (5-.5,.5);
	\end{tikzpicture}}^{a_{9}},
	\; \overbrace{
	\begin{tikzpicture}[scale = 1/3,baseline={(current bounding box.center)},xscale = -1] 
	\clip (-.5,.5) -- (5.5, .5) -- (5.5,3.5) -- (-.5,3.5) -- (-.5,.5);
	\draw[black, line width = 1pt] (1,1) .. controls (1,2) and (2,2) .. (2,1);
	\draw[black, line width = 1pt] (3,1) .. controls (3,2) and (4,2) .. (4,1);
	\draw[black, line width = 1pt] (1,3) .. controls (1,2) and (0,2) .. (0,3);
	\draw[black, line width = 1pt] (2,3) .. controls (2,1.5) and (5,1.5) .. (5,3);
	\draw[black, line width = 1pt] (3,3) .. controls (3,2) and (4,2) .. (4,3);
	\draw[black, line width = 2pt] (.5,1) -- (4.5,1);
	\draw[black, line width = 2pt] (.5,3) -- (4.5,3);
	\filldraw[white] (-.5,.5) -- (.5,.5) -- (.5,3.5) -- (-.5,3.5) -- (-.5,.5);
	\filldraw[white] (5-.5,.5) -- (5.5,.5) -- (5.5,3.5) -- (5-.5,3.5) -- (5-.5,.5);
	\end{tikzpicture}}^{a_{10}},
\end{equation}
\begin{equation}\label{eq:appendix.diag.fusion.2}
	\begin{tikzpicture}[scale = 1/3,baseline={(current bounding box.center)}] 
	\draw[black, line width = 1pt] (1,1) .. controls (1,2) and (2,2) .. (2,1);
	\draw[black, line width = 1pt] (3,1) .. controls (3,2.2) and (0,2.2) .. (0,1);
	\draw[black, line width = 1pt] (4,1) .. controls (4,2) and (5,2) .. (5,1);
	\draw[black, line width = 1pt] (1,3) .. controls (1,2) and (2,2) .. (2,3);
	\draw[black, line width = 1pt] (3,3) .. controls (3,2) and (4,2) .. (4,3);
	\draw[black, line width = 2pt] (.5,1) -- (4.5,1);
	\draw[black, line width = 2pt] (.5,3) -- (4.5,3);
	\filldraw[white] (-.5,.5) -- (.5,.5) -- (.5,3.5) -- (-.5,3.5) -- (-.5,.5);
	\filldraw[white] (5-.5,.5) -- (5.5,.5) -- (5.5,3.5) -- (5-.5,3.5) -- (5-.5,.5);
	\end{tikzpicture}, 
	\begin{tikzpicture}[scale = 1/3,baseline={(current bounding box.center)}] 
	\draw[black, line width = 1pt] (1,1) .. controls (1,2) and (2,2) .. (2,1);
	\draw[black, line width = 1pt] (3,1) .. controls (3,2.2) and (0,2.2) .. (0,1);
	\draw[black, line width = 1pt] (4,1) .. controls (4,2) and (5,2) .. (5,1);
	\draw[black, line width = 1pt] (1,3) .. controls (1,2) and (0,2) .. (0,3);
	\draw[black, line width = 1pt] (5,3) .. controls (5,2) and (4,2) .. (4,3);
	\draw[black, line width = 1pt] (3,3) .. controls (3,2) and (2,2) .. (2,3);
	\draw[black, line width = 2pt] (.5,1) -- (4.5,1);
	\draw[black, line width = 2pt] (.5,3) -- (4.5,3);
	\filldraw[white] (-.5,.5) -- (.5,.5) -- (.5,3.5) -- (-.5,3.5) -- (-.5,.5);
	\filldraw[white] (5-.5,.5) -- (5.5,.5) -- (5.5,3.5) -- (5-.5,3.5) -- (5-.5,.5);
	\end{tikzpicture},
	\begin{tikzpicture}[scale = 1/3,baseline={(current bounding box.center)}] 
	\draw[black, line width = 1pt] (1,1) .. controls (1,2) and (2,2) .. (2,1);
	\draw[black, line width = 1pt] (3,1) .. controls (3,2.1) and (0,2.1) .. (0,1);
	\draw[black, line width = 1pt] (4,1) .. controls (4,2) and (5,2) .. (5,1);
	\draw[black, line width = 1pt] (1,3) .. controls (1,1.75) and (4,1.75) .. (4,3);
	\draw[black, line width = 1pt] (2,3) .. controls (2,2) and (3,2) .. (3,3);
	\draw[black, line width = 2pt] (.5,1) -- (4.5,1);
	\draw[black, line width = 2pt] (.5,3) -- (4.5,3);
	\filldraw[white] (-.5,.5) -- (.5,.5) -- (.5,3.5) -- (-.5,3.5) -- (-.5,.5);
	\filldraw[white] (5-.5,.5) -- (5.5,.5) -- (5.5,3.5) -- (5-.5,3.5) -- (5-.5,.5);
	\end{tikzpicture},
	\begin{tikzpicture}[scale = 1/3,baseline={(current bounding box.center)}] 
	\draw[black, line width = 1pt] (1,1) .. controls (1,2) and (2,2) .. (2,1);
	\draw[black, line width = 1pt] (3,1) .. controls (3,2.2) and (0,2.2) .. (0,1);
	\draw[black, line width = 1pt] (4,1) .. controls (4,2) and (5,2) .. (5,1);
	\draw[black, line width = 1pt] (1,3) .. controls (1,2) and (0,2) .. (0,3);
	\draw[black, line width = 1pt] (2,3) .. controls (2,1.5) and (5,1.5) .. (5,3);
	\draw[black, line width = 1pt] (3,3) .. controls (3,2) and (4,2) .. (4,3);
	\draw[black, line width = 2pt] (.5,1) -- (4.5,1);
	\draw[black, line width = 2pt] (.5,3) -- (4.5,3);
	\filldraw[white] (-.5,.5) -- (.5,.5) -- (.5,3.5) -- (-.5,3.5) -- (-.5,.5);
	\filldraw[white] (5-.5,.5) -- (5.5,.5) -- (5.5,3.5) -- (5-.5,3.5) -- (5-.5,.5);
	\end{tikzpicture},
	\begin{tikzpicture}[scale = 1/3,baseline={(current bounding box.center)}] 
	\clip (-.5,.5) -- (5.5, .5) -- (5.5,3.5) -- (-.5,3.5) -- (-.5,.5);
	\draw[black, line width = 1pt] (1,1) .. controls (1,2) and (2,2) .. (2,1);
	\draw[black, line width = 1pt] (3,1) .. controls (3,2.2) and (0,2.2) .. (0,1);
	\draw[black, line width = 1pt] (4,1) .. controls (4,2) and (5,2) .. (5,1);
	\draw[black, line width = 1pt] (1,3) .. controls (1,2) and (0,2) .. (0,3);
	\draw[black, line width = 1pt] (2,3) .. controls (2,1.85) and (-1,1.85) .. (-1,3);
	\draw[black, line width = 1pt] (3,3) .. controls (3,1.5) and (6,1.5) .. (6,3);
	\draw[black, line width = 1pt] (4,3) .. controls (4,2) and (5,2) .. (5,3);
	\draw[black, line width = 2pt] (.5,1) -- (4.5,1);
	\draw[black, line width = 2pt] (.5,3) -- (4.5,3);
	\filldraw[white] (-1-.5,.5) -- (.5,.5) -- (.5,3.5) -- (-1-.5,3.5) -- (-1-.5,.5);
	\filldraw[white] (5-.5,.5) -- (7.5,.5) -- (7.5,3.5) -- (5-.5,3.5) -- (5-.5,.5);
	\end{tikzpicture},
	\begin{tikzpicture}[scale = 1/3,baseline={(current bounding box.center)}] 
	\clip (-.5,.5) -- (5.5, .5) -- (5.5,3.5) -- (-.5,3.5) -- (-.5,.5);
	\draw[black, line width = 1pt] (1,1) .. controls (1,1.85) and (2,1.85) .. (2,1);
	\draw[black, line width = 1pt] (3,1) .. controls (3,2.15) and (0,2.15) .. (0,1);
	\draw[black, line width = 1pt] (4,1) .. controls (4,2) and (5,2) .. (5,1);
	\draw[black, line width = 1pt] (1,3) .. controls (1,2.25) and (2,2.25) .. (2,3);
	\draw[black, line width = 1pt] (3,3) .. controls (3,1.85) and (0,1.85) .. (0,3);
	\draw[black, line width = 1pt] (4,3) .. controls (4,2) and (5,2) .. (5,3);
	\draw[black, line width = 2pt] (.5,1) -- (4.5,1);
	\draw[black, line width = 2pt] (.5,3) -- (4.5,3);
	\filldraw[white] (-.5,.5) -- (.5,.5) -- (.5,3.5) -- (-.5,3.5) -- (-.5,.5);
	\filldraw[white] (5-.5,.5) -- (5.5,.5) -- (5.5,3.5) -- (5-.5,3.5) -- (5-.5,.5);
	\end{tikzpicture}.
\end{equation}
However, we then have
\begin{equation}
	\begin{tikzpicture}[scale = 1/3,baseline={(current bounding box.center)}] 
	\draw[black, line width = 1pt] (1,1) .. controls (1,2) and (2,2) .. (2,1);
	\draw[black, line width = 1pt] (3,1) .. controls (3,2.2) and (0,2.2) .. (0,1);
	\draw[black, line width = 1pt] (4,1) .. controls (4,2) and (5,2) .. (5,1);
	\draw[black, line width = 1pt] (1,3) .. controls (1,2) and (2,2) .. (2,3);
	\draw[black, line width = 1pt] (3,3) .. controls (3,2) and (4,2) .. (4,3);
	\draw[black, line width = 1pt] (1,-1) .. controls (1,0) and (2,0) .. (2,-1);
	\draw[black, line width = 1pt] (3,-1) .. controls (3,0) and (5,0) .. (5,-1);
	\draw[white, line width = 4pt] (4,-1) -- (4,1);
	\draw[black, line width = 1pt] (4,-1) -- (4,1);
	\draw[black, line width = 1pt] (1,1) .. controls (1,0) and (0,0) .. (0,1);
	\draw[black, line width = 1pt] (2,1) .. controls (2,0) and (3,0) .. (3,1);
	\draw[black, line width = 2pt] (.5,-1) -- (4.5,-1);
	\draw[black, line width = 2pt] (.5,1) -- (4.5,1);
	\draw[black, line width = 2pt] (.5,3) -- (4.5,3);
	\filldraw[white] (-.5,-1.5) -- (.5,-1.5) -- (.5,3.5) -- (-.5,3.5) -- (-.5,-1.5);
	\filldraw[white] (5-.5,-1.5) -- (5.5,-1.5) -- (5.5,3.5) -- (5-.5,3.5) -- (5-.5,-1.5);
	\end{tikzpicture} = - (-\q)^{3/2} 
	\begin{tikzpicture}[scale = 1/3,baseline={(current bounding box.center)}] 
	\draw[black, line width = 1pt] (1,1) .. controls (1,2) and (2,2) .. (2,1);
	\draw[black, line width = 1pt] (3,1) .. controls (3,2) and (4,2) .. (4,1);
	\draw[black, line width = 1pt] (1,3) .. controls (1,2) and (2,2) .. (2,3);
	\draw[black, line width = 1pt] (3,3) .. controls (3,2) and (4,2) .. (4,3);
	\draw[black, line width = 2pt] (.5,1) -- (4.5,1);
	\draw[black, line width = 2pt] (.5,3) -- (4.5,3);
	\filldraw[white] (-.5,.5) -- (.5,.5) -- (.5,3.5) -- (-.5,3.5) -- (-.5,.5);
	\filldraw[white] (5-.5,.5) -- (5.5,.5) -- (5.5,3.5) -- (5-.5,3.5) -- (5-.5,.5);
	\end{tikzpicture},
\end{equation}
where we used the closed braid identity
\begin{equation}
\begin{tikzpicture}[scale = 1/3,baseline={(current bounding box.center)},yscale = -1]
	\draw[black, line width = 1pt] (0,0) -- (2,2);
	\draw[white, line width = 3pt] (0,2) -- (2,0);
	\draw[black, line width = 1pt] (0,2) -- (2,0);	
	\draw[black, line width = 1pt] (2,0) .. controls (3,0) and (3,2) .. (2,2);
\end{tikzpicture} \; = \; -(-\q)^{3/2} \;
\begin{tikzpicture}[scale = 1/3,baseline={(current bounding box.center)}]
	\draw[black, line width = 1pt] (2,0) .. controls (3,0) and (3,2) .. (2,2);
\end{tikzpicture} \;.
\end{equation}
Using equation \eqref{eq:appendix.fusion.twist}, it follows that
\begin{equation}
	\begin{tikzpicture}[scale = 1/3,baseline={(current bounding box.center)}] 
	\draw[black, line width = 1pt] (1,1) .. controls (1,2) and (2,2) .. (2,1);
	\draw[black, line width = 1pt] (3,1) .. controls (3,2.2) and (0,2.2) .. (0,1);
	\draw[black, line width = 1pt] (4,1) .. controls (4,2) and (5,2) .. (5,1);
	\draw[black, line width = 1pt] (1,3) .. controls (1,2) and (2,2) .. (2,3);
	\draw[black, line width = 1pt] (3,3) .. controls (3,2) and (4,2) .. (4,3);
	\draw[black, line width = 2pt] (.5,1) -- (4.5,1);
	\draw[black, line width = 2pt] (.5,3) -- (4.5,3);
	\filldraw[white] (-.5,.5) -- (.5,.5) -- (.5,3.5) -- (-.5,3.5) -- (-.5,.5);
	\filldraw[white] (5-.5,.5) -- (5.5,.5) -- (5.5,3.5) -- (5-.5,3.5) -- (5-.5,.5);
	\end{tikzpicture} y = - z(-\q)^{3/2} 
	\begin{tikzpicture}[scale = 1/3,baseline={(current bounding box.center)}] 
	\draw[black, line width = 1pt] (1,1) .. controls (1,2) and (2,2) .. (2,1);
	\draw[black, line width = 1pt] (3,1) .. controls (3,2) and (4,2) .. (4,1);
	\draw[black, line width = 1pt] (1,3) .. controls (1,2) and (2,2) .. (2,3);
	\draw[black, line width = 1pt] (3,3) .. controls (3,2) and (4,2) .. (4,3);
	\draw[black, line width = 2pt] (.5,1) -- (4.5,1);
	\draw[black, line width = 2pt] (.5,3) -- (4.5,3);
	\filldraw[white] (-.5,.5) -- (.5,.5) -- (.5,3.5) -- (-.5,3.5) -- (-.5,.5);
	\filldraw[white] (5-.5,.5) -- (5.5,.5) -- (5.5,3.5) -- (5-.5,3.5) -- (5-.5,.5);
	\end{tikzpicture} y.
\end{equation}
Applying the same reasoning to the other diagrams in \eqref{eq:appendix.diag.fusion.2}, one concludes that $M_0$ is generated entirely by elements of the form $a y$ with $a$ one of the diagrams in \eqref{eq:appendix.diag.fusion.1}. Furthermore, these elements are clearly linearly independent since the bottom side of the diagrams in \eqref{eq:appendix.diag.fusion.1} are all identical, and any relation of the form $a y = a b y$ for some $b \in \atl{4}$ cannot affect the top side of the diagram $a$, since it acts on its bottom (provided that $a$ has no through line, which is the case here). We thus conclude that this set of elements spans $M_0$.

We now consider $\bar{M}_{2}$, the subspace of $\mathsf{W}_{1/2,z}(3) \times^{o}_{f} \mathsf{S}_{1/2}(1)$ generated by elements of the form $a y$, where $a$ is a diagram with exactly two through lines, and such that $ae_1 = \beta a$. Such diagrams are all those which can be written in the form $u^{k} e_{j_m} e_{j_{m-1}} \hdots e_{j_{0}} $, where $k \in \mathbb{Z}$ and $\lbrace j_{i} \rbrace $ is a set of $m \in \mathbb{N}$ consecutive integers such that $j_0 = 1$. 

However, from \eqref{eq:appendix.fusion.twist},
\begin{align}
	e_1 y & = z^{-1}e_1\left((-\q)^{-1/2}u e_1 + (-\q)^{1/2} u e_1e_3 \right) y,\\
	& = z^{-1}(-\q)^{-1/2} u e_4 e_1 y + a_0y,
\end{align}
where $a_0$ is a diagram with zero through lines (so $a_0y \in M_0$). It thus follows that
\begin{equation}
u^{-1} e_1 y = w^{-1} e_4e_1 y + b_1 y,
\end{equation}
where $w = z(-\q)^{1/2}$ and $b_1$ is a diagram with no through lines. Repeating this argument, we find
\begin{equation}
u^{-2} e_1 y = w^{-2} e_3 e_4 e_1 y + b_2 y,
\end{equation}
\begin{equation}
u^{-3} e_1 y = w^{-3} e_2 e_3 e_4 e_1 y + b_3 y,
\end{equation}
\begin{equation}
u^{-4} e_1 y = w^{-4} e_1 e_2 e_3 e_4 e_1 y + b_4 y,
\end{equation}
where $b_j$, $j = 1, \hdots , 4$ are diagrams with no through lines. Note that from the algebra relations \eqref{eq:period-gen-1} one has $e_1 e_2 e_3 = u^2 e_3 $, so that
\begin{align}
	u^{-4} e_1 y & = w^{-4} u^2 e_3 e_4 e_1 y + b_4 y, \\
	& = w^{-2} e_1 y + b_4 y - w^{-2} b_2 y.
\end{align}
It follows that $M_{2}$, the quotient of $\mathsf{W}_{1/2,z}(3) \times^{o}_{f} \mathsf{S}_{1/2}(1)$ by the submodule $M_0$, is spanned by the set of equivalence classes $\lbrace u^{-k} e_1 ( y + M_0)| k = 0,1,2,3 \rbrace. $

We thus conclude that a basis of $\mathsf{W}_{1/2,z}(3) \times^{o}_{f} \mathsf{S}_{1/2}(1)$ can be chosen as $\lbrace a_i y| i = 5, \hdots, 10 \rbrace \cup \lbrace u^{-k} e_1 y| k = 0, \hdots, 3 \rbrace $, where  the diagrams $a_i$ are given in equation \eqref{eq:appendix.diag.fusion.1}. Note that it is often more convenient to work with the set $ e_1y, e_2e_1y, \hdots $ instead of $e_1 y, u^{-1} e_1 y, \hdots$ as the resulting diagrams are easier to draw.

\end{document}